\documentclass[fleqn,usenatbib]{mnras}

\usepackage{newtxtext,newtxmath}

\usepackage[T1]{fontenc}

\DeclareRobustCommand{\VAN}[3]{#2}
\let\VANthebibliography\thebibliography
\def\thebibliography{\DeclareRobustCommand{\VAN}[3]{##3}\VANthebibliography}

\usepackage{graphicx}
\usepackage{array}
\usepackage{amsmath}	
\usepackage{mathtools}
\usepackage{gensymb}
\usepackage{hhline}
\usepackage{float}
\usepackage{longtable,booktabs,etoolbox,multirow,tabularx}
\usepackage{CJK}
\usepackage[whole]{bxcjkjatype} 
\usepackage{CJKutf8}

\title[SN\,2018evt]{The Interaction of Supernova 2018evt with a Substantial Amount of Circumstellar Matter --- An SN\,1997cy-like Event}

\author[Y. Yang et al.]{Yi Yang\begin{CJK*}{UTF8}{gbsn}
(杨轶)
\end{CJK*}$^{1,2\thanks{E-mail: yi.yang@berkeley.edu}}$,
Dietrich Baade$^{3}$, 
Peter Hoeflich$^{4}$, 
Lifan Wang$^{5}$, 
Aleksandar Cikota$^{6}$, 
Ting-Wan Chen$^{7,8}$, \newauthor
Jamison Burke$^{9,10}$,
Daichi Hiramatsu$^{11,12,9,10}$,
Craig Pellegrino$^{9,10}$,
D. Andrew Howell$^{9,10}$, 
Curtis McCully$^{9}$, \newauthor
Stefano Valenti$^{13}$, 
Steve Schulze$^{14,2}$, 
Avishay Gal-Yam$^{2}$, 
Lingzhi Wang$^{15,16}$, 
Alexei V. Filippenko$^{1,17}$, \newauthor
Keiichi Maeda$^{18}$, 
Mattia Bulla$^{8}$, 
Yuhan Yao$^{19}$, 
Justyn R. Maund,$^{20}$, 
Ferdinando Patat$^{3}$, 
Jason Spyromilio$^{3}$, \newauthor
J. Craig Wheeler$^{21}$, 
Arne Rau$^{7}$, 
Lei Hu$^{22}$, 
Wenxiong Li$^{23}$, 
Jennifer.~E. Andrews$^{24,25}$, 
Ll\'{u}is Galbany$^{26, 27}$, \newauthor
David.~J. Sand$^{24}$, 
Melissa Shahbandeh$^{4}$, 
Eric Y. Hsiao$^{4}$, and
Xiaofeng Wang$^{28,29}$ \\
\\
$^{1}$Department of Astronomy, University of California, 
Berkeley, CA 94720-3411, USA \\
$^{2}$Department of Particle Physics and Astrophysics, 
Weizmann Institute of Science, Rehovot 76100, Israel \\
$^{3}$European Organisation for Astronomical Research in the Southern Hemisphere (ESO), Karl-Schwarzschild-Str. 2, 85748 Garching b. M{\"u}nchen, Germany \\
$^{4}$Department of Physics, Florida State University, Tallahassee, Florida 32306-4350, USA \\
$^{5}$George P. and Cynthia Woods Mitchell Institute for Fundamental Physics $\&$ Astronomy, Texas A.$\&$M. University, 4242 TAMU, College Station, TX 77843, USA \\
$^{6}$European Organisation for Astronomical Research in the Southern Hemisphere (ESO), Alonso de Cordova 3107, Vitacura, Casilla 19001, Santiago de Chile, Chile \\
$^{7}$Max-Planck-Institut f{\"u}r Extraterrestrische Physik, Giessenbachstra\ss e 1, 85748, Garching, Germany \\
$^{8}$The Oskar Klein Centre, Department of Astronomy, Stockholm University, AlbaNova, SE-10691 Stockholm, Sweden \\
$^{9}$Las Cumbres Observatory, 6740 Cortona Drive, Suite 102, Goleta, CA 93117-5575, USA \\
$^{10}$Department of Physics, University of California, Santa Barbara, CA 93106-9530, USA \\
$^{11}$Center for Astrophysics \textbar{} Harvard \& Smithsonian, 60 Garden Street, Cambridge, MA 02138-1516, USA \\
$^{12}$The NSF AI Institute for Artificial Intelligence and Fundamental Interactions \\
$^{13}$Department of Physics, University of California, Davis, CA 95616, USA \\
$^{14}$The Oskar Klein Centre, Physics Department, Stockholm University, Albanova University Center, SE 106 91 Stockholm, Sweden \\
$^{15}$Chinese Academy of Sciences, South America Center for Astronomy, National Astronomical Observatories, CAS, Beijing 100101, People's Republic of China \\
$^{16}$CAS Key Laboratory of Optical Astronomy, National Astronomical Observatories, Chinese Academy of Sciences, Beijing 100101, People's Republic of China \\
$^{17}$Miller Institute for Basic Research in Science, University of California, Berkeley, CA 94720, USA \\
$^{18}$Department of Astronomy, Kyoto University, Kitashirakawa-Oiwake-cho, Sakyo-ku, Kyoto 606-8502, Japan \\
$^{19}$Cahill Center for Astrophysics, California Institute of Technology, MC 249-17, 1200 E. California Boulevard, Pasadena, CA 91125, USA \\
$^{20}$Department of Physics and Astronomy, University of Sheffield, Hicks Building, Hounsfield Road, Sheffield S3 7RH, UK \\
Affiliations continue at the end of the paper
}

\date{Accepted XXX. Received YYY; in original form ZZZ}

\pubyear{2021}

\begin{document}
\label{firstpage}
\pagerange{\pageref{firstpage}--\pageref{lastpage}}
\maketitle

\begin{abstract}
A rare class of supernovae (SNe) is characterized by strong  interaction 
between the ejecta and several solar masses of circumstellar matter (CSM) as 
evidenced by strong Balmer-line emission. Within the first few weeks after the 
explosion, they may display spectral features similar to overluminous Type Ia 
SNe, while at later phase their observation properties exhibit remarkable 
similarities with some extreme case of Type IIn SNe that show strong Balmer 
lines years after the explosion.
We present polarimetric observations of 
SN\,2018evt obtained by the ESO Very Large Telescope from 172 to 219 days 
after the estimated time of peak luminosity to study the geometry of the 
CSM. The nonzero continuum polarization decreases over time, suggesting 
that the mass loss of the progenitor star is aspherical. The prominent 
H$\alpha$ emission can be decomposed into a broad, time-evolving component 
and an intermediate-width, static component. The former shows polarized 
signals, and it is likely to arise from a cold dense shell (CDS) within the 
region between the forward and reverse shocks. The latter is significantly 
unpolarized, and it is likely to arise from 
\textcolor{black}{shocked, fragmented gas clouds in the H-rich CSM.} 
We infer that SN\,2018evt 
exploded inside a massive and aspherical circumstellar cloud. The symmetry 
axes of the CSM and the SN appear to be similar. 
\textcolor{black}{SN\,2018evt shows observational properties common to events that display strong interaction between the ejecta and CSM, implying 
that they share similar circumstellar configurations.} Our preliminary 
estimate also suggests that the circumstellar environment of SN\,2018evt has 
been significantly enriched at a rate of $\sim 0.1$\,M$_\odot$\,yr$^{-1}$ 
over a period of $> 100$\,yr. 
\end{abstract}




\section{Introduction} \label{sec_intro}
Type Ia supernovae (SNe\,Ia) originate from an exploding white dwarf (WD) 
after mass transfer from a donor star (see, e.g., \citealp{Hoyle_etal_1960, 
Nomoto_etal_1997, Howell_2011, Hillebrandt_etal_2013, Maoz_etal_2014, 
Branch_Wheeler_2017, Hoeflich_2017} for reviews). The threshold in mass 
for the explosion may be reached by accretion from a non-WD companion star 
(single-degenerate channel [SD]; \citealp{Whelan_Iben_1973}) or by the 
merger of two degenerate objects (double-degenerate channel 
[DD]; \citealp{Iben_Tutukov_1984, Webbink_1984}). Direct, head-on 
collisions of two WDs in triple systems provide another possibility for 
triggering SNe\,Ia \citep{Katz_etal_2012}. Two-dimensional high-resolution 
hydrodynamical simulations also show that such a shock-ignition process is 
able to reproduce the major observational 
properties \citep{Kushnir_etal_2013}. 

Optical spectra of SNe\,Ia are typically characterised by the absence of 
hydrogen and the presence of intermediate-mass elements ($9 \leq Z \leq 20$) 
such as silicon and sulfur in the first weeks after the explosion (see, 
e.g., \citealp{Filippenko_1997} for a review). Except for very few historical 
Galactic transients (e.g., Tycho SN\,1572, \citealp{Rest_etal_2008b}; Kepler 
SN\,1604, \citealp{Kerzendorf_etal_2014_kepler}), the extragalactic nature of 
SNe\,Ia hinders any direct identification of their progenitor systems. 
Because the environment of an SN can provide an archive of the evolution of 
its progenitor system, substantial effort has gone into searching for 
circumstellar matter (CSM) to help discriminate between different models. 
Most SNe\,Ia reveal no evidence of CSM as predicted for the DD channel 
(however, see \citealp{Shen_etal_2013} for possible CSM enrichment when a He 
WD surrounded by an H-rich layer interacts with a C/O WD companion). Detailed 
observations of the nearby Type Ia SN\,2011fe \citep{Li_etal_2011} have 
excluded a luminous red-giant companion and concluded that the companion of 
the exploding WD is a compact object consistent with a 
WD \citep{Nugent_etal_2011}. Both circumstances have been used to infer a DD 
origin for these SNe. 

Efforts to search the CSM around normal SNe\,Ia have detected some evidence 
of the presence of moderate amount of circumstellar dust for some events 
(see, e.g., \citealp{Patat_etal_2007, Wang_etal_2008, Wang_etal_2019, Yang_etal_2018_pol}). An extreme case of SN\,2002ic has established a new 
variety of SNe\,Ia that explode inside a dense circumstellar 
envelope \citep{Hamuy_etal_2003, Deng_etal_2004, Kotak_etal_2004, 
Wood-Vasey_etal_2004, Wang_etal_2004}. Such a configuration is demonstrated 
by the presence of strong Balmer emission lines and X-ray 
emission \citep{Bochenek_etal_2018}, and these objects are overluminous by 
a factor of $\sim 100$ compared with normal SNe\,Ia several months after the 
explosion. Often, the initially narrow H$\alpha$ line dramatically broadens 
and also strengthens in the first $\sim$100--150 
days \citep{Dilday_etal_2012, Silverman_etal_2013}. 
Modeling of the late-time spectroscopic evolution of such events
shows that a few solar masses (M$_{\odot}$) of CSM are involved in the 
emission processes \citep{Chugai_Yungelson_2004, Fox_etal_2015, 
Inserra_etal_2016}. As far as we know, no such event has ever been detected 
at radio wavelengths. The first detection of X-ray emission from a 
strongly-interacting SNmight be the case of SN\,2012ca, which clearly 
indicates an interaction between the explosion ejecta and dense 
CSM \citep{Bochenek_etal_2018}. Although interaction has also been suggested 
by optical observations \citep{Inserra_etal_2014, Fox_etal_2015,
Inserra_etal_2016}, the data favour the interpretation of SN\,2012ca as an 
SN\,IIn triggered by core collapse of a massive star rather than an 
a thermonuclear explosion \citep{Inserra_etal_2016}. An excess of infrared 
emission has been observed in SNe\,2012ca and 2013dn, suggesting the presence 
of circumstellar dust \citep{Szalai_etal_2019}. 

At early phases, the spectra of such strongly-interacting SNe show 
similarities to the spectra of SN\,1991T-like events, a subclass characterised 
by overluminous and slowly declining light curves, strong \ion{Fe}{{\sc III}} 
absorption, and weak or no \ion{Ca}{{\sc II}} and \ion{Si}{{\sc II}} absorption 
around one week after the 
explosion \citep{Filippenko_etal_1992_91T, Phillips_etal_1992}. 
Due to such spectroscopic similarities at early phases, these events are often 
denoted as `Type Ia-CSM SNe' in some literature. 
In their spectra, Balmer emission lines can be identified at early 
phases. They start to dominate the spectra after $\sim 2$ weeks past maximum, 
suggesting that SN-CSM interaction contributes more flux than the radioactive 
decay of Ni$^{56}$ and Co$^{56}$ (\citealp{Hamuy_etal_2003, 
Silverman_etal_2013, Fox_etal_2015}). The spectral features of SNe\,Ia-CSM 
exhibit a resemblance to those of SNe\,IIn, in which the Balmer lines are 
considered to arise from ionised CSM previously expelled by the massive 
progenitors of core-collapse SNe. A systematic search for SNe\,Ia-CSM among 
the spectra of 226 SNe\,IIn suggests that $\sim 11$\% of Type IIn events have 
observational signatures similar to those of the Type Ia-CSM 
SN\,2002ic \citep{Silverman_etal_2013}. However, apart from SN\,2002ic, only 
very few SNe~Ia-CSM have been studied in detail: 
SNe\,1997cy \citep{Turatto_etal_2000, Germany_etal_2000}, 
1999E \citep{Rigon_etal_2003}, 2005gj \citep{Aldering_etal_2006, 
Prieto_etal_2007, Silverman_etal_2013}, PTF11kx \citep{Dilday_etal_2012, 
Silverman_etal_2013_PTF11kx, Graham_etal_2017}, 2013dn \citep{Fox_etal_2015}, 
and 2015cp \citep{Graham_etal_2019}.

In previous literature, classifications of SNe\,Ia-CSM are generally based on 
similarities of their late-time spectra to those of previous events 
(e.g., \citealp{Silverman_etal_2013}). It remains to be seen if all such events 
are of thermonuclear origin. On the one hand, early-time spectral sequences 
of some SNe\,Ia-CSM exhibit a striking resemblance to those of thermonuclear 
SNe without evidence of circumstellar interaction (e.g., 
PTF\,11kx; \citealp{Dilday_etal_2012}. A near-ultraviolet (NUV) survey with 
the {\it Hubble Space Telescope (HST)} designed to search for the UV signals 
of SN\,Ia ejecta-CSM interaction identified only one such case, namely 
SN\,2015cp at day 664 \citep{Graham_etal_2019}. This SN has also been 
classified as an overluminous SN\,1991T-like object. On the other hand, an 
SN\,Ic embedded in a gas-rich environment might also account for the 
observational features of SN\,2002ic \citep{Benetti_etal_2006}. Support for 
a nonthermonuclear nature of SNe\,Ia-CSM could also be derived from the 
agreement between the mass-loss profiles of SN\,2005gj and luminous blue 
variables \citep[LBVs;][]{Trundle_etal_2008}. A large energy budget and/or 
high kinetic-luminosity conversion efficiency are additionally 
required \citep{Inserra_etal_2016}. 

Owing to the late-time spectral similarities between SNe\,Ia-CSM and 
core-collapse SNe showing prominent ejecta-CSM interaction (Type IIn), the 
two populations are likely contaminated by each 
other \citep{Silverman_etal_2013, Inserra_etal_2016, Leloudas_etal_2015}. 
More effort is required to unveil the progenitor systems that lead to SN 
explosions within substantial CSM.  This is not helped by the rarity of 
SNe\,Ia-CSM and the scarcity of high-quality datasets. It is remarkable 
that even though a substantial amount of H-rich CSM is involved in the 
interaction with the SN ejecta, the mechanism for establishing such a 
circumstellar environment still remains unclear. The most widely accepted 
single- or double-degenerate models do not predict such large amounts of 
CSM (i.e., $\lesssim 0.03\,{\rm M}_{\odot}$; \citealp{Lundqvist_etal_2013}).

Outside of mainstream models, several M$_{\odot}$ of H may correspond to the 
integrated mass loss from a massive (3--7\,M$_{\odot}$) asymptotic giant 
branch (AGB) star before the SN explosion. The binary scenario of a C/O WD 
merging with the C/O core of a red supergiant has been suggested 
by \citet{Hamuy_etal_2003} to explain the substantial CSM in SN\,2002ic, but 
it does not provide a clear explanation for the origin of such strong mass 
loss just prior to the SN explosion \citep{Chugai_Yungelson_2004}. An 
alternative interpretation is suggested by the single-star scenario. For 
some initially massive AGB stars ($\lesssim 8$\,$M_{\odot}$), mass loss may 
not reduce the mass of the star below the Chandrasekhar mass limit 
($M_{\rm Ch} \approx 1.4$\,M$_{\odot}$) before carbon ignites in the core. 
The high energy needed to lift the degeneracy in the core will trigger a 
thermonuclear explosion \citep{Iben_etal_1983}. The designation ``Type 
I$\frac{1}{2}$\,SN'' is derived from the simultaneous resemblance of such a 
model to SNe~Ia, in which the explosion of the core liberates a substantial 
amount of radioactive Ni and Co, as well as to SNe\,IIn, with ionised H-rich 
environments from heavy pre-explosion mass loss. 

Besides the uncertain nature of the progenitor system, the origin of the 
enormous width of the Balmer emission lines is also unclear. They typically 
consist of a broad ($\sim 7000$\,km\,s$^{-1}$), an intermediate 
($\sim 2000$\,km\,s$^{-1}$), and a narrow ($\sim 100$\,km\,s$^{-1}$) 
component, they persist for a long time, and they dominate the late-time 
spectra. The narrow central core of the H$\alpha$ emission is mostly 
produced by the ionisation of H in the CSM by the SN photons. The 
intermediate-width wings of the H$\alpha$ profile can result from multiple 
scattering of photons in the narrow line by thermal electrons in optically 
thick circumstellar gas \citep{Chugai_etal_2001, Wang_etal_2004}. The 
parameter dependence of the line profiles, including the optical depth, 
density, and velocity profile of the circumstellar gas, was carefully 
investigated by \citet{Huang_etal_2018}. 
\textcolor{black}{
Alternatively, a broad velocity 
distribution may be caused by shear flows around radiatively shocked 
circumstellar clouds (see, e.g., \citealp{Chugai_etal_1994}, and a more detailed discussion by \citealp{Chugai_1997}).} 
In this case, the 
broadening of the H$\alpha$ profile would be brought about by recombination 
in the shocked CSM.

The pre-explosion mass-loss history of SNe\,Ia-CSM should be encoded in the 
geometry of the CSM: mass loss in a binary system is likely to develop a 
disk/ring-like profile, while an AGB wind from a single star would produce 
a (probably multiple) shell profile. In direct imaging of AGB stars by 
{\it HST} \citep{Morris_etal_2006} and ALMA \citep{Kim_etal_2017}, thin 
spiral patterns with multiple windings were found that probably result from 
thermal mass-loss pulses. Radiation from a relatively spherical structure 
is expected to show little to moderate polarization, while a more disk-like 
CSM geometry leads to a $\sim 10$\% continuum polarization. The continuum 
polarization is expected to be low if a disk/torus geometry is viewed 
face-on. 

SN\,2018evt stands out as one of the nearest events compared to the 
$\sim 25$ SNe\,Ia-CSM in the sample compiled by \citet{Silverman_etal_2013}. 
It was discovered at $V \approx 16.5$\,mag in Aug. 
2018 \citep{Nicholls_etal_2018}, in the outskirts of the sprial galaxy 
MCG-01-35-011 (redshift 
$z = 0.025352\pm0.000133$; \citealp{daCosta_etal_1998}). 

The classification spectrum is the only publicly available spectrum from 
the early phases of SN\,2018evt. It exhibits hybrid characteristics: narrow 
Balmer emission lines superimposed on an overluminous SN\,1991T-like 
spectrum \citep{Stein_etal_2018}. Direct follow-up observations were not 
possible since the SN was discovered as an evening-twilight object and soon
was too close to the Sun in the sky. In Dec. 2018, when SN\,2018evt was 
again observable, the brightness was still at a surprising 
$r \approx 16.4$\,mag (absolute magnitude 
$M_{r} \approx -18.8$\,mag; \citealp{Dong_etal_2018}). The relative strength 
of the H$\alpha$ emission had increased dramatically compared to the first 
spectrum obtained at an early phase \citep{Stein_etal_2018}. These two 
pieces of evidence suggest a violent interaction with the CSM, which 
efficiently converts kinetic energy of the ejecta into radiation. Both early 
and late spectra closely resemble those of the Type Ia-CSM 
SN\,2002ic \citep{Hamuy_etal_2003} and the possible Type Ia-CSM (though 
perhaps SN\,IIn) SN\,2012ca \citep{Inserra_etal_2014}. Owing to the ambiguity 
in the classification and the separation  between the classes of SNe\,Ia-CSM 
and SNe\,IIn, we refer to SN\,2018evt as an SN\,1997cy-like event throughout 
the paper.

This paper presents optical and near-infrared (NIR) photometry, as well as 
optical spectroscopy and spectropolarimetry, of SN\,2018evt. It is organised 
as follows. Observations and data reduction are outlined in 
Section~\ref{sec_obs}. Sections~\ref{sec_lc} and \ref{sec_spec} describe the 
photometric and spectroscopic evolution, respectively. The 
spectropolarimetric behaviour of the SN is investigated in 
Section~\ref{sec_specpol}. Section~\ref{sec_obs_summary} provides a summary 
of the major observational properties. Our discussion and final remarks are 
given in Sections~\ref{sec_discussion} and ~\ref{sec_summary}, respectively. 

\begin{figure*}
\includegraphics[width=1.0\linewidth]{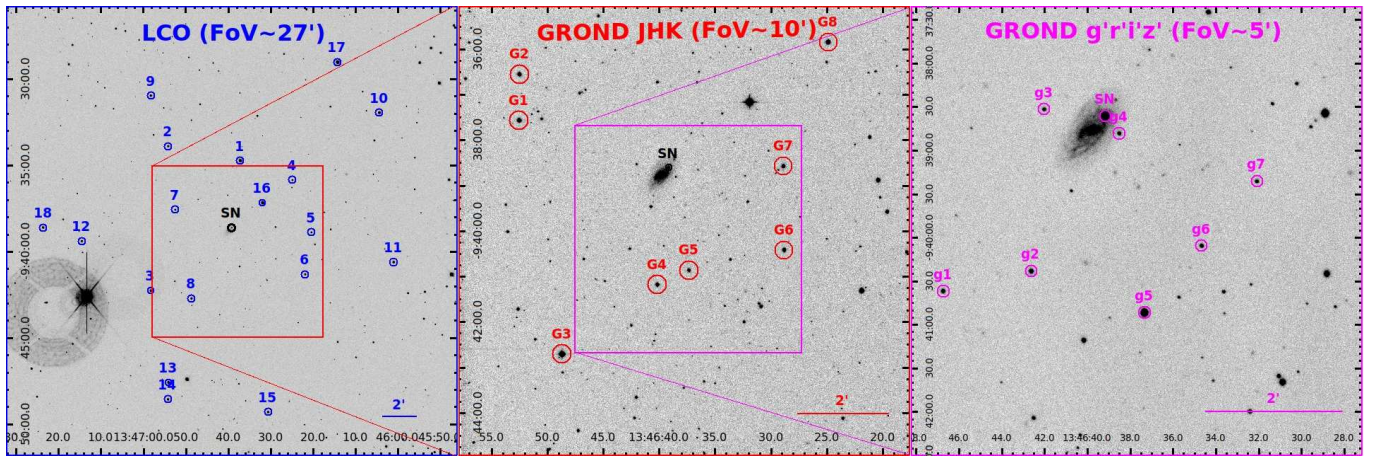}
\caption{
{\it Left:} LCO $V$-band image from 2019-06-30 showing the location of 
SN\,2018evt. Local photometric reference stars are marked with blue circles. 
The red square outlines the GROND $JHK$ field of view (FoV) shown in the 
middle panel. {\it Middle:} GROND $J$-band image obtained on 2019-01-10 
indicating the location of SN\,2018evt. Local photometric reference stars 
used for the $JHK$ bands are marked with red circles and labeled with prefix 
``G''. The magenta square shows the size of the GROND $g'r'i'z'$-band FoV. 
{\it Right:} GROND $g'$-band image of SN\,2018evt obtained on 2019-01-10. 
Local photometric reference stars used for the $g'r'i'z$ bands are marked 
with magenta circles and labeled with prefix ``g''. The FoV and the angular 
scale (2\arcmin) are also marked in each panel. North is up, east is to the 
left. 
\label{Fig_image}
}
\end{figure*} 

\section{Observations and Data Reduction} \label{sec_obs}
SN\,2018evt (ASASSN-18ro) was discovered by the All-Sky Automated Survey for 
Supernovae (ASAS-SN; \citealp{Shappee_etal_2014}) on 2018-08-11 (UT dates are 
used throughout this paper; MJD 58341.005) at $V \approx 16.5$\,mag (absolute 
magnitude $M_{V} \approx -18.7$\,mag; \citealp{Nicholls_etal_2018}). 
Follow-up spectroscopy was obtained by the extended Public ESO Spectroscopic 
Survey for Transient Objects (ePESSTO; \citealp{Smartt_etal_2015}) with the 
New Technology Telescope (NTT) + ESO Faint Object Spectrograph and Camera 2 
(EFOSC2; \citealp{Buzzoni_etal_1984}) on 2018-08-12 23:59 (MJD 
58343.000;  \citealp{Stein_etal_2018}). Cross-correlation with a library of 
SN spectra using the ``Supernova Identification code'' 
(SNID; \citealp{Blondin_etal_2007}) suggests that the spectrum matches 
SN\,1991T-like templates at $-9$ days relative to the $B$-band maximum. 
Because of the lack of early-time data, we adopt a peak-light epoch at MJD 
58352 based on the best match from SNID but do not attempt to estimate the 
uncertainty. All phases are given relative to the roughly estimated $B$-band 
peak luminosity at MJD 58352 or 2018-08-22 (see Sec.~\ref{sec_lc}) 
throughout the paper. 

Astrometric measurements on the images obtained by the Sinistro cameras of 
the Las Cumbres Observatory (LCO) global network of 1\,m telescopes (see 
Sec.~\ref{sec_obs_lco}) has been derived by using 
Astrometry.net\footnote{\url{http://astrometry.net/}} \citep{Lang_etal_2010}. 
The world coordinate system (WCS) was solved for each frame and calibrated to 
the {\it GAIA} DR2 catalog \citep{Gaia_2016, Gaia_2018}. We chose a total of 
six exposures obtained by LCO in $g'r'i'$ under very good conditions on 
2019-05-09 (MJD 58612) to calculate the centroids of the SN and the host 
nucleus. We selected $\sim 100$ bright (signal-to-noise ratio [SNR] $>50$), 
isolated field stars within a $20\arcmin \times 20\arcmin$ box around the SN, 
cross-correlated their coordinates against the {\it GAIA} DR2 catalog, and 
deduced a median offset of 
$\rm \alpha^{LCO} - \alpha^{GAIA} = -0\farcs{110} \pm 0\farcs{164}$ and 
$\rm \delta^{LCO} - \delta^{GAIA} =  0\farcs{103} \pm 0\farcs{125}$.

Adopting the median value of the measurements obtained on the six frames and 
correcting for the offset between the LCO images and the {\it GAIA} DR2 
catalog, we estimate the position of SN\,2018evt as 
$\rm (\alpha, \delta)_{SN} = (13^{h}46^{m}39\farcs{181} \pm 0\farcs{003} \pm 0\farcs{164},  \ -09^{\circ}38'36\farcs{042} \pm 0\farcs{040} \pm 0\farcs{125}$).
The coordinates of the nucleus of the host spiral galaxy MCG-01-35-011 are 
$\rm (\alpha, \delta)_{Host} = (13^{h}46^{m}39\farcs{779} \pm 0\farcs{004} \pm 0\farcs{164}, \ -09^{\circ}38'45\farcs{641} \pm 0\farcs{070} \pm 0\farcs{125})$; 
see Fig.~\ref{Fig_image}. For each quantity, the first and the second 
uncertainties represent the errors due to filter-to-filter differences and 
the 1$\sigma$ deviation of the coordinate differences among the stars used in 
the cross-calibration, respectively. 

The heliocentric radial velocity of the host galaxy amounts to 
$7600 \pm 40$\,km\,s$^{-1}$ \citep{daCosta_etal_1998}. From the peak 
wavelength of the well-resolved narrow H$\alpha$ P~Cygni profile in the flux 
spectrum obtained with a higher spectral resolution, we deduce the redshift 
of SN\,2018evt to be $z = 0.02523 \pm 0.00015$ (for more details, see 
Sec.~\ref{sec_pcygni}). This value is consistent with the reported redshift 
of the host galaxy; it is used throughout the paper. We interpret the radial 
velocity as exclusively due to redshift and adopt a Hubble constant of 
H$_0 = 73.24 \pm 1.74$\,km\,s$^{-1}$\,Mpc$^{-1}$\citep{Riess_etal_2016}. We 
derive a distance for the host of $103.3 \pm 2.5$\,Mpc. We measure the angular 
separation between SN\,2018evt and the nucleus of its host as 13\farcs{0}, 
corresponding to a sky-projected separation of $6.5 \pm 0.2$\,kpc. 

\subsection{Optical and NIR Photometry}
\subsubsection{Las Cumbres Optical Photometry \label{sec_obs_lco}}
Extensive $Bg'Vr'i'$ photometry was acquired with the Sinistro cameras of the 
LCO network of 1\,m telescopes. The data were taken as part of the Global 
Supernova Project. The pixel size is $0\farcs{389}$ pixel$^{-1}$, and most of 
the measured full widths at half maximum (FWHMs) of the point-spread function 
(PSF) fall within the range of $1\farcs{4}$ to $2\farcs{5}$. The images were 
preprocessed, including bias subtraction and flat-field correction, with the 
BANZAI automatic pipeline \citep{McCully_etal_2018}. Figure~\ref{Fig_image} 
(left panel) shows the field around SN\,2018evt. For each frame, the PSF was 
determined from bright, isolated field stars and matched to the SN and local 
comparison stars. The PSF model fitting radius was chosen as the FWHM. Owing 
to the lack of template images obtained by LCO before the SN exploded, we 
estimate the galaxy contribution by fitting a median pixel value of an 
annulus around the SN with an inner radius of 4\farcs{0} and an outer radius 
of 5\farcs{5}. The background was determined and subtracted iteratively during 
the fitting of the PSF using the ALLSTAR task under the IRAF\footnote{{IRAF} 
is distributed by the National Optical Astronomy Observatories, which are 
operated by the Association of Universities for Research in Astronomy, Inc., 
under cooperative agreement with the National Science Foundation (NSF).} 
DAOPHOT package \citep{Stetson_1987}. The choice of the inner radius is 
justified by the fact that the residuals measured from the PSF-subtracted 
field stars were consistent with the noise beyond $\sim 1.5 \times$ FWHM. The 
small inner and outer radii and the iteration procedure provide a realistic 
estimate of the local background of the SN. Employing magnitudes of 15 local 
comparison stars from the AAVSO Photometric All Sky Survey (APASS) DR9 
Catalogue (\citealp{Henden_etal_2016}), we calibrated the instrumental $BV$ 
and $g'r'i'$ magnitudes of SN\,2018evt in the Johnson $BV$ 
system \citep{Johnson_1966} in Vega magnitudes and in the SDSS photometric 
system \citep{Fukugita_etal_1996} in AB magnitudes \citep{Oke_etal_1983}, 
respectively. The final $Bg'Vr'i'$-band calibrations were derived from the 
median of the difference between catalogue and instrumental magnitudes. The 
$Bg'Vr'i'$ comparison stars are identified in the left panel of 
Figure~\ref{Fig_image}. We list the photometry of SN\,2018evt in 
Table~\ref{Table_phot}. 

\subsubsection{GROND Optical and NIR Photometry~\label{sec_obs_grond}}
Simultaneous 7-band photometry in $g'r'i'z'JHK$ was obtained with the 
Gamma-Ray burst Optical/Near-infrared Detector 
(GROND; \citealp{Greiner_etal_2008}) mounted on the 2.2\,m MPG/ESO telescope 
at the La Silla Observatory (Chile). The plate scales of the GROND optical 
($g'r'i'z$) and NIR ($JHK$) images are $0\farcs{389}$ pixel$^{-1}$ and 
$0\farcs{60}$ pixel$^{-1}$, respectively. In the NIR, the field of view (FoV) 
of $10' \times 10'$ is imaged onto a $1024 \times 1024$ pixeo Rockwell 
HAWAII-1 array (pixel size 18.5 $\mu$m, plate scale 
$0\farcs{60}$ pixel$^{-1}$). The GROND pipeline resamples the NIR frame to 
2k $\times$ 2k, yielding a pixel scale of $0\farcs{30}$ in the reduced images. 
The measured FWHM of the PSF from the GROND images ranges from $0\farcs{8}$ 
to $1\farcs{9}$. The median value of the measured FWHM during the entire 
observing sequence for different bandpasses is $1\farcs{2}$ with small 
variations (i.e., $1\farcs{1}$ for the $g'$ band and $1\farcs{3}$ for the $H$ 
band). The fluxes of the SN and local reference stars were determined 
following a similar PSF-fitting procedure as for the LCO photometry. The inner 
and outer radii of the annulus were chosen to be 3\farcs{0} and 4\farcs{5}, 
respectively. No images of the SN\,2018evt field were obtained by GROND before 
the SN exploded, so template subtraction could not be performed. 

The GROND $g'r'i'z'$ photometry was calibrated relative to 
PanSTARRS1 \citep{Chambers_etal_2016} field stars in the AB system. The 
different sets of photometric standards for the GROND and LCO observations are 
mandated by the different FoVs: 5\arcmin\ for GROND optical, 10\arcmin\ for 
GROND NIR ($JHK$), and 27\arcmin\ for the Sinistros. NIR magnitudes ($JHK$) 
were derived with respect to Two Micron All Sky Survey 
(2MASS; \citealp{Cutri_etal_2003}) field stars in the Vega system. The final 
calibration of the GROND photometry is based on the median difference between 
catalogue and instrumental magnitudes of seven and eight field stars in the 
$g'r'i'z'$ and $JHK$ bands, respectively. These stars are identified by the 
purple and red circles in the right panel of Figure~\ref{Fig_image}. The GROND 
photometry is tabulated in Table~\ref{Table_grond_phot}. 

The resulting LCO $g'$ and $r'$ light curves display offsets from the 
corresponding GROND photometry (Fig.~\ref{Fig_lc}). The difference between the 
two light curves may be caused by the different calibration catalogues. 
Querying the APASS and PanSTARRS1 catalogues centred on SN\,2018evt with a box 
size of the LCO FoV, we found from more than 100 stars in common to both 
catalogues the median of the differences in magnitude and associated 1$\sigma$ 
ranges of $g_{\rm APASS} - g_{\rm Panstarrs1} \approx 0.05 \pm 0.06$\,mag and 
$r_{\rm APASS} - r_{\rm Panstarrs1} \approx -0.02 \pm 0.06$\,mag. 

\subsection{Optical Spectroscopy} \label{sec_obs_spec}
A journal of the spectroscopic observations of SN\,2018evt can be found in 
Table~\ref{Table_log_spec}. In addition to the early NTT classification 
spectrum (day $-9$), the late-time spectral sequence of SN\,2018evt spans days 
129 to 365. Apart from the EFOSC flux spectra described in 
Section~\ref{sec_obs}, the spectral database consists of LCO optical spectra 
taken with the FLOYDS spectrographs mounted on the 2\,m Faulkes Telescopes 
North and South at Haleakala, USA (FTN) and Siding Spring, Australia (FTS), 
through the Global Supernova Project \citep{Brown_etal_2013_lcogt}. A 2$\arcsec$ 
slit was placed on the target at the parallactic angle \citep{Filippenko_1982}. 
One-dimensional spectra were extracted, reduced, and calibrated following 
standard procedures using the FLOYDS 
pipeline\footnote{\url{https://github.com/svalenti/FLOYDS_pipeline}} \citep{Valenti_etal_2014}.
The $Bg'Vr'i'$ light curves and FLOYDS/LCO spectra were obtained as part of the 
Global Supernova Project. 
All photometry and spectroscopy will become available 
via WISeREP \footnote{https://wiserep2.weizmann.ac.il/} 
\citep{Yaron_Gal-Yam_2012}. 

\subsection{VLT Imaging Polarimetry} \label{sec_obs_impol}
SN\,2018evt was observed with FORS at the Cassegrain focus of UT1 at the VLT in 
imaging polarimetric mode (IPOL) as part of the Type Ia SN imaging polarimetry 
survey (Prog. ID 0102.D-0163(A), PI Cikota). The observations were obtained 
through the standard b\_HIGH (on 2019-01-09/day 140) and v\_HIGH (on 
2019-01-10/day 141) FORS2 filters, with half-wave retarder plate angles of 
$\theta = 0^\circ$, $22.5^\circ$, $45^\circ$, and $67.5^\circ$ at each epoch. 

All frames were bias subtracted using dedicated bias frames, and we removed 
particle events using LACosmic \citep{van_Dokkum_etal_2001}. Aperture 
photometry with a radius of 2 times the FWHM [$2 \times 0\farcs{44}$ 
($2 \times 3.5$ pixels) in the b\_HIGH images and $2 \times 0\farcs{45}$ 
($2 \times 3.6$ pixels) in the v\_HIGH images] was performed in the ordinary 
and extraordinary beams using the DAOPHOT.PHOT package \citep{Stetson_1987}. 
The linear polarization and the polarization angle were derived following 
the FORS2 manual \citep{Anderson_etal_2018}. The Stokes $Q$ and $U$ values 
and the polarization angle were corrected for the chromatism of the half-wave 
plate, and the polarization was debiased following \citet{Wang_etal_1997}. 
In order to study the intrinsic geometry of the SN, the interstellar 
polarization estimated in Section~\ref{sec_isp} was subtracted from both the 
imaging and the spectropolarimetry. 

On 2019-01-10/day 141, we measured a high linear polarization of $\sim 1.4$\% 
in the $B$ and $V$ bands. Since the high polarization level of SN\,2018evt 
suggested significant contributions intrinsic to the SN, we requested 
Director's Discretionary Time observations with FORS2 on the VLT to obtain 
multi-epoch spectropolarimetry (Prog. ID 2102.D-5031, PI Wang) for the 
geometric characterisation of the SN ejecta, the massive CSM, and the 
ejecta-CSM interaction region. 

\subsection{VLT Spectropolarimetry \label{vlt_obs_specpol}}
Spectropolarimetry of SN\,2018evt was conducted with the FOcal Reducer and 
low-dispersion Spectrograph 2 (FORS2; \citealp{Appenzeller_etal_1998}) on 
Unit Telescope~1 (UT1, Antu) of the ESO Very Large Telescope (VLT). 
Observations were carried out in the Polarimetric Multi-Object Spectroscopy 
(PMOS) mode at four epochs: days 172/2019-02-10, 195/2019-03-05, 
198/2019-03-08, and 219/2019-03-29. For each epoch, a flux standard star was 
observed at half-wave plate angle $0^\circ$. Grism 300V and a 1$\arcsec$ 
slit were used at epochs 1, 2, and 4. According to the VLT FORS2 user 
manual \citep{Anderson_etal_2018}, this configuration provides a spectral 
resolving power of $R \approx 440$ (or 13\,\AA\ FWHM) at a central 
wavelength of 5849\,\AA. VLT observations at epoch 3 were obtained with 
grism 1200R and a $1\arcsec$ slit, providing a spectral resolving power 
$R \approx 2140$ (or 3\,\AA\ FWHM) at a central wavelength of 6530\,\AA). 
A log of the VLT spectropolarimety is presented in 
Table~\ref{Table_log_specpol}. 

\begin{table*}
\caption{Log of spectroscopic observations of SN\,2018evt. \label{Table_log_spec}}
\begin{small}
\begin{tabular}{ccccccc}
\hline
  UT Time        &      MJD      &  Phase$^a$ &        Range       & Resolving Power &   Exp. Time    &  Instrument/Telescope  \\
(yy-mm-dd hh:mm)      &               &   (days)   &        (\AA )    & (blue/red)  &      (s)      &                        \\
\hline
18-08-12 23:59  &  58343.00  &  $-$9.0 &  3600$-$9000  &  $\sim$18 \AA$^{b}$ &   300  &  EFOSC2+gm13/NTT 3.6 m \\
18-12-24 15:16  &  58476.64  &  124.6  &  3400$-$9800  &  619/500  &  1800  &  FLOYDS/LCO 2.0 m FTN  \\
19-01-01 16:45  &  58484.70  &  132.7  &  3400$-$9800  &  497/398  &  1600  &  FLOYDS/LCO 2.0 m FTS \\
19-01-11 13:56  &  58494.58  &  142.6  &  3400$-$9800  &  413/513  &  1600  &  FLOYDS/LCO 2.0 m FTN \\
19-01-21 15:15  &  58504.64  &  152.6  &  3400$-$9800  &  627/498  &  1600  &  FLOYDS/LCO 2.0 m FTN \\
19-02-10 06:37  &  58524.28  &  172.3  &  4100$-$9100  &  440 & 480$\times$4  &  FORS2/PMOS+300V/VLT 8.2 m \\
19-03-04 10:21  &  58546.43  &  194.4  &  3400$-$9800  &  626/510  &  1800  &  FLOYDS/LCO 2.0 m FTN \\
19-03-05 06:00  &  58547.25  &  195.3  &  3400$-$9100  &  440 & 640$\times$4  &  FORS2/PMOS+300V/VLT 8.2 m \\
19-03-08 05:23  &  58550.22  &  198.2  &  5700$-$7100  &  2140 & 570$\times$4 &  FORS2/PMOS+1200R/VLT 8.2 m \\
19-03-17 11:39  &  58559.48  &  207.5  &  3400$-$9000  &  639/502  &  1800  &  FLOYDS/LCO 2.0 m FTN \\
19-03-29 04:53  &  58571.20  &  219.2  &  3400$-$9100  &  440 & 570$\times$4  &  FORS2/PMOS+300V/VLT 8.2 m \\
19-03-30 11:45  &  58572.49  &  220.5  &  3400$-$9800  &  622/504  &  1800  &  FLOYDS/LCO 2.0 m FTN \\
19-04-23 15:47  &  58596.66  &  244.7  &  3400$-$9800  &  469/396  &  2700  &  FLOYDS/LCO 2.0 m FTS \\
19-05-11 09:19  &  58614.39  &  262.4  &  3400$-$9800  &  382/540  &  2700  &  FLOYDS/LCO 2.0 m FTN \\
19-06-09 08:48  &  58643.37  &  291.4  &  3400$-$9800  &  604/542  &  2700  &  FLOYDS/LCO 2.0 m FTN  \\
19-07-15 06:06  &  58679.25  &  327.3  &  3400$-$9800  &  641/553  &  3600  &  FLOYDS/LCO 2.0 m FTN  \\
19-08-22 08:54  &  58717.37  &  365.4  &  3700$-$9800  &  641/553  &  3600  &  FLOYDS/LCO 2.0 m FTN  \\
\hline
\end{tabular}\\
{$^a$}{Days after $B$-band maximum on MJD 58352 / 2018 Aug. 22.} \\
{$^b$}{Resolution in \AA\ (FWHM).}
\end{small}
\end{table*}

\begin{table*}
\caption{Log of spectropolarimetic observations. \label{Table_log_specpol}}
\begin{small}
\begin{tabular}{ccccccc}
\hline
\hline 
Epoch &  Object        &    Date    &  Phase$^a$ &  Exposure    & Grism / Resol.\ Power &  Mean   \\
      &                &    (UT)    &  (day)     &   (s)                    &       & Airmass \\
\hline
1     & SN\,2018evt    & 2019-02-10 & 172.3      & $4 \times 480$ &     300V/440     &  1.23   \\
      & CD-32d9927$^b$ & 2019-02-10 & --         & $1 \times 10$  &     300V/440     &  1.16   \\
\hline
2     & SN\,2018evt    & 2019-03-05 & 195.3      & $4 \times 640$ &     300V/440     &  1.09   \\
      & L595-22$^b$    & 2019-03-05 & --         & $1 \times 60$  &     300V/440     &  1.01   \\
\hline
3     & SN\,2018evt    & 2019-03-08 & 198.2      & $4 \times 570$ &    1200R/2140    &  1.14   \\
      & CD-32d9927$^b$ & 2019-03-08 & --         & $4 \times 20$  &    1200R/2140    &  1.08   \\
\hline
4     & SN\,2018evt    & 2019-03-29 & 219.2      & $4 \times 570$ &     300V/440     &  1.06   \\
      & CD-32d9927$^b$ & 2019-03-29 & --         & $1 \times 20$  &     300V/440     &  1.03   \\

\hline
\end{tabular}\\
{$^a$}{Relative to the estimated peak on MJD 58352.} \\
{$^b$Flux standard, observed at a half-wave plate angle of $0^\circ$.} \\
\end{small}
\end{table*}

The high spectral resolution configuration in epoch 3 enabled us to measure 
more details of the spectropolarimetric properties across the H$\alpha$ 
profile, which mostly fall in the spectral range 5750--7310\,\AA. Only at epochs 
1 and 3 was the GG435 filter used, which has a cutoff at $\sim 4350$\,\AA\ and 
serves to prevent shorter-wavelength second-order contamination. The effect of 
second-order contamination on spectropolarimetry is mostly negligible unless the 
source is very blue (see the Appendix of \citealp{Patat_etal_2010}). The absence 
of the GG435 filter at epochs 2 and 4 is deliberate to extend the blue coverage 
as the SN aged, and any contamination by second-order light was considered 
negligible in extracting the true polarization signal. The slit position angle, 
$\chi$, was aligned with the north celestial meridian (i.e., $\chi=0$). Since 
all  observations were conducted at small airmass ($\lesssim1.2$), the loss of 
blue light can be well compensated by the linear atmospheric dispersion 
compensator (LADC; \citealp{Avila_etal_1997}). Therefore, we consider any effect 
on the spectral energy distribution (SED) caused by the misalignment between 
$\chi$ and the parallactic angle to be negligible.

For each epoch of observation, four exposures were carried out at 
retarder-plate angles of $0^\circ$, $22.5^\circ$, $45^\circ$, and $67.5^\circ$. 
The data were bias subtracted and flat-field corrected. Extraction of the 
ordinary (o) and extraordinary (e) beams was achieved following standard 
procedures within IRAF. Wavelength calibration was carried out separately for 
the o-ray and e-ray in each individual exposure (all four retarder-plate 
angles) using He-Ne-Ar arc-lamp exposures. A typical root-mean-square (RMS) 
accuracy of $\sim 0.25$\,\AA\ was achieved. Calculation of the Stokes 
parameters, as well as the determination of the bias-corrected polarization 
and associated errors, were performed with our own routines, following the 
recipes of \citet{Patat_etal_2006_polerr} and \citet{Maund_etal_2007_05bf}. 
A wavelength-dependent instrumental polarization in FORS2 ($\lesssim 0.1$\%) 
was further corrected based on the quantification 
by \citet{Cikota_etal_2017_fors2}. More detailed descriptions of the reduction 
of FORS spectropolarimetry can be found in a recent FORS2 Spectropolarimetry 
Cookbook and Reflex Tutorial\footnote{\url{ftp://ftp.eso.org/pub/dfs/pipelines/instruments/fors/fors-pmos-reflex-tutorial-1.3.pdf}}, 
as well as in \citet{Cikota_etal_2017_fors2} and Appendix A 
of \citet{Yang_etal_2020}. 

We write the observed polarization degree and position angle ($p_{\rm obs}$, 
{\it PA}$_{\rm obs}$) and the true values after bias correction ($p$, {\it PA}) 
in terms of the intensity ($I$)-normalised Stokes parameters ($Q$, $U$) as 
\begin{equation}
\begin{aligned}
p_{\rm obs} = \sqrt{Q^2 + U^2}, \ 
p = (p_{\rm obs} - \sigma_{p}^2 / p_{\rm obs}) \times h(p_{\rm obs} - \sigma_p); \\
PA_{\rm obs} = \frac{1}{2} {\rm arctan} \bigg{(} \frac{U}{Q} \bigg{)}, 
\ {\rm \ and \ } PA = PA_{\rm obs}. 
\end{aligned}
\label{Eqn_stokes0}
\end{equation}

Correction of the polarization bias followed the equations 
of \citet{Simmons_etal_1985} and \citet{Wang_etal_1997}, where $\sigma_{p}$ 
and $h$ give the $1\sigma$ uncertainty in $p_{\rm obs}$ and the Heaviside 
step function, respectively. 

\section{Light Curves of SN\,2018\lowercase{evt}} \label{sec_lc} 
\begin{figure}
\includegraphics[width=1.0\linewidth]{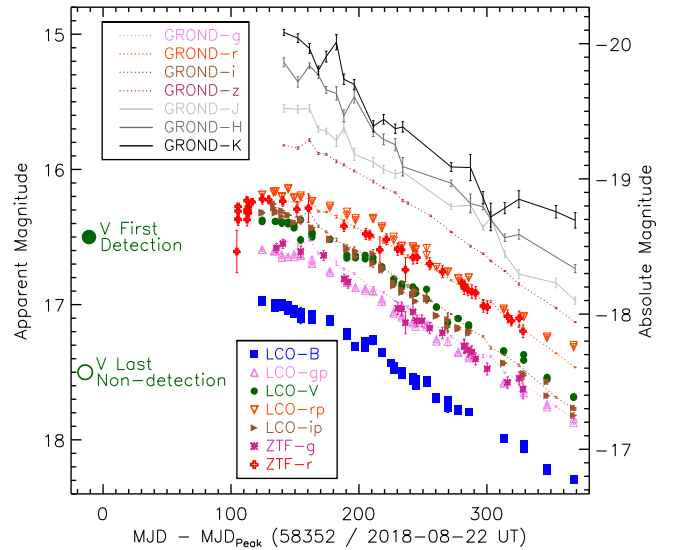}
\caption{The optical and NIR light curves of SN\,2018evt. 
\label{Fig_lc}
}
\end{figure}

In Figure~\ref{Fig_lc}, we show the $Bg'Vr'i'$-band light curves without 
correction for interstellar extinction. The $Bg'Vr'i'$ light curves were 
sampled during the period $t \approx$ 124 to 368 days. We list the calibrated 
LCO $Bg'Vr'i'$ photometry in Table~\ref{Table_phot}; the magnitudes are not 
corrected for extinction in the host galaxy or the Milky Way. We also present 
the Zwicky Transit Facility (ZTF; \citealp{Bellm_etal_2019}) $g$ and $r$ 
light curves of SN\,2018evt obtained with the forced-PSF photometry based on 
the pipeline developed by \citet{Yao_etal_2019}. The results are shown in 
Table~\ref{Table_ztf_phot}. 

\subsection{Interstellar Extinction Correction}\label{sec_extinction}
The Galactic reddening along the line of sight to SN\,2018evt was estimated 
as $E(B-V)^{\rm MW}_{\rm 18evt} = 0.051$\,mag by means of the NASA/IPAC NED 
Galactic Extinction Calculator\footnote{https://ned.ipac.caltech.edu/forms/calculator.html} and the extinction map by \citet{Schlafly_etal_2011}. 
Although an empirical relation between dust extinction and strength of the 
Na~{\sc I}~D $\lambda\lambda 5890$, 5896 absorption doublet has been proposed
\citep{Munari_etal_1997} and is widely applied in extinction estimations, the 
validity of the method has been questioned for use with low-resolution 
spectra \citep{Poznanski_etal_2011}. Since all spectroscopic observations 
discussed in this study were carried out in the low- and medium-resolution 
regime, we do not consider extinction corrections based on interstellar 
Na~{\sc I}~D lines. In the epoch-3 VLT spectrum ($R \approx 2140$), the 
spectral resolution is $\sim 2.8$\,\AA\ at Na~{\sc I}~D. Limited by the 
insufficiently high spectral resolution, we calculate upper limits of the 
equivalent widths (EWs) of the two features as 0.47\,\AA\ and 0.37\,\AA\ for 
the Milky Way, and 0.66\,\AA\ and 0.50\,\AA\ for the host galaxy. Adopting 
an empirical relation between Na~{\sc I}~D line width and dust 
reddening \citep{Poznanski_etal_2012}, we place upper limits on the 
extinction from the Milky Way and the host galaxy of  
$E(B-V)_{\rm MW}^{\rm Na~I~D} \textless 0.14\pm0.02$ and 
$E(B-V)_{\rm Host}^{\rm Na~I~D} \textless 0.32\pm0.05$, respectively. 
The intrinsically depolarized narrow H$\alpha$ emission line as measured 
from the high-resolution polarization spectrum at epoch 3 also suggests a 
low level of host reddening. See Section~\ref{sec_isp} for more details. 

\subsection{Pseudobolometric Light Curves} \label{sec_bolo} 
To better quantify the luminosity evolution of SN\,2018evt, we computed its 
pseudobolometric luminosity over a range of wavelengths 
($\sim 3870$--23,200\,\AA\ using the LCO $Bg'Vr'i'$ optical and GROND $zJHK$ 
NIR photometry. The steps of the procedure are detailed in 
Appendix~\ref{app_bolo}. 

\begin{figure}
\includegraphics[width=1.0\linewidth]{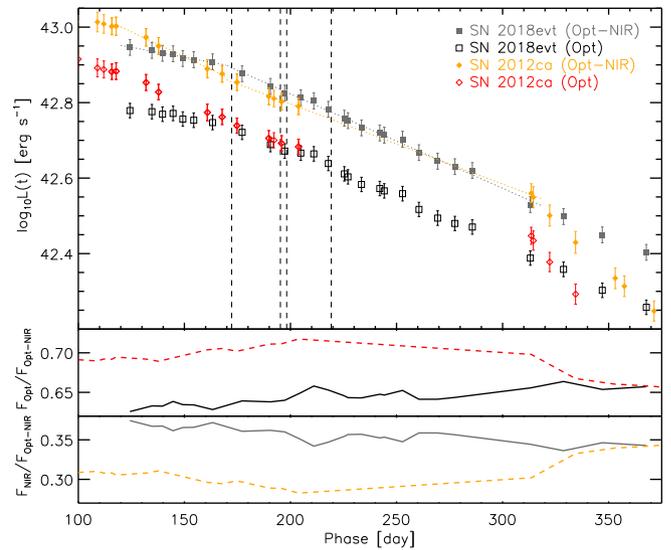}
\caption{Pseudobolometric light curves of SNe\,2018evt and 2012ca in the 
upper panel. Open black squares represent the bolometric luminosity 
integrated over the optical wavelength range (3870--9000\,\AA) while filled 
grey squares present the total bolometric luminosity within the optical-NIR 
wavelength range (3870--23,200\,\AA). Open red and filled orange diamonds 
give these quantities calculated for SN\,2012ca. The dotted grey line 
segments fit the optical-NIR bolometric decline rate of SN\,2018evt from 
days 120 to 170 and from days 170 to 320, respectively. The dotted orange 
line segments present the similar fit to SN\,2012ca. Vertical dashed lines 
mark the times of spectropolarimetric observations of SN\,2018evt. The 
middle and lower panels present the fractions of the optical and the NIR 
(9000--23,200\,\AA) fluxes of the total bolometric flux for SNe\,2018evt 
(solid line) and 2012ca (dashed line), respectively. 
\label{Fig_pseudo_bolo}
}
\end{figure}

The optical and optical-NIR pseudobolometric light curves of SN\,2018evt are 
plotted in Figure~\ref{Fig_pseudo_bolo}. For comparison, we also applied the 
same procedure to the $Bg'Vr'i'zJHK$ photometry of SN\,2012ca 
\citep{Inserra_etal_2016}. We adopt a distance modulus of 
$39.454 \pm 0.014$\,mag for SN\,2012ca \citep{Shappee_etal_2011}, which is 
the same as the distance applied in the bolometric luminosity calculation 
conducted by \citet{Inserra_etal_2016}. The calculated pseudobolometric light 
curve of SN\,2012ca is also shown in Figure~\ref{Fig_pseudo_bolo}. The 
integration of the SN\,2012ca SED was performed over the same wavelength 
ranges as for SN\,2018evt. The middle panel presents the ratio of the optical 
(Opt, 3870--9000\,\AA) to optical--NIR (Opt--NIR, 3870--23,200\,\AA) flux. 
The Opt/Opt--NIR flux ratio ($F_{\rm Opt}/F_{\rm Opt-NIR}$) of SN\,2018evt is 
lower than that of SN\,2012ca. 

We tabulate the optical and optical--NIR pseudobolometric luminosities of 
SN\,2018evt in Table~\ref{Table_bolo_18evt}. As a sanity check, a comparison 
was carried out between the pseudobolometric luminosity of SN\,2012ca 
derived by \citet{Inserra_etal_2016} and our calculation. We found that 
SN\,2012ca has a maximum pseudobolometric luminosity of 
$L_{\rm Opt} = (1.37 \pm 0.07) \times 10^{43}$\,erg\,s$^{-1}$ and a peak 
optical--NIR luminosity of 
$L_{\rm Opt-NIR} = (1.84 \pm 0.10) \times 10^{43}$\,erg\,s$^{-1}$. These 
values are consistent with those published by \citet{Inserra_etal_2016}: 
$L_{\rm pseudobol} \approx 1.29 \times 10^{43}$\,erg\,s$^{-1}$ and 
$L_{\rm bol} \approx 1.90 \times 10^{43}$\,erg\,s$^{-1}$, respectively. 
Since no data were obtained from days $\sim -10$ to 120, we do not attempt 
to estimate the peak bolometric luminosity of SN\,2018evt. As presented in 
Figure~\ref{Fig_pseudo_bolo}, we suggest that the luminosity of SN\,2018evt 
is similar to that of SN\,2012ca at similar phases.  

About 170 days after the estimated time of peak luminosity, the decline rate  
of the bolometric luminosity of SN\,2018evt changed from $-0.111 \pm 0.010$ 
to $-0.226 \pm 0.002$\,dex\,(100\,{\rm days})$^{-1}$ as shown in 
Figure~\ref{Fig_pseudo_bolo}. Between days 170 and 320, SN\,2018evt exhibited 
a similar decline rate as SN\,2012ca. The steeper decline at later phases 
(days $\sim 300$--400) observed in all three events presented 
by \citet{Inserra_etal_2016} with data after a year from the peak (i.e., 
SNe\,1997cy, 1999E, and 2012ca) was not followed by SN\,2018evt. Conversely, 
none of these three SNe showed an earlier break at a similar phase as 
SN\,2018evt. The logarithmic luminosity decline rates of SNe\,2018evt and 
2012ca over different phases are listed in Table~\ref{Table_decline}. 

Since a break in the bolometric light curve was found around day 170 
(Sec.~\ref{sec_bolo}), before the epoch of the optical SED template used in 
the above calculations, we also performed the same analysis using the 
FLOYDS/LCO spectrum on day 125 as the template SED of SN\,2018evt in the 
optical. This spectrum was obtained before the bolometric luminosity break. 
It has a similar profile to all the late-time spectra, but with relatively 
weaker line-emission features. The pseudobolometric flux calculated from the 
day-125 spectrum is overall $\sim 1.5$\% lower than that from the day-219 spectrum. This systematic difference between the different SED converts to 
0.007 in log\,$L$ or $\sim 1/3$ of the total uncertainty of the 
pseudobolometric luminosity. Therefore, we suggest that the early break in 
the bolometric flux has no significant effect on the late-time spectral 
evolution. 

\begin{table}
\begin{center}
\caption{Luminosity decline rates of SNe\,2018evt and 2012ca. \label{Table_decline}}
\begin{small}
\begin{tabular}{ccc}
\hline
SN & Phase$^a$ &      Decline rate, log\,$L$/time \\
           &   [days]  & [log\,(erg\,s$^{-1}$)/100\,d] \\
\hline
2018evt    & 120$-$170 &  $-$0.105$\pm$0.009 \\
           & 170$-$320 &  $-$0.250$\pm$0.004 \\
\hline
2012ca     & 120$-$170 &  $-$0.266$\pm$0.008 \\
           & 170$-$320 &  $-$0.213$\pm$0.002 \\
           & 120$-$320 &  $-$0.229$\pm$0.005 \\
           & 380$-$560 &  $-$0.460$\pm$0.011 \\
\hline
\end{tabular}\\
{$^a$}{Relative to the estimated peak at MJD 58352.} \\
\end{small}
\end{center}
\end{table}

\section{Spectroscopy} \label{sec_spec}
\begin{figure}
\includegraphics[width=1.0\linewidth]{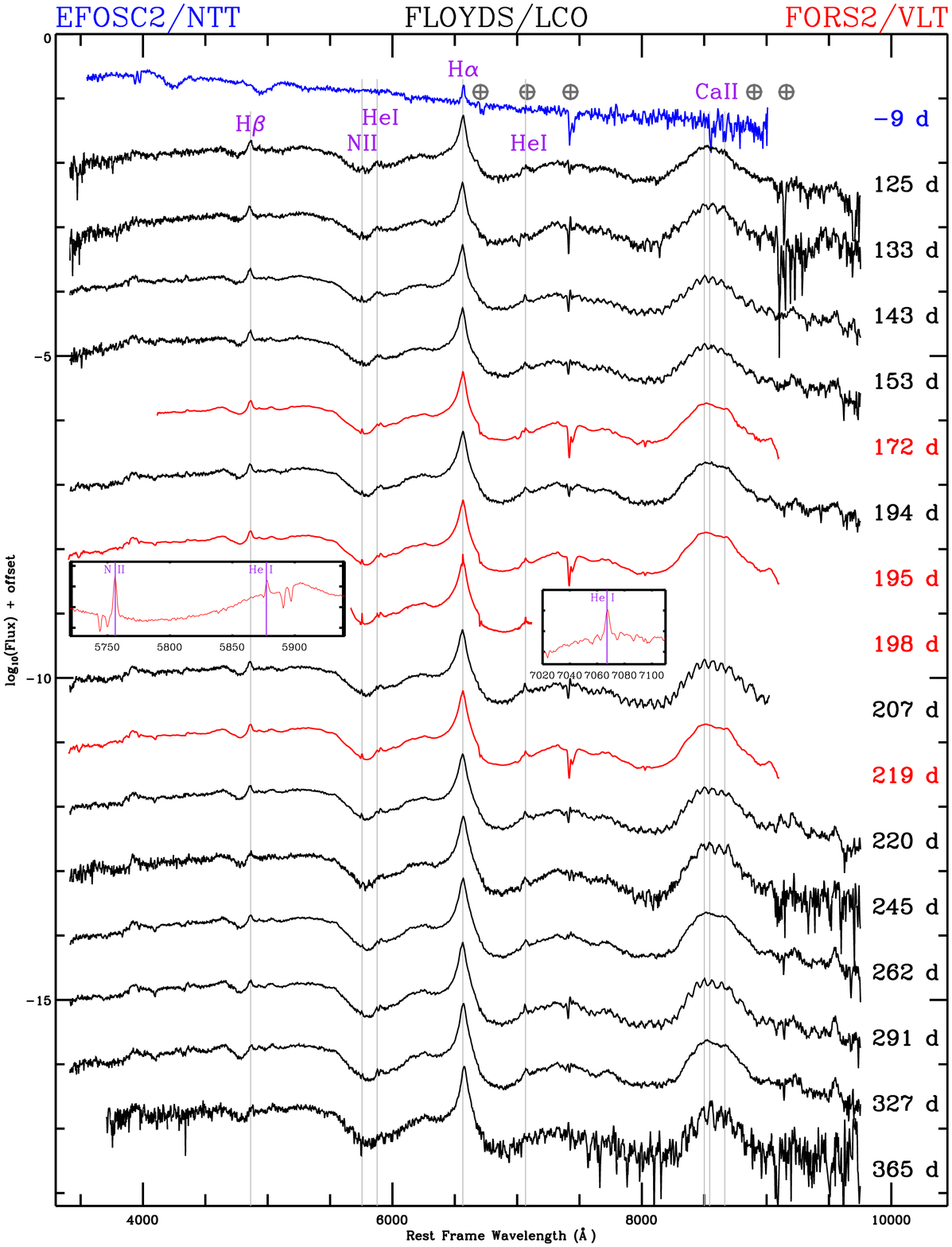}
\caption{Spectral time series of SN\,2018evt observed by NTT, LCO, and VLT 
(solid curves, phases are marked on the right; colours distinguish the 
spectrographs as shown at the top). Several prominent spectral lines are 
labeled, and some telluric lines are marked by crossed circles. The 
semi-regular fluctuations above $\sim 8000$\,\AA\ in some of the spectra 
are caused by fringing in the detectors; at still longer wavelengths, 
atmospheric water-vapor absorption sets in. 
\label{Fig_spec}
}
\end{figure}

Figure~\ref{Fig_spec} presents our spectral sequence of SN\,2018evt. 
In addition to the initial classification spectrum obtained with EFOSC2 
on the NTT, the dataset consists of 16 further optical spectra spanning 
the interval from approximately days 125 to 365. All wavelengths were 
corrected for the redshift of the host galaxy.

\subsection{Evolution of the H$\alpha$ and H$\beta$ Lines} \label{sec_balmer} 
The prominent Balmer emission features in the late-time spectra indicate that 
the spectral evolution of SN\,2018evt is slow and resembles that of other 
known SN\,1997cy-like events. It has been suggested that the H$\alpha$ region 
in these objects can be characterised by a pseudocontinuum plus multiple emission 
components (e.g., \citealp{Hamuy_etal_2003, Deng_etal_2004, Kotak_etal_2004}). 
We suggest that the late-time H$\alpha$ emission of SN\,2018evt is 
satisfactorily described by a combination of a broad and an intermediate 
Gaussian component. The details of the fitting procedure is described in 
Appendix~\ref{app_balmer_fit}.

\begin{figure}
\includegraphics[width=1.0\linewidth]{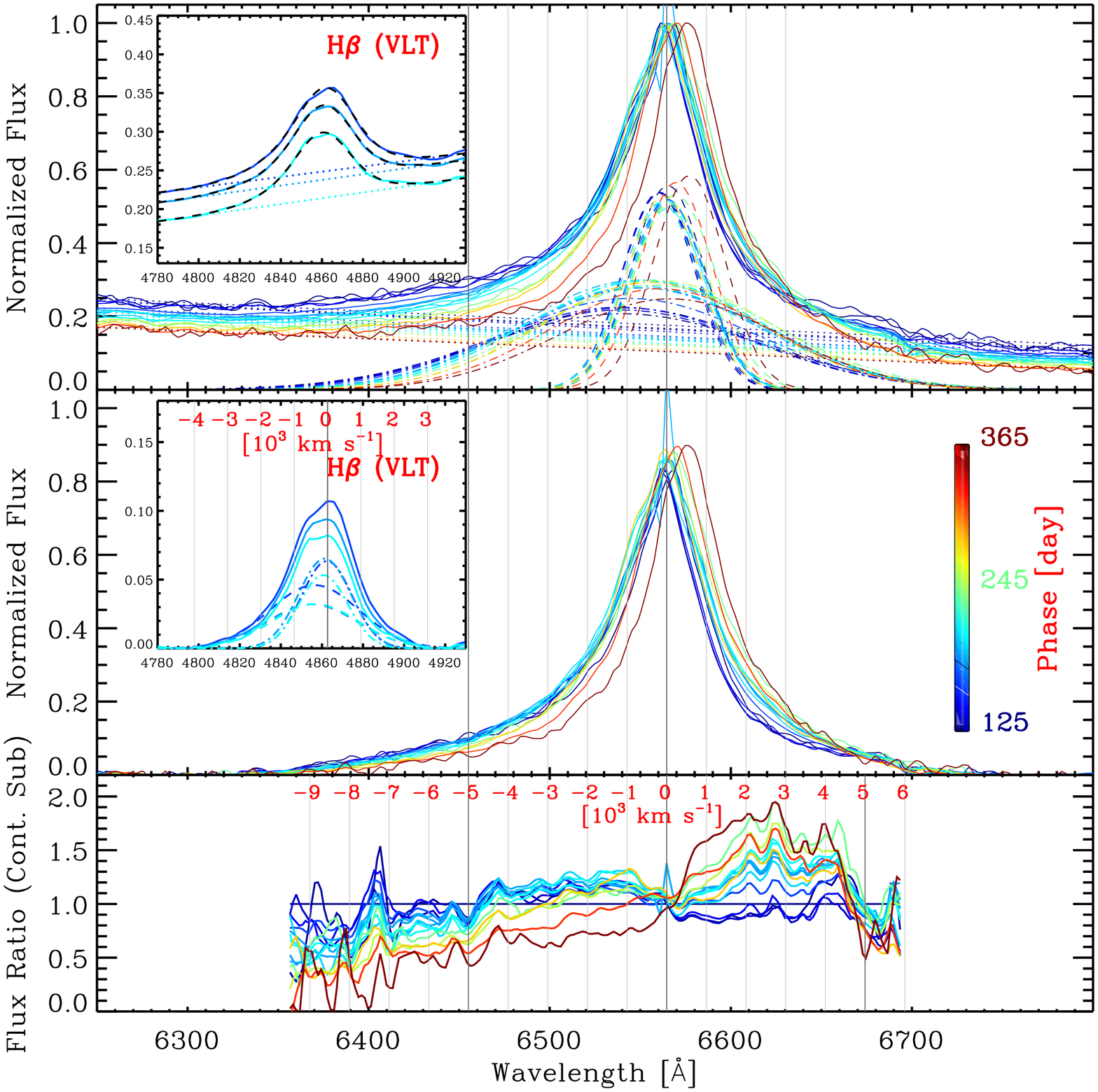}
\caption{
Temporal evolution of the H$\alpha$ profile of SN\,2018evt. All spectra were 
corrected for Galactic extinction and are shown in the rest frame. The phases 
are colour coded as indicated by the vertical colour bar. In the upper panel, 
solid lines present the peak-normalised spectra. Dotted straight lines, 
dot-dashed curves, and dashed curves represent the pseudocontinuum and 
Gaussian fits to the broad and the intermediate components, respectively. The 
middle panel displays the H$\alpha$ profiles after subtracting the 
pseudocontinuum. Upper and middle insets show similar diagrams for H$\beta$ 
in the high-SNR VLT observations, indicating that the H$\beta$ profile can 
also be described with two Gaussian components like those of H$\alpha$. The 
H$\alpha$ profiles in the middle panel but divided by that of the first epoch 
on day 125 are plotted in the bottom panel. Wavelengths corresponding to 
different velocities relative to the H$\alpha$ peak in the rest frame are 
shown as vertical grey lines and labeled. 
\label{Fig_ha_evolve2}
}
\end{figure}

Figure~\ref{Fig_ha_evolve2} illustrates the evolution of the H$\alpha$ 
emission of SN\,2018evt from days $\sim$125 to 365. The profile 
exhibits a prominent emission on an underlying quasi-continuum. The 
latter could be formed by the blending of relatively narrow lines of 
iron-group elements from fragmented cool shocked ejecta, mostly 
\ion{Fe}{{\sc II}} lines \citep{Chugai_etal_2004}. In the upper panel of 
Figure~\ref{Fig_ha_evolve2}, we present the H$\alpha$ profiles with the 
peak normalised to unity. Spectra after subtracting the 
pseudocontinuum are shown in the middle panel. To better expose the 
temporal evolution of the H$\alpha$ profile, we divided all late-phase 
spectra by our first late-time spectrum taken at day 125. Flux ratios 
were calculated from the normalised, pseudocontinuum-subtracted spectra 
and are shown in the bottom panel of Figure~\ref{Fig_ha_evolve2}.

\begin{figure*}
\includegraphics[width=0.9\linewidth]{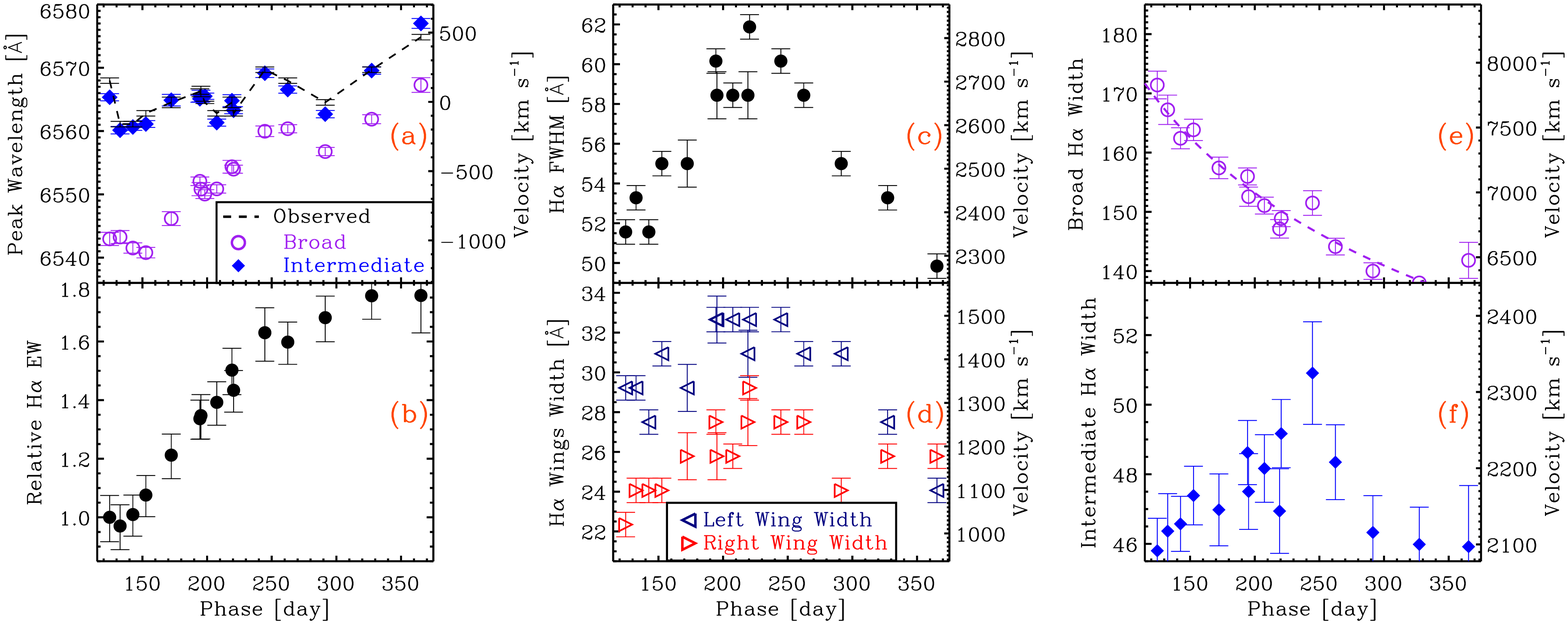}
\caption{Evolution of H$\alpha$ emission parameters. Panel (a) shows the 
measured wavelength of the peak, together with the fitted central 
wavelengths of the broad and intermediate components, as a function of 
phase. Panel (b) depicts the ratio between the time series of the H$\alpha$ 
equivalent width and that measured at day $+$125 when the first late-time 
spectrum of SN\,2018evt was obtained. 
Panels (c) and (d) present the time evolution of the FWHM, and the widths 
measured in the blue and the red emission wings only (all three in 
velocity units). Panels (e) and (f) give the width of the broad and 
the intermediate H$\alpha$ components, respectively. Except for panel 
(b), the corresponding velocities can be read off the right-hand ordinates.
} 
\label{Fig_ha_evolve3}
\end{figure*}

In Figure~\ref{Fig_ha_evolve3}, we plot the temporal evolution of the 
H$\alpha$ profile as encoded in several of its properties. We show the 
central wavelengths of the observed peak, the fitted broad and 
intermediate components (Fig.~\ref{Fig_ha_evolve3}a), the total FWHM 
(Fig.~\ref{Fig_ha_evolve3}c), and the widths of the blue and the red 
wing alone (Fig.~\ref{Fig_ha_evolve3}d). The absolute value of the 
pseudo-equivalent width (pseudo-EW; $W_{\lambda}$) of the H$\alpha$ 
emission in Figure~\ref{Fig_ha_evolve3}b is defined as 
$W_{\lambda} = \int_{\lambda_{\rm blue}^{\rm H\alpha}}^{\lambda_{\rm red}^{\rm H\alpha}} \left| \frac{f_{c}(\lambda) - f(\lambda)}{f_{c} (\lambda)} \right| d\lambda$, 
where $f(\lambda)$ and $f_{c}(\lambda)$ denote the flux across the emission 
line and the underlying continuum (respectively) at wavelength $\lambda$. 
Figures~\ref{Fig_ha_evolve3}e and \ref{Fig_ha_evolve3}f present the width 
of the broad and the intermediate H$\alpha$ components, respectively. The 
1$\sigma$ uncertainty of $W_{\lambda}$ was derived by error propagation of 
the uncertainty in the pseudocontinuum fitting. The absolute value of the 
pseudo-EW of an emission line measures how large a (pseudo)continuum range 
would have to be integrated over in order to obtain the same energy flux as 
contained in the emission line. To quantify the temporal evolution of the 
width of the H$\alpha$ profile, we divided $W_{\lambda}$ computed for 
different phases by that calculated for the first late-time spectrum acquired 
at day $+$125 and presented the result in Figure~\ref{Fig_ha_evolve3}b. 

In Figure~\ref{Fig_ha_evolve3}a, one can see that the peak wavelength of 
both the continuum-subtracted flux spectrum (the black dashed line labelled 
``Observed'') and the intermediate component (blue filled diamonds) were 
confined to a narrow range of $\sim \pm 200$\,km\,s$^{-1}$ before day 300, 
and started to shift toward longer wavelengths afterward. By contrast, the 
central wavelength of the broad component drifted from $-1200$\,km\,s$^{-1}$ 
to $\sim +100$\,km\,s$^{-1}$ between days 125 and 365. In 
Figure~\ref{Fig_ha_evolve3}c, the FWHM (in velocity units) of the H$\alpha$ 
profile shows a monotonic increase from $\sim 2350$\,km\,s$^{-1}$ and 
reached a maximum of $\sim 2800$\,km\,s$^{-1}$ at day $\sim 220$, and then 
decreased to $\sim 2300$\,km\,s$^{-1}$ at day $\sim 365$. This general trend 
is also shared by the widths of the blue and the red wings characterised by 
the absolute value of $W_{\lambda}$ as presented in the bottom-right panel. 
In Figure~\ref{Fig_ha_evolve3}d, the absolute value of the H$\alpha$ 
pseudo-EW increases continuously until day $\sim 240$ by as much as 80\%.

\begin{figure}
\includegraphics[width=1.0\linewidth]{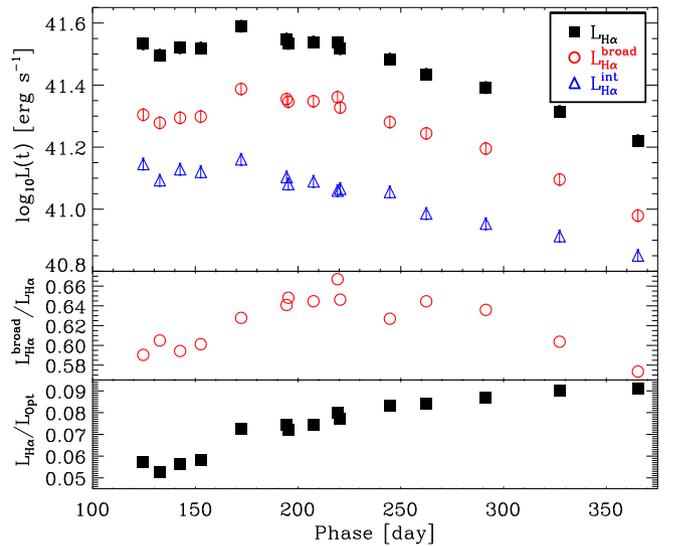}
\caption{
Temporal evolution of the H$\alpha$ luminosity and the contribution of its 
different components of SN\,2018evt. The top panel compares the H$\alpha$ 
luminosity with the flux from the Gaussian-decomposed broad and intermediate 
components. The middle panel presents the flux ratios between the broad 
component and the H$\alpha$ luminosity. The bottom panel shows the ratio 
between the H$\alpha$ and the optical bolometric luminosities. 
} 
~\label{Fig_ha_lumi}
\end{figure}

\textcolor{black}{ 
H$\alpha$ luminosity of SN\,2018evt from days 125 to 365 
is shown in Figure~\ref{Fig_ha_lumi}. Flux contribution from the 
pseudocontinuum has been subtracted. The temporal evolution of the broad 
and the intermediate components based on the Gaussian decomposition, the 
ratio between the broad component and the H$\alpha$ luminosity, 
$\rm L_{H\alpha}^{broad} / L_{H\alpha}$, and the ratio between the H$\alpha$ 
and the optical luminosities, $\rm L_{H\alpha}/L_{opt}$ are also presented. 
From day 125, the H$\alpha$ and the flux of the broad component increase 
with time, while the flux of the intermediate component stays roughly 
constant. After reaching the peak at around days $\approx$170--220, both 
the broad and the intermediate components decrease monotonically. The ratio 
between the luminosity of the broad component and the H$\alpha$ profile, 
except for the first two epochs where it is more difficult to measure. The 
FWHM of the central Gaussian is rather constant, but it decreases in the 
last two spectra. The luminosities of the two components behave similarly, 
rising to a peak and then declining. SN\,2018evt exhibits comparable 
strength of H$\alpha$ emission to SNe\,1997cy \citep{Turatto_etal_2000}, 
1999E \citep{Hamuy_etal_2003}, and 2002ic \citep{Wang_etal_2004} at similar 
phases. The primary energy source of the broad H$\alpha$ is the interaction 
between the SN ejecta and the CSM, and its luminosity is proportional to 
the dissipation rate of the kinetic energy across the shock 
front \citep{Kotak_etal_2004}. The intermediate component is most likely 
arises from the preionized gas in the unshocked, optically thick 
CSM \citep{Taddia_etal_2020}. The origin of the different H$\alpha$ 
components will be discussed in Section~\ref{sec_interact}.}

\subsection{The H$\alpha$ P~Cygni Profile} \label{sec_pcygni} 
\begin{figure*}
\includegraphics[width=0.8 \linewidth]{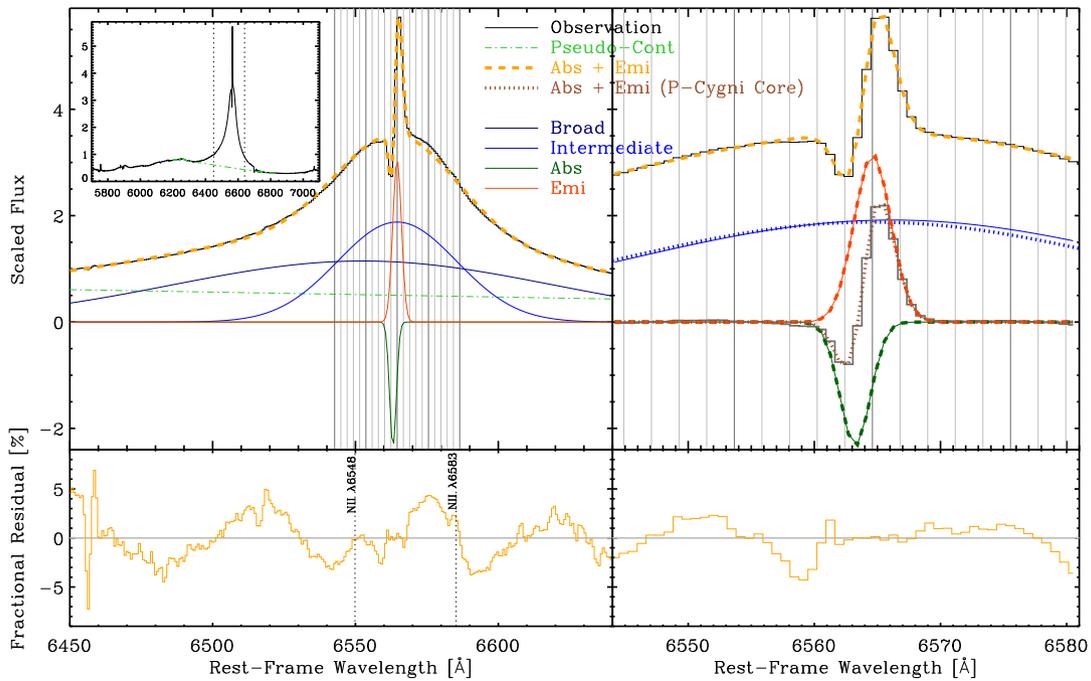}
\caption{The H$\alpha$ profile of SN\,2018evt in the VLT epoch-3 (day 198) 
spectrum fitted with multiple Gaussian functions. In the top-left panel, 
the observed spectrum (black histograms) is well approximated by the 
orange-dashed line, which consists of a pseudocontinuum (green 
dot-dashed line) and four Gaussian components, namely the broad (solid 
navy line), intermediate (solid blue line), and narrow (solid orange-red 
line) emission components and the narrow absorption (solid dark green line). 
The fractional residuals are provided in the bottom panel. The upper inset 
covers the full spectral range of the epoch-3 spectrum;  the two vertical 
dashed lines mark the spectral range of the main panel. The top-right panel 
presents a three-component Gaussian fit to the spectrum near the H$\alpha$ 
core after subtracting the pseudocontinuum and the broad component. 
Blue-dotted, red-dashed, and green-dashed lines identify the intermediate 
and narrow emission components and the absorption, respectively. After 
further intermediate-component subtraction (grey histogram), the spectrum 
is well fitted by the narrow Gaussian absorption and emission components 
(brown dotted line). The fit to the core region is given by the orange 
dashed line. The fractional residuals are shown in the lower panel. Vertical 
grey lines are drawn at $\pm 1000$\,km\,s$^{-1}$ relative to the H$\alpha$ 
narrow peak; 100\,km\,s$^{-1}$ intervals are denoted by grey lines. 
\label{Fig_pcygni}
}
\end{figure*}

VLT/FORS2 epoch-3 observations obtained with Grism 1200R provides a higher 
spectral resolving power ($R \approx 2140$) than the rest of the spectra 
presented in this work. The corresponding resolution element 
$\delta \lambda = \lambda/R \approx 3$\,\AA\ at the central wavelength of 
6530\,\AA\ enables a detailed study of the narrow P~Cygni core of the 
H$\alpha$ emission (rest-frame wavelength 
$\lambda_{0}^{\rm H\alpha}= 6564.614$\,\AA) that is only revealed at this 
higher resolution. Therefore, a multiple-component Gaussian fitting process 
similar to that used with the lower-resolution spectra was applied to this 
P~Cygni core. The pseudocontinuum was approximated by a low-order 
polynomial fitted to the spectrum between 5700 and 7300 \AA\ with the 
H$\alpha$-dominated range 6300--6700\,\AA\ excluded. Apart from the broad 
and the intermediate components, two additional functions characterising 
the narrow absorption and the narrow emission components were included to 
fit the P~Cygni core. The results are illustrated in the left panel of 
Figure~\ref{Fig_pcygni}.

A more physical description of the H$\alpha$ profile would result from 
Monte Carlo simulations of selected structures of the electron-scattering 
zone (e.g., \citealp{Huang_etal_2018}). Moreover, the H$\alpha$ emission of 
CSM-interacting SNe may also be approximated with Lorentzian or 
exponent-modified Lorentzian profiles \citep{Leonard_etal_2000, 
Smith_etal_2011}. For example, the sum of a narrow Gaussian and a broad 
modified Lorentzian yielded a plausible fit in the case of the Type IIn 
SN\,1998S \citep{Shivvers_etal_2015}. However, the physical interpretation 
of such a profile fitting is still not clear (see, 
e.g., \citealp{Jerkstrand_2017}). There is no intuition for expecting that 
the H$\alpha$ emission can be represented by a superposition of a few 
simple functions. In our analysis, we also fitted the broad and the 
intermediate components with two Lorentzian functions as well as with a 
Gaussian plus a Lorentzian function, but found no significant improvement 
in the achieved quality. In view of the very limited knowledge of the CSM 
configuration and the arbitrarily defined pseudocontinuum, we descoped 
the decomposition and fitting of the H$\alpha$ profile to characterise the 
temporal evolution of the overall appearance of the feature. 

In order to better separate the absorption and emission components of the 
P~Cygni profile and determine the redshift and the wind velocity, a 
two-stage analysis was carried out in addition to a four-component Gaussian 
fit. In this latter process, the broad Gaussian component was removed along 
with the pseudocontinuum. Thereafter a three-component Gaussian fit to the 
spectrum near the H$\alpha$ core was performed over the wavelength interval 
6544--6581\,\AA, roughly corresponding to a velocity range from $-950$ to 
$+750$\,km\,s$^{-1}$. The results can be found in the right-hand panel of 
Figure~\ref{Fig_pcygni}. The narrow absorption and emission components 
together achieve an acceptable fit to the spectrum after further removal of 
the intermediate component. Assuming the narrow emission component has its 
peak at the rest-frame wavelength of the line, we measured a redshift 
$z = 0.02523 \pm 0.00015$ based on the three-component Gaussian fit over a 
narrow range near the H$\alpha$ emission core. This is consistent with the 
value deduced from the four-component Gaussian fitting 
($z = 0.02523 \pm 0.00037$) and the redshift of the host galaxy reported by 
\citet{daCosta_etal_1998}. The wind velocity inferred from the narrow, 
blueshifted absorption minimum is $v_{\rm wind}=63\pm17$\,km\,s$^{-1}$ (see
Fig.~\ref{Fig_pcygni}). Note that the covariance between the fitted 
positions of the narrow emission and absorption components has been taken 
into account. We also conducted fits to the H$\alpha$ profile by adding 
additional Gaussian components and observed no improvement in the results. 

\subsection{Narrow Emission Lines} \label{sec_narrow_line}
Several very narrow emission lines were identified in the late-time spectra 
of SN\,2018evt. The most prominent feature is the 
[\ion{N}{{\sc II}}] $\lambda$5755 forbidden emission. The line appears to 
have the same redshift as indicated by the narrow P~Cygni H$\alpha$ 
component. Based on the VLT epoch-3 observations, we found that the FWHM of 
the line is consistent with the size of the spectral resolution element, 
$\Delta \lambda \approx 2.7$\,\AA. Therefore, the upper limit of the 
corresponding velocity is smaller than 140\,km\,s$^{-1}$ within the N-rich 
matter, which is in agreement with the wind velocity inferred from the 
H$\alpha$ profile. We suggest that the N line forms in the CSM. The presence 
of the [\ion{N}{{\sc II}}] $\lambda 5755$ line indicates that it originates 
from a cooler component of the CSM (i.e., less than a 
few $\times 10^{5}$\,K; \citealp{Salamanca_etal_2002}). 

The density of the CSM can be constrained by the intensity ratio of certain 
narrow lines. We measure the intensity ratio 
$I$([\ion{N}{{\sc II}} $\lambda\lambda$6548, 6583])/$I$(\ion{N}{{\sc II}} $\lambda 5755$) 
$\approx$0.18--0.28 from the VLT epoch-3 observations after subtracting 
either a pseudocontinuum or the multiple-component Gaussian fitting 
estimated from its ambient spectral region. The uncertainty of the intensity 
ratio depends mainly on the quality of the continuum removal and the 
wavelength ranges selected for the flux integration. Therefore, we calculated 
the intensity ratio for a series of wavelength bounds from $\pm 2$ to 
$\pm 4$\,\AA\ relative to the central wavelength. As the final value for
$I$([\ion{N}{{\sc II}} $\lambda\lambda$6548, 6583])/$I$(\ion{N}{{\sc II}} $\lambda 5755$), we adopted a range of the values obtained with the aforementioned 
series of bounds. Following Equation~5.5 of \citet{Osterbrock_1989}, for 
temperatures in the range $10^{4} < T < 10^{7}$\,K, the electron number 
density $n_e$ is between $\sim 2.9^{+0.8}_{-0.5} \times 10^{6}$\,cm$^{-3}$ 
and $\sim 2.5^{+0.7}_{-0.5} \times 10^{7}$\,cm$^{-3}$, with the lower bound 
of $n_{e}$ generally obtained at $T_e \approx 5\times 10^{4}$\,K. The 
estimated high density in the CSM around SN\,2018evt is broadly similar to 
those derived for SNe\,IIn; see, e.g., \citet{Salamanca_etal_2002} 
and \citet{Hoffman_etal_2008}. 

\section{Polarimetry} \label{sec_specpol}
In the next subsections, we present the VLT spectropolarimetry of SN\,2018evt. 
First, we focus on the interstellar polarization. Second, we discuss the 
global intrinsic polarization properties of the SN and thereafter continue 
with the time evolution of the continuum polarization. Finally, we undertake 
a more detailed investigation of the polarization spectra across the most 
prominent H$\alpha$ emission at the late phases. 

\subsection{Interstellar Polarization} \label{sec_isp}
When light passes through the interstellar medium (ISM), it is polarized 
through dichroic absorption by nonspherical paramagnetic dust grains 
partially aligned by the large-scale magnetic field of the galaxy. The 
removal of this interstellar polarization (ISP) is essential for deriving 
the polarization vector intrinsic to the source. This step is challenging 
and requires a beacon shining through the ISM and tracing the ISP. The 
estimation of the ISP toward SNe usually relies on assumptions of certain 
parts of the intrinsic optical spectrum of the SN being unpolarized. 
Spectral signatures often considered unpolarized include (1) the late-time 
spectra of SNe ($\sim 40$ days after peak luminosity) when electron 
scattering in the substantially diluted ejecta is much reduced 
\citep [e.g.,][]{Wang_etal_2001, Howell_etal_2001}; (2) the blanketing by 
iron-group absorption features over certain wavelength ranges, e.g., 
$\sim 4800$--5600\,\AA\ \citep{Howell_etal_2001, Chornock_etal_2006, 
Patat_etal_2009, Maund_etal_2010_05hk}, where the electron-scattering 
opacity is dominated by line-blanketing opacity; and (3) the emission 
components of strong P~Cygni profiles because the line emission is 
dominated by recombination photons \citep[][see also below]{Wang_etal_2004}. 
The assumption underlying all three cases is that any residual electron 
scattering does not contribute significantly to the total flux. 

The narrow H$\alpha$ emission feature in the late-time spectra of 
SN\,1997cy-like events is produced by the recombination of H in the nearly stationary 
CSM. Since the mechanism by which a proton captures an electron is distinct 
from the process causing a net polarization due to incomplete cancellation 
of electric vectors in the photosphere, H recombination lines are 
intrinsically unpolarized in the absence of strong magnetic fields. By 
contrast, the broad wings of the H$\alpha$ emission indicate the presence of 
fast-moving electrons in the CSM gas that is optically thick in H$\alpha$. 
Accordingly, following an approach similar to that of \citet{Wang_etal_2004}, 
we adopt the value of the polarization at the narrow H$\alpha$ peak for our 
estimate of the ISP, which gives $p_{\rm ISP} = 0.14 \pm 0.08$\%. The method 
is detailed in Appendix~\ref{app_isp}. 

An empirical rule for the dichroic extinction-induced ISP by Milky Way-like 
dust grains derived from observations of supposedly intrinsically unpolarized 
Galactic stars stipulates that 
$p_{\rm ISP} < 9\%\times E(B-V)$ \citep{Serkowski_etal_1975}. For a standard 
Galactic $R_V=3.1$ extinction law \citep{Cardelli_etal_1989}, the upper limit 
on the ISP derived from the Milky Way reddening only, $E(B-V)=0.051$\,mag, 
yields $p_{\rm ISP} < 0.46$\%. Accordingly, our ISP estimate passes this 
sanity check even in the case of vanishing extinction in the host galaxy. 
Therefore, we subtracted the above deduced $q_{\rm ISP}$ and $u_{\rm ISP}$ 
from all Stokes $Q$ and $U$ measurements in our VLT observations. The low 
line-of-sight polarization level of $\sim 0.14$\% also suggests a low level 
of host reddening. Because of the relatively low ISP suggested by the low 
extinction toward SN\,2018evt, we adopted a wavelength-independent ISP 
correction.

\begin{figure}
\includegraphics[width=1.0\linewidth]{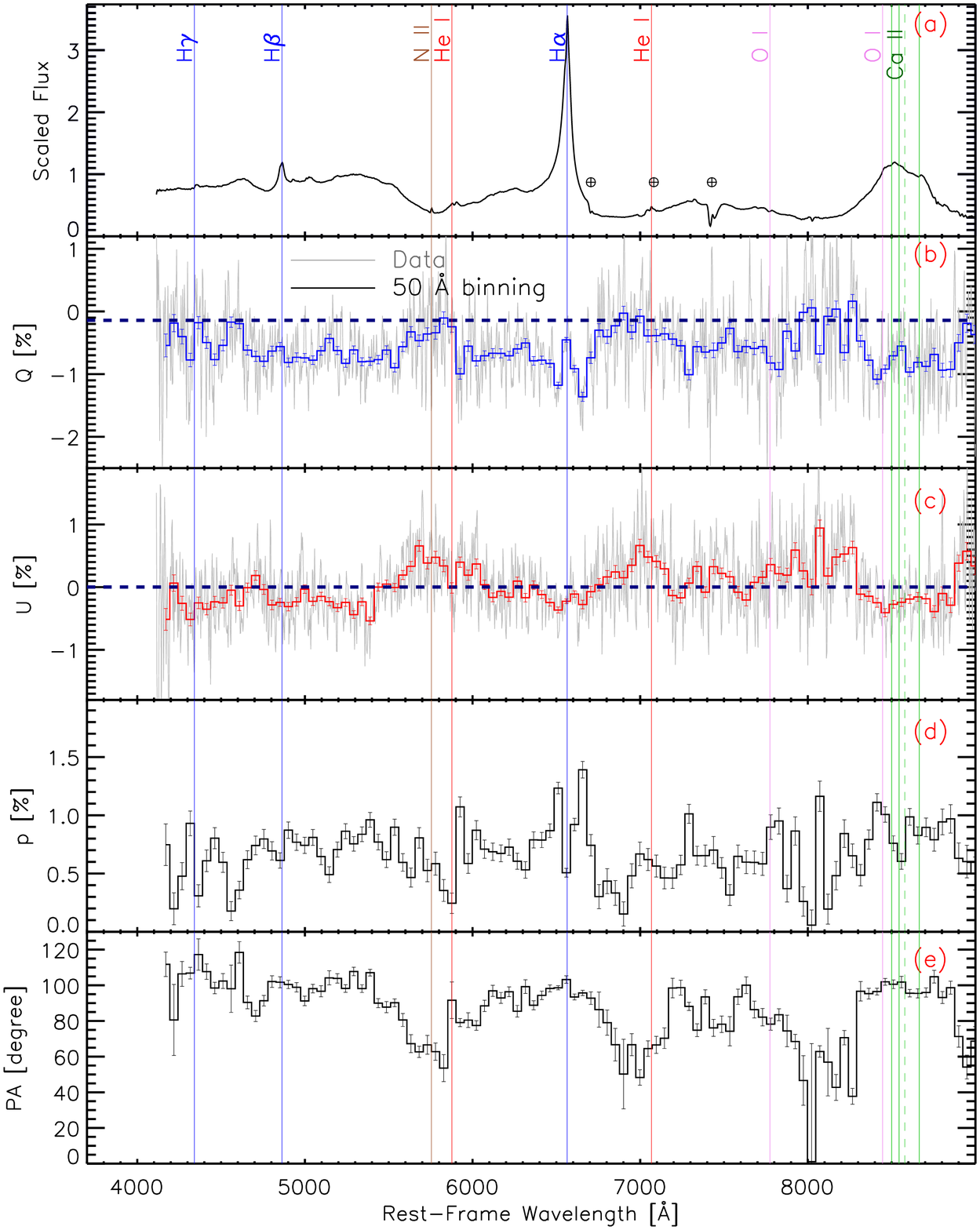}
\caption{Spectropolarimetry of SN\,2018evt on day 172 (epoch 1). The five 
panels (from top to bottom) give (a) the scaled flux spectrum; (b,c) the 
normalised Stokes parameters $Q$ and $U$, respectively; (d) the polarization 
spectrum ($p$); and (e) the polarization position angle {\it PA}. Line 
identifications are provided in the top panel. The data have been rebinned 
to 50\,\AA\ for clarity. 
\label{Fig_iqu_ep1}
}
\end{figure}

\begin{figure}
\includegraphics[width=1.0\linewidth]{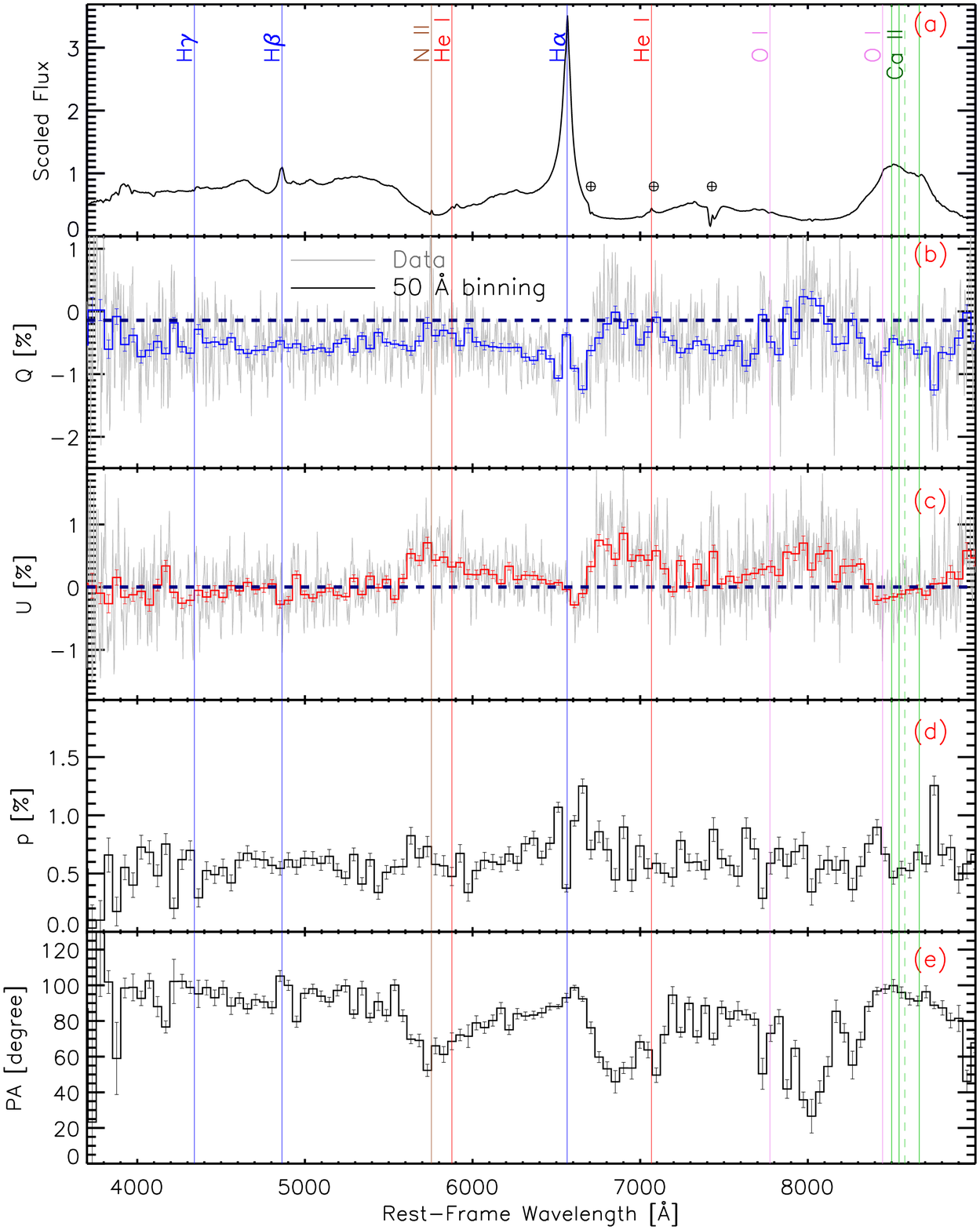}
\caption{Same as Fig.~\ref{Fig_iqu_ep1}, but for spectropolarimetry of 
SN\,2018evt on day 195 (epoch 2).  
\label{Fig_iqu_ep2}
}
\end{figure}

\begin{figure*}
\includegraphics[width=0.9\linewidth]{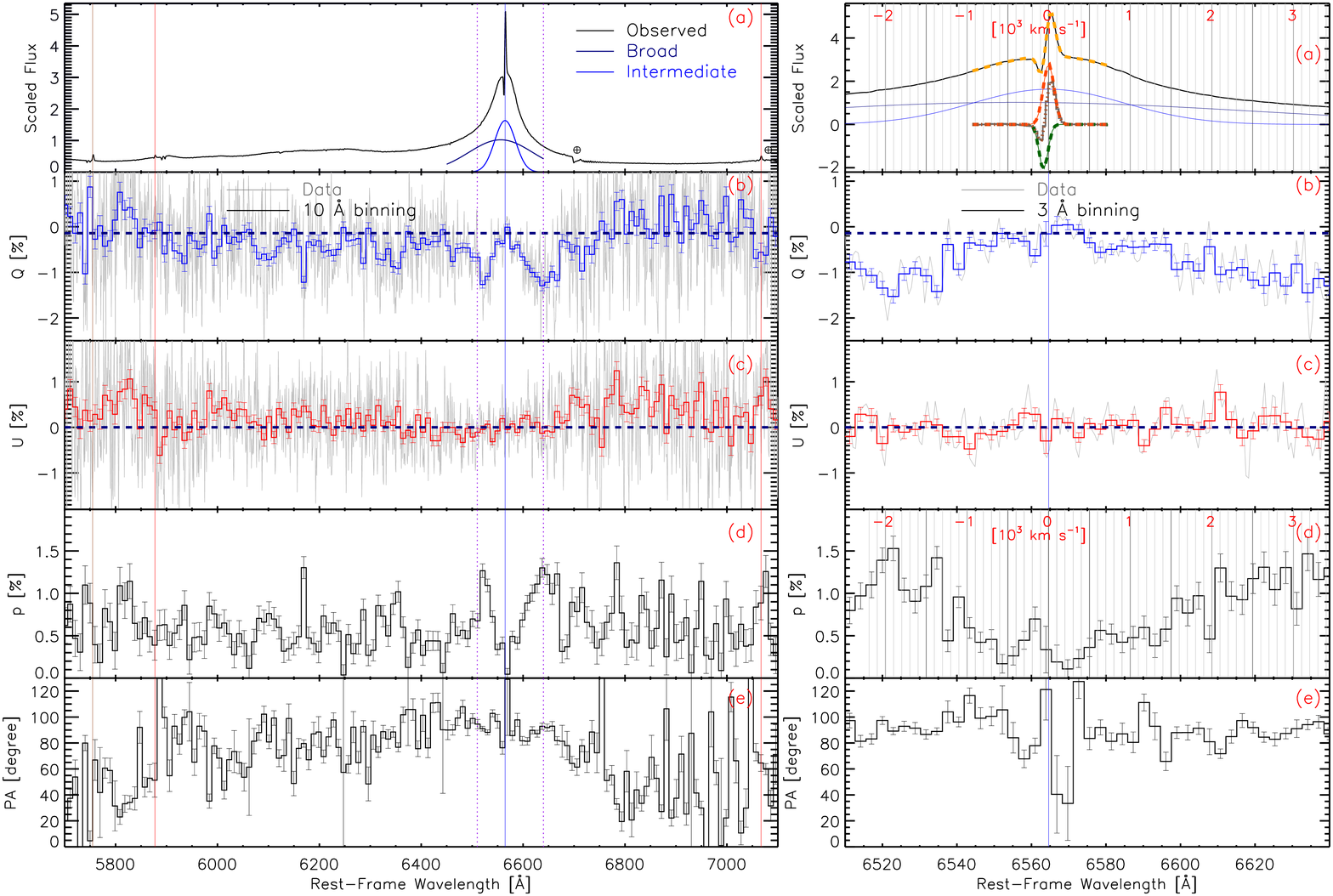}
\caption{Spectropolarimetry of SN\,2018evt obtained on day 198 (epoch 3) and 
at higher spectral resolution presented similarly as in Fig.~\ref{Fig_iqu_ep1}. 
The vertical blue line marks H$\alpha$ at zero velocity. The right column 
portrays the H$\alpha$ emission core between the two vertical purple-dotted 
lines in the left panel. Vertical grey lines indicate velocities relative to 
H$\alpha$ in the rest frame in steps of 100\,km\,s$^{-1}$. As shown in the 
fourth panels (d), the polarization reached local maxima at 
$\sim -2000$\,km\,s$^{-1}$ and $\sim 3200$\,km\,s$^{-1}$, where flux 
contributions from the intermediate component are negligible as can be deduced 
from the profile decompositions in the top panels. 
\label{Fig_iqu_ep3}
}
\end{figure*}

\begin{figure}
\includegraphics[width=1.0\linewidth]{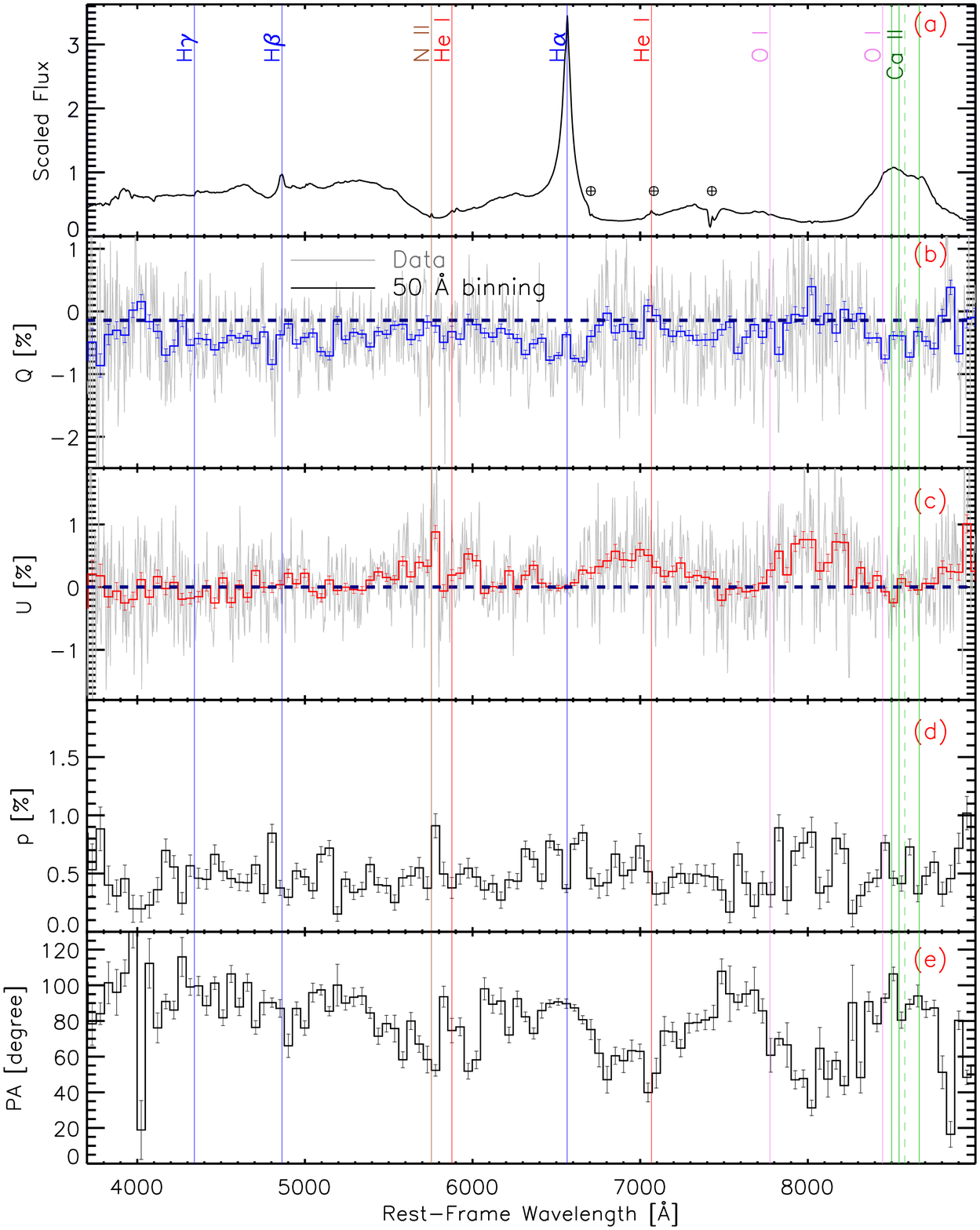}
\caption{Same as Fig.~\ref{Fig_iqu_ep1}, but for spectropolarimetry of 
SN\,2018evt on day 219 (epoch 4). 
\label{Fig_iqu_ep4}
}
\end{figure}

\begin{figure*}
\includegraphics[width=0.9\linewidth]{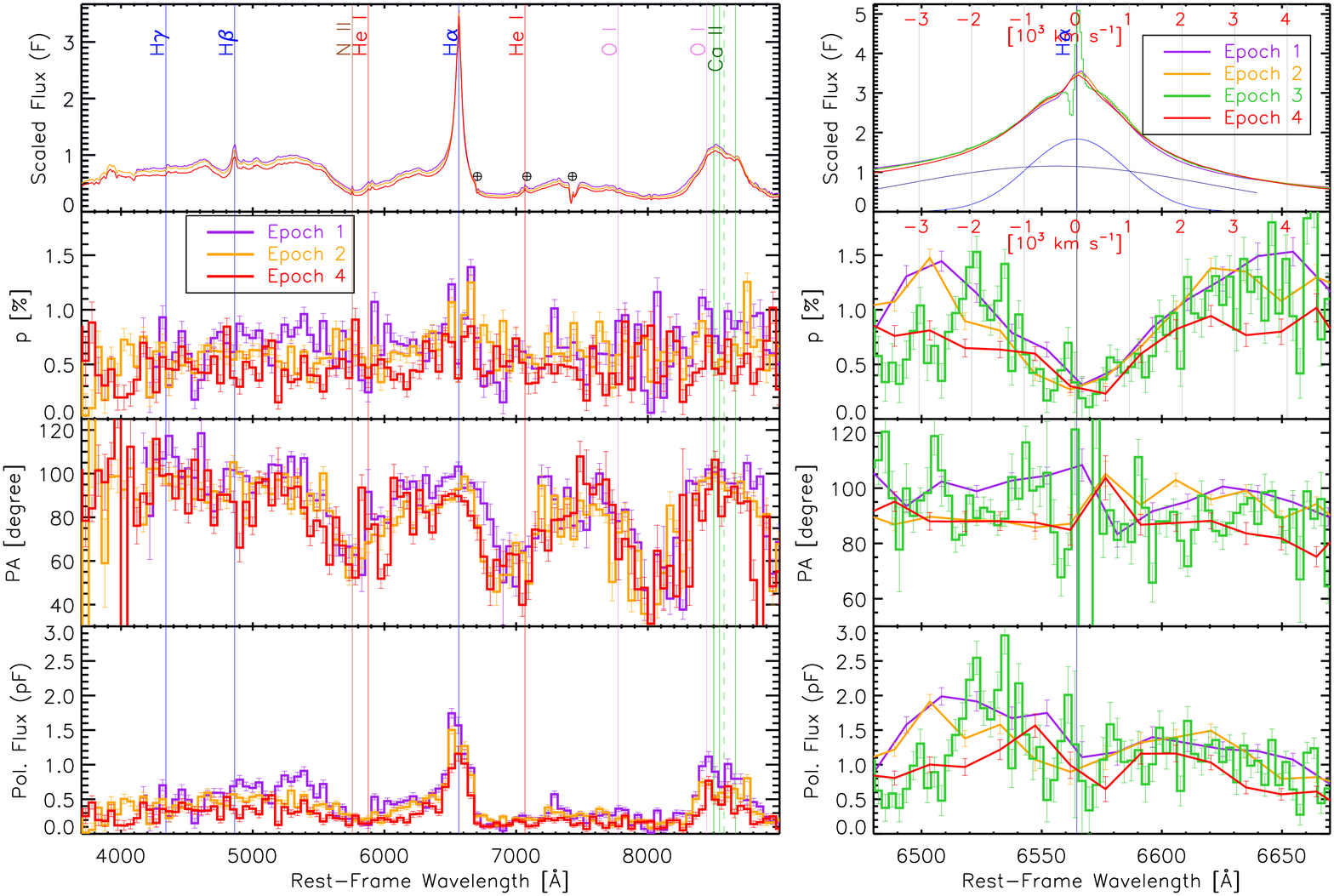}
\caption{Comparision of the spectropolarimetry of SN\,2018evt at different 
epochs. The left panel shows the observations on epochs 1, 2, and 4 at days 
172 (purple), 195 (orange), and 219 (red) in the rest frame, respectively. 
The four rows present (from top to bottom) the scaled flux spectrum ($I$) 
with major lines from several species labeled at zero velocity, the degree 
of polarization ($p$), the polarization position angle ({\it PA}), and the 
polarized flux (the scaled product of $p$ and $I$), respectively. The $p$ 
and {\it PA} data in the left panel have been binned to 50\,\AA\ for clarity. 
The right panel portrays the polarization profile of the H$\alpha$ core on 
epoch 3 / day 198 (green) with a 3\,\AA\ bin size while 15\,\AA\ binning was 
used for the lower-resolution data on epochs 1, 2, and 4. Vertical grey lines 
indicate velocities in steps of 1000\,km\,s$^{-1}$ relative to H$\alpha$. 
\label{Fig_ipth}
}
\end{figure*}

The Stokes $Q$ and $U$ values measured by the imaging polarimetry after 
correcting for the ISP are listed in Table~\ref{Table_impol}. 
Spectropolarimetry of SN\,2018evt obtained at days 172, 195, 198, and 219, 
together with the associated flux spectra in the rest frame, is visualised 
in Figures~\ref{Fig_iqu_ep1} to \ref{Fig_iqu_ep4}, respectively. 
Additionally, we compare the $p$ and $PA$ at different epochs in 
Figure~\ref{Fig_ipth}, where the amount of polarized flux ($p\times I$) is 
also presented. 

\begin{table*}
\begin{center}
\caption{VLT/FORS2 imaging polarimetry of SN\,2018evt. \label{Table_impol}}
\begin{normalsize}
\begin{tabular}{ccc|cc|cc|cc}
\hline
\hline
UT of Obs. & Phase & Band & $Q_0$ & $U_0$ & $Q$  &  $U$ & $p$  &  {\it PA}    \\
          & (day) &      &  (\%) &  (\%) & (\%) & (\%) & (\%) & ($^{\circ}$) \\
\hline
 2019-01-09 08:26:34 & 140.4 & $b_{\rm HIGH}$ & -1.43$\pm$0.09 & -0.38$\pm$0.09 & -1.29$\pm$0.09 & -0.38$\pm$0.09 & 1.34$\pm$0.12 & 98.3$\pm$2.5 \\
 2019-01-10 08:38:52 & 141.4 & $v_{\rm HIGH}$ & -1.49$\pm$0.09 &  0.02$\pm$0.09 & -1.35$\pm$0.09 &  0.02$\pm$0.09 & 1.35$\pm$0.12 & 89.6$\pm$2.5 \\
\hline
\end{tabular}\\
\end{normalsize}
\end{center}
\end{table*}

\subsection{The $Q-U$ Plane and Dominant Axes} \label{sec_quplane} 
Similarly to the difference between polar and rectangular coordinates, we 
present the observations in the $Q-U$ Plane, which is a mathematically 
convenient alternative to the degree of polarization and polarization 
position angle. The $Q-U$ plane defined by the Stokes parameters offers an 
intuitive visualisation of the polarization of the continuum as well as the 
spectral features \citep{Wang_etal_2001}. Each point represents the $Q$ and 
$U$ values measured in the chosen wavelength bin. Distances to the origin 
give the degree of polarization, i.e., $p = \sqrt{Q^2 + U^2}$. The azimuth 
of each data point is directly related to 
{\it PA}$ = (1/2)\,{\rm arctan}\,(U/Q)$. Depending on the departure from 
spherical and axial symmetries, spectral features representing particular 
chemical distributions may form specific patterns on the $Q-U$ plane.

If the data points (roughly) form a straight line, the {\it PA} is about 
the same at all wavelengths covered, indicating a common symmetry axis in 
the plane of the sky. Such a straight line on the $Q-U$ plane is known as 
the dominant axis \citep{Wang_etal_2003_01el,  Maund_etal_2010_05hk}. Any 
deviation from the dominant axis would be caused by (combinations of) 
regions of different composition, opacity, or velocity not (fully) sharing 
the symmetry axis. The dominant axis can be described as 
\begin{equation}
U = \alpha + \beta Q. 
\label{Eqn_daxis}
\end{equation}
In general, the SN spectral features arise from a variety of depths in 
the moving atmosphere and by a variety of processes. The polarization is, 
therefore, often decomposed into the dominant component $P_{d}$ and the 
orthogonal one, $P_{o}$, in the perpendicular direction. More details 
regarding this procedure are described by \citet{Wang_etal_2003_01el} 
and \citet{Stevance_etal_2017}.

\begin{figure*}
\includegraphics[width=0.9\linewidth]{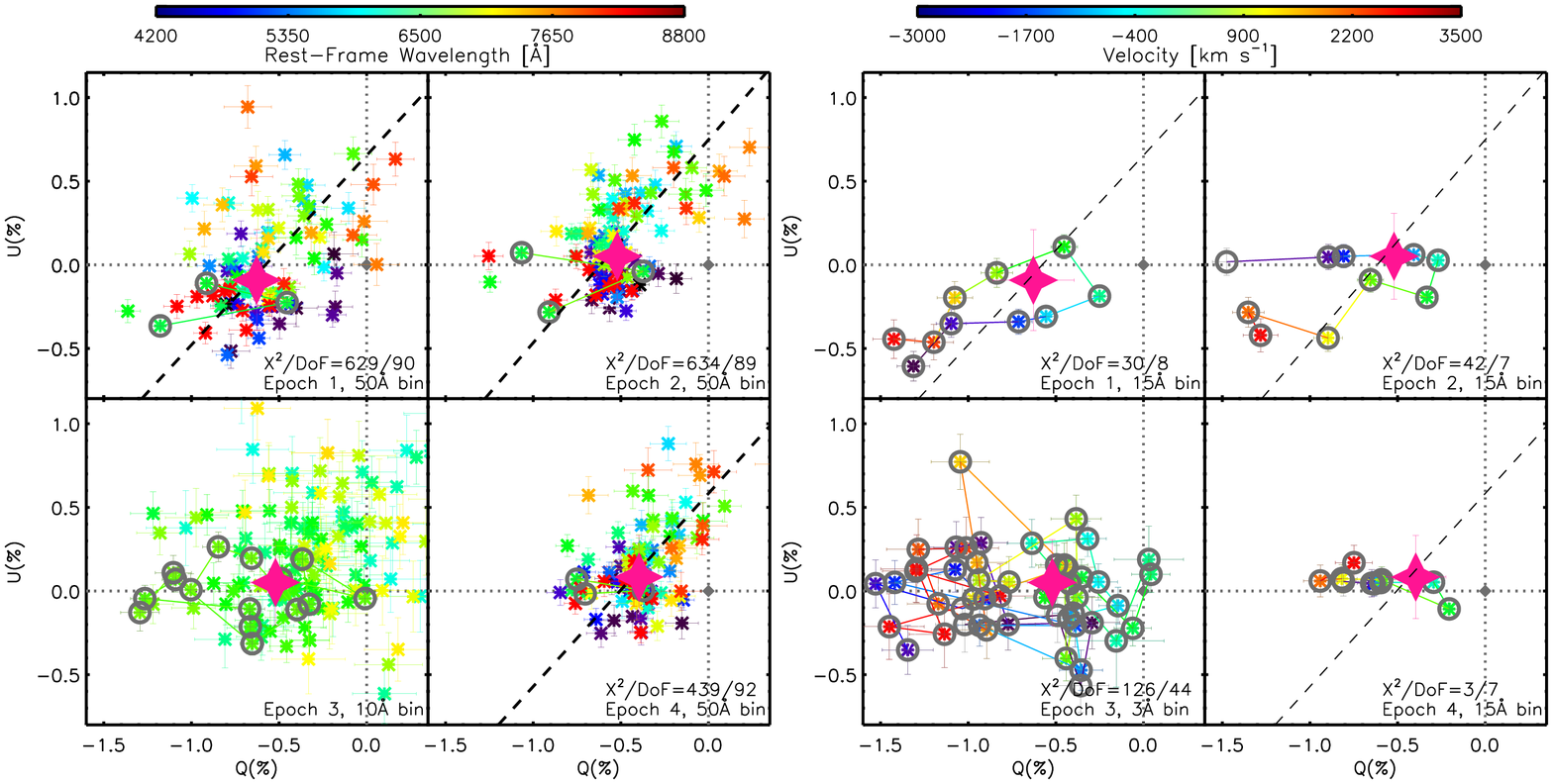}
\caption{
{\it Left:} Stokes parameters of SN\,2018evt displayed on the $Q-U$ plane. 
The data have been rebinned to 10\,\AA\ for epoch 3 and to 50\,\AA\ for 
the other epochs. In each panel, the wavelength of each bin can be read 
from the colour bar at the top. The black lines trace the dominant axis 
computed from the data between 4200\,\AA\ and 8800\,\AA. The solid pink 
four-point stars identify the continuum polarization deduced over the 
wavelength range of 4300--6250\,\AA. The continuum polarization at epoch 3 
has been assumed to be the same as at epoch 2 because the observations do 
not fully cover the continuum reference wavelength region. The larger open 
grey circles mark the H$\alpha$ region over the velocity range from 
$-3000$ to $+3500$\,km\,s$^{-1}$. {\it Right:} same as the left panel but 
on the H$\alpha$ profile from $-3000$ to $+3500$\,km\,s$^{-1}$ with smaller 
spectral binning as labelled. For the left and the right panels, all data 
points are coloured according to their wavelength and velocities 
(respectively), as encoded in the horizontal colour bars. 
\label{Fig_qu}
}
\end{figure*}

The left panel of Figure~\ref{Fig_qu} shows the ISP-corrected Stokes 
parameters on the $Q-U$ plane for all four epochs of our VLT observations. 
The dominant axis of the SN\,2018evt ejecta was determined by performing an 
inverse-error-weighted linear least-squares fitting of the data. The black 
long-dashed lines present the dominant polarization axes determined over 
the wavelength range from 4200\,\AA\ to 8800\,\AA\ for epochs 1, 2, and 4. 
Their common slope, $\beta = {\rm tan}^{-1} (\theta_{d})$, indicates that 
the direction on the sky of the symmetry axis tends to be constant from days 
173 to 219 ($\theta_d \approx 50^{\circ}$; see Table~\ref{Table_pol}). 
However, although a dominant axis seems to be present at all epochs, we 
suggest that the large $\chi^2$ values per degree of freedom (DoF) as 
labeled in the lower-right corner of each subpanel imply that the geometry 
of the ejecta-CSM interaction cannot be well described by a single axial 
symmetry. Substantial departures on the $Q-U$ plane from a dominant axis can 
be recognised especially toward shorter wavelengths. This behaviour on the 
$Q-U$ plane from days 173 to 219 indicates that SN\,2018evt 
belongs to the spectropolarimetic type D1 \citep{Wang_Wheeler_2008}, in which 
a dominant axis is present with significant scatter of the data points. 

\begin{figure*}
\includegraphics[width=0.9\linewidth]{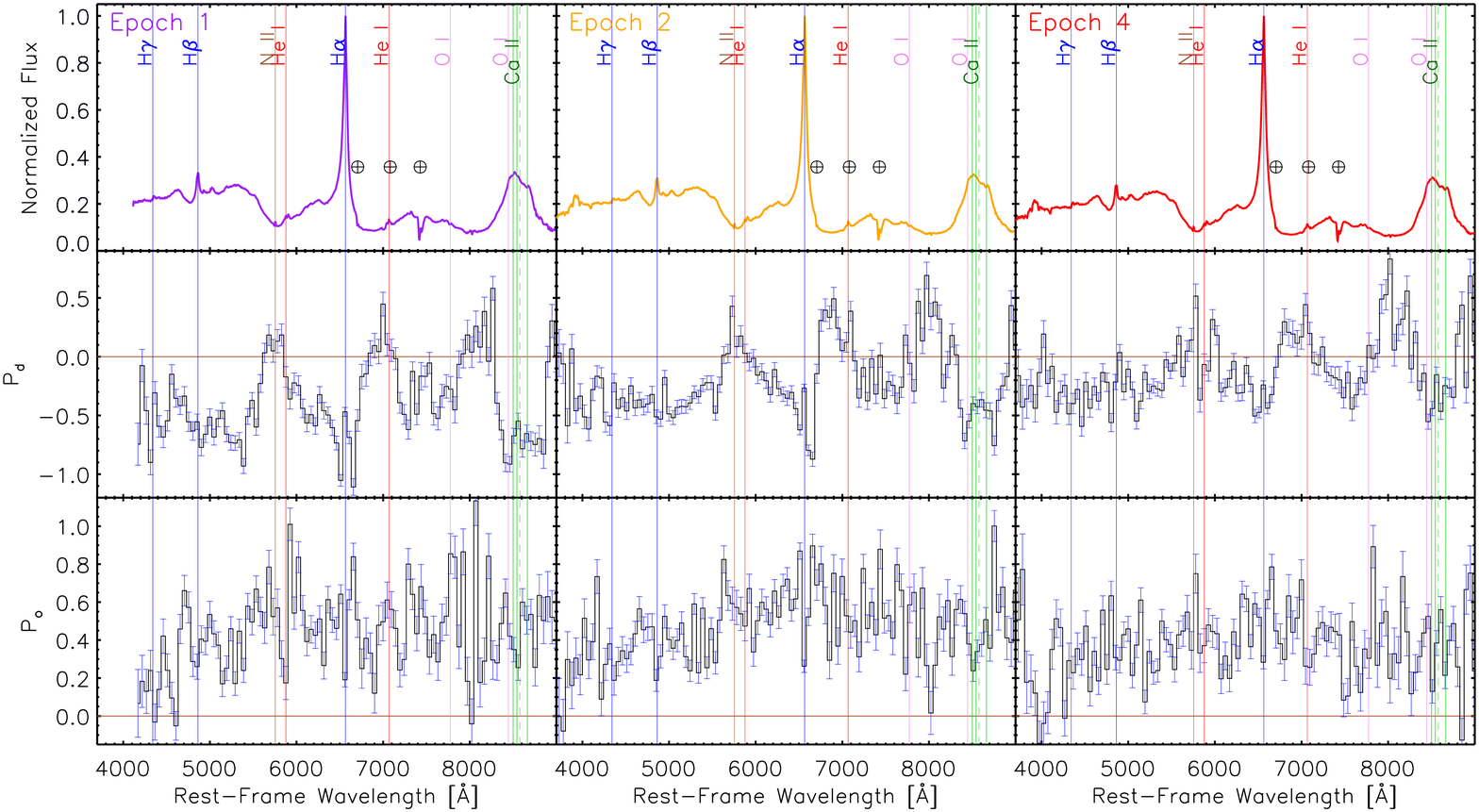}
\caption{
The normalised flux spectra together with the principle-components analysis 
of the SN\,2018evt spectropolarimetry at days 172 (left), 195 (middle), and 
219 (right). The top row gives the flux spectra normalised to the maximum 
value within the observed range. The middle and the bottom rows illustrate 
the polarization spectra projected onto the dominant axis ($P_d$) and the 
orthogonal axis ($P_o$), respectively. The vertical solid lines identify 
selected  spectral features at zero rest-frame velocity. The $\earth$ 
symbols mark the major telluric features. 
\label{Fig_stokes_pca}
}
\end{figure*}

The polarization can therefore be decomposed into two components, parallel 
($P_{\rm d}$) and orthogonal ($P_{\rm o}$) to the dominant axis. The 
orthogonal component carries information about departures from the axial 
symmetry defined by the parallel component. This procedure is equivalent to 
determining the first two principal components of the polarization (see, for 
example, \citealp{Wang_etal_2003_01el, Maund_etal_2010_05hk, 
Stevance_etal_2017}). The projected $P_{\rm d}$ and $P_{\rm o}$ at different 
epochs are shown in Figure~\ref{Fig_stokes_pca}. Over the three epochs with 
polarization measurements, the dominant component decreased across the 
H$\alpha$ wings, the H$\beta$ region and the \ion{Ca}{\sc II} NIR triplet 
profile. The overall declining $P_{\rm d}$ indicates an increasingly 
spherically symmetric geometry of the ejecta as more circumstellar H 
recombines. 

\subsection{Intrinsic Continuum Polarization} \label{sec_contpol}
On 2019-01-10 (day $\sim 141$) we detected a linear polarization of 
$\sim 1.4$\% in the $B$ and $V$ bands. This is at a higher level than in the 
only other SN\,1997cy-like event (SN\,2002ic) having polarimetry \citep{Wang_etal_2004}, 
which exhibited a $\sim 0.8$\% continuum polarization and a $\sim 0.8$\% 
polarization difference between H$\alpha$ and its ambient spectral region. 
Continuum polarization results from Thomson scattering of free electrons, 
which is independent of wavelength. After subtracting the ISP as determined 
in Section~\ref{sec_isp}, we arbitrarily selected the wavelength range of 
4300--6250\,\AA, which appears to have no strongly polarized lines, to 
characterise the continuum polarization. The continuum polarization 
($p^{\rm Cont}$) of SN\,2018evt at epochs 1, 2, and 4 and the associated 
position angle ({\it PA}$^{\rm Cont}$) were derived from the mean of 
50\,\AA\ binned Stokes spectra weighted by the inverse-squared 1$\sigma$ 
uncertainties. This error was estimated by adding the uncertainties in the 
weighted mean (the standard deviation calculated from the same 
50\,\AA\ binned spectra over the same continuum wavelength range) and the 
uncertainties in the ISP, in quadrature. Bias correction to the nonnegative 
$p^{\rm Cont}$ was carried out using Equation~\ref{Eqn_stokes0}. Continuum 
polarizations at the three epochs are listed in Table~\ref{Table_pol}. 

\begin{figure}
\includegraphics[width=0.95\linewidth]{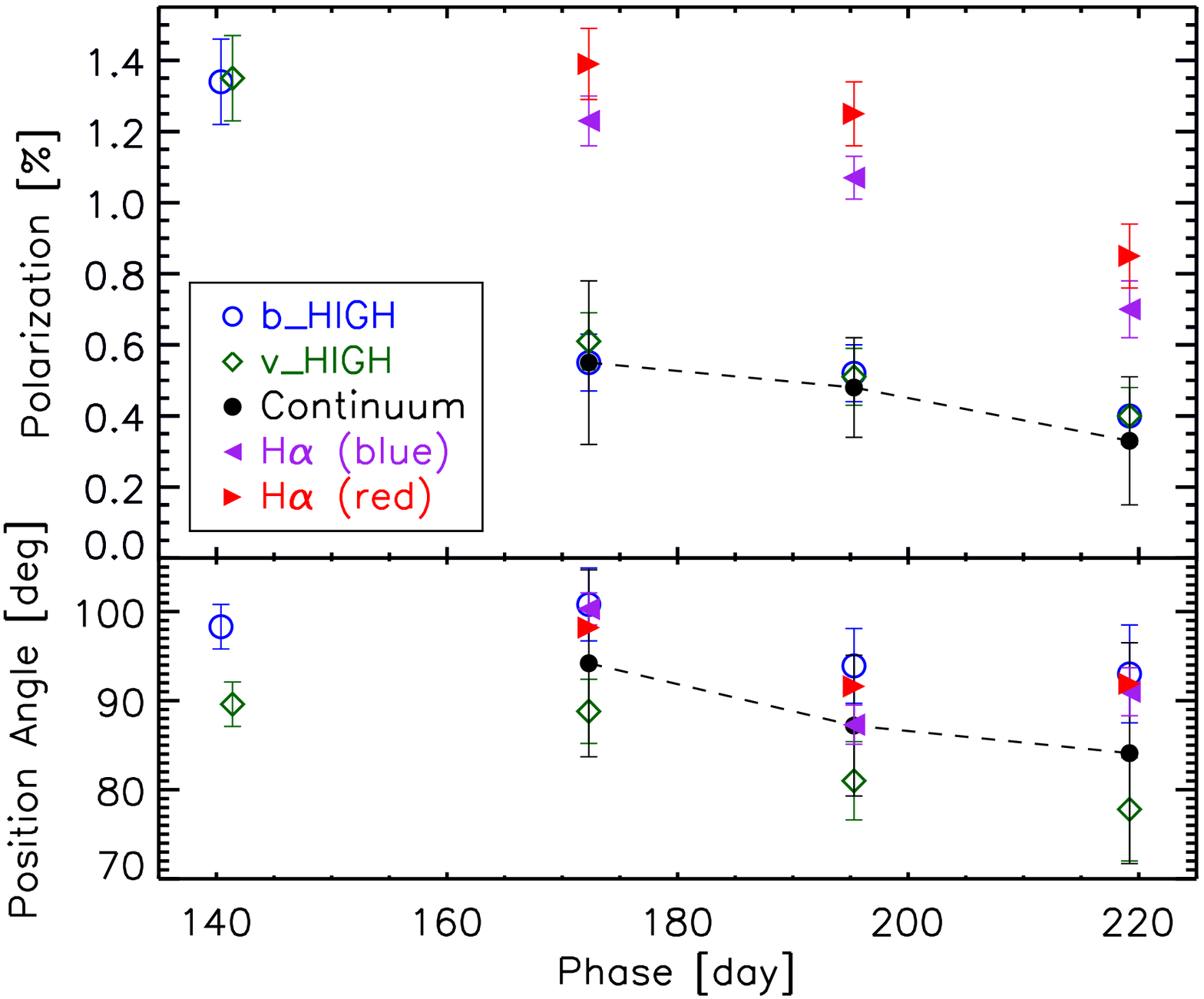}
\caption{
{\it Upper panel:} Time evolution of the intrinsic broad-band polarization 
of SN\,2018evt. The presented values include the polarization determined 
across the VLT b\_HIGH and v\_HIGH filters, an arbitrarily defined continuum 
region (4300--6250\,\AA), as well as the blue and the red wings of the 
H$\alpha$ emission. {\it Lower panel:} The associated temporal evolution of 
the polarization position angle. 
\label{Fig_pol_time}
}
\end{figure}

We also binned the spectropolarimetry at days 172, 195, and 219 over the 
b\_HIGH and v\_HIGH filter passbands that have been used with the imaging 
polarimetry at days 140/141. This process determines the equivalent imaging 
polarimetry data points in the b\_high and v\_high bandpasses. In this way, 
we formed the polarimetric dataset with the largest possible time baseline. 
The broad-band polarization was calculated through the integration over 
wavelength of the filter-transmission-weighted polarized flux. We only 
consider the uncertainty from the ISP estimation since the 
spectropolarimetric observations were carried out at very high SNR. The 
results are given in Table~\ref{Table_pol}. We also present the time 
evolution of the broad-band polarization and the continuum polarization in 
Figure~\ref{Fig_pol_time}. The b\_HIGH and v\_HIGH polarizations have 
decreased substantially in the time interval days 140/141 to 172 during 
which the break in the pseudobolometric light curve occurred (see 
Sec.~\ref{sec_bolo}). As shown in the bottom panel of 
Figure~\ref{Fig_pol_time}, we see no strong evidence of time evolution in 
the polarization position angle, indicating a consistent geometry of the 
continuum-emitting zone. 

\subsection{Intrinsic Polarization of the H$\alpha$ Emission Line} \label{sec_pol_ha}
SN\,2018evt offers only the second opportunity to date to study the geometry 
of the H-rich matter of SN\,1997cy-like events by spectropolarimetry. 
Compared to the first case, SN\,2002ic \citep{Wang_etal_2004}, high-SNR data 
with a much higher spectral resolution are available for it (Fig.~\ref{Fig_iqu_ep3}). 
This dataset reveals more details of the prominent H$\alpha$ emission and 
further enables a more careful interpretation of the nature of the H-rich 
CSM component. 

In addition, one can see that the polarization signal rises more rapidly 
from the central narrow emission core toward shorter than to longer 
wavelengths. Such a behaviour is more clearly visible in the right panel 
of Figure~\ref{Fig_iqu_ep3}, which depicts the wavelength range of 
6510--6640\,\AA\ and presents the data with a bin size of 3\,\AA. In the 
upper panels of Figure~\ref{Fig_iqu_ep3}, we also show the broad and 
intermediate H$\alpha$ flux components. Furthermore, the narrow absorption 
and emission components are overplotted in the top-right panel of 
Figure~\ref{Fig_iqu_ep3}. The maximum polarization level within the narrow 
absorption component amounts to $\sim 0.6$\% in the blue wing of the 
absorption minimum at around $-$100 to $-$200\,km\,s$^{-1}$. This can be 
understood as the blocking of unpolarized forward-scattered photons from 
the photosphere along the line of sight. Absorbing material blocks the 
unpolarized flux, which leads to an increased fraction of the 
scattering-polarized flux from the asymmetric limb \citep{McCall_etal_1984}. 

On day 198, the most prominent H$\alpha$ features in the polarization 
spectra are significantly polarized line wings and an essentially 
depolarized core. The latter spans a narrow wavelength range as can be seen 
in Figures~\ref{Fig_iqu_ep3} and \ref{Fig_ipth}. The widths of the broad 
and intermediate components are 
FWHM$_{\rm broad}^{\rm H\alpha} = 7198\pm34$\,km\,s$^{-1}$ and 
FWHM$_{\rm int}^{\rm H\alpha} = 2291\pm33$\,km\,s$^{-1}$, respectively. 
Outside the essentially depolarized intermediate component, the peak 
polarizations were attained as $p_{\rm blue}^{\rm H\alpha}=1.53\pm0.20$\% at 
$\sim -2000$\,km\,s$^{-1}$ ($\sim2.1\sigma_{\rm broad}^{\rm H\alpha}$) and 
$p_{\rm blue}^{\rm H\alpha}=1.99\pm0.34$\% at $\sim 3200$\,km\,s$^{-1}$ 
($\sim3.3\sigma_{\rm broad}^{\rm H\alpha}$) in the blue and the red wings, 
respectively. Similar structures can be identified in all other VLT 
observations (see right panels of Fig.~\ref{Fig_ipth}). These peak 
polarization levels in H$\alpha$ are significantly higher than the continuum 
polarization, i.e., $p^{\rm Cont}=0.48\pm0.14$\%, as estimated in 
Section~\ref{sec_contpol}. The corresponding values are also listed in 
Table~\ref{Table_pol}. 

In Figure~\ref{Fig_pol_time} we also compare the polarization position angle 
measured in the blue and the red wings of the H$\alpha$ profile. The adopted 
position angle was that exhibited by the local maximally polarised emission
after 50\,\AA\ binning. We see neither significant time evolution of the $PA$ 
nor strong deviation of the $PA$ in the H$\alpha$ wings from the continuum. 
Therefore, we infer that the ejecta of SN\,2018evt and its ambient CSM 
exhibit similar axial symmetry. Unlike the case of the Type IIn 
SNe\,1997eg \citep{Hoffman_etal_2008} and 
2010jl \citep{Patat_etal_2011_sn2010jl}, the symmetry axes of the SN ejecta 
and the CSM in SN\,2018evt may not be substantially misaligned. 

\begin{table*}
\caption{Polarization properties of SN\,2018evt$^{(a)}$. \label{Table_pol}}
\begin{small}
\begin{tabular}{cc|cc|cc|cc|cc|c}
\hline
\hline
Epoch &  Phase & $p^{\rm Cont}$     &  {\it PA}$^{\rm Cont}$ & $p^{\rm b\_HIGH}$     &  {\it PA}$^{\rm Cont}$ & $p^{\rm v\_HIGH}$     &  {\it PA}$^{\rm v\_HIGH}$ & $p^{\rm H\alpha}_{\rm blue}$ &  $p^{\rm H\alpha}_{\rm red}$ &  $\theta^{\rm H\alpha}_{d}$     \\
      & (days) & (\%)          & (degree)      & (\%)          & (degree)  &   (\%)                   &                      (degree)  & (\%)     & (\%)                 & (degree)            \\
\hline
1$^{(a)}$ & 172.3  & 0.55$\pm$0.23 & 94.2$\pm$10.5 & 0.55$\pm$0.08 & 100.8$\pm$4.1 & 0.61$\pm$0.08 & 88.8$\pm$3.6 & 1.23$\pm$0.07 & 1.39$\pm$0.10 & 48.6$\pm$2.0  \\
2$^{(a)}$ & 195.3  & 0.48$\pm$0.14 & 87.2$\pm$7.9  & 0.52$\pm$0.08 & 93.9$\pm$4.2  & 0.51$\pm$0.08 & 81.0$\pm$4.4  & 1.07$\pm$0.06 & 1.25$\pm$0.09 & 50.5$\pm$2.3  \\
3$^{(b)}$     & 198.2  & --            & --            & --            & --            & --            & --            & 1.53$\pm$0.20 & 1.99$\pm$0.34 &           --  \\
4$^{(a)}$ & 219.2  & 0.33$\pm$0.18 & 84.1$\pm$12.4 & 0.40$\pm$0.08 & 93.0$\pm$5.5 & 0.40$\pm$0.08 & 77.8$\pm$5.8 & 0.70$\pm$0.08 & 0.85$\pm$0.09 & 49.1$\pm$3.1  \\
\hline
\end{tabular}\\
{$^{(a)}$}{Measurements of epochs 1, 2, and 4 are based on spectra binned to 
50\,\AA\ except column $\theta^{\rm H\alpha}_{d}$}. \\
{$^{(b)}$}{Measurements of epoch 3 are based on spectra binned to 3\,\AA.} 
\end{small}
\end{table*}

\section{Overview of the Main Observational Properties} \label{sec_obs_summary}
The observing campaign on SN\,2018evt started at around 110 days past the 
estimated peak luminosity. At these late times, the SN showed similar 
multiband absolute magnitudes, decline rates, and spectral evolution as 
other SN\,1997cy-like events (Figs.~\ref{Fig_lc} and \ref{Fig_spec}). In the 
following, we give a concise overview of the main observational signatures of 
SN\,2018evt. Some of them may not have been identified in earlier 
SN\,1997cy-like events, most prominently the early break in the bolometric light 
curve, the evolution of the H$\alpha$ and H$\beta$ profiles after day $\sim 100$, 
the variability of the polarization, and details of the polarization profile 
of H$\alpha$. 

(1) The decline rate of the optical-NIR pseudobolometric luminosity of 
SN\,2018evt increased after day $\sim 170$, which can be seen as a break in 
Figure~\ref{Fig_pseudo_bolo}. To our knowledge, such an early break has not 
yet been identified in other SN\,1997cy-like events, 
which are only known to exhibit a rapid drop between days $\sim 300$ and 400. 
Their light curves can be fitted with the equations formulated 
by \citet{Nicholl_etal_2014}, which are based on a semianalytic model for the 
case of ejecta colliding with optically thick CSM \citep{Chatzopoulos_etal_2012}. 
A change in the late-time bolometric luminosity decline rate can be expected if 
there is a transition of the CSM radial profile from a denser inner region to an 
outer region with a steep drop in its density. Alternatively, such changes may 
also be caused by the reverse shock becoming ineffective, so that the forward 
shock is significantly decelerated. This can happen when the mass of the shocked 
CSM is comparable to that of the ejecta and the reverse shock is no longer 
propagating through the ejecta \citep{Svirski_etal_2012}.

We point out that the spectropolarimetric observations were only conducted 
after the first break of the bolometric luminosity; hence, they may not 
provide insights into the nature of this change. However, the synthesised 
broad-band polarization at these late epochs is significantly lower than 
the broad-band polarization measured on day $\sim 140$, obtained $\sim$30 
days before the break. Therefore, we do not rule out that the initial 
luminosity break at day $\sim 170$ could be associated with a significant 
change of the geometry of the interaction zone between the ejecta and the 
CSM. Such a break can be caused by the uneven diminishing of the reverse 
shock if the shocked shell reaches the boundary of the dense part of the 
CSM \citep{Moriya_2014}. For instance, this can be expected when the shock 
front has crossed the volume defined by the semiminor axis of a 
hypothetical dense ellipsoid but has not yet fully traversed the range 
spanned by the semimajor axis. Follow-up photometry has shown a secondary 
break in the multiband light curves of SN\,2018evt at day $\sim$480 (see 
Sec.~\ref{sec_mass} and Wang, Lingzhi et al., in prep.). The time and 
amplitude of bolometric decline-rate variations are essential for modeling 
the mass and spatial extent of the CSM. 

(2) The late-time H$\alpha$ emission can be satisfactorily described by the 
superposition of a pseudocontinuum and a broad and an intermediate Gaussian 
component, with widths at day 198 of  
FWHM$_{\rm broad}^{\rm H\alpha} = 7198\pm34$\,km\,s$^{-1}$ and 
FWHM$_{\rm int}^{\rm H\alpha} = 2291\pm33$\,km\,s$^{-1}$, respectively. The 
broad component exhibits conspicuous time evolution while the intermediate 
component is relatively stationary until day 300 (Figs.~\ref{Fig_ha_evolve2} 
and \ref{Fig_ha_evolve3}). After that, the intermediate component shifts 
toward longer wavelengths.

(3) The VLT spectrum with higher spectral resolution obtained on day 198 
reveals the P~Cygni nature of the inner H$\alpha$ profile. The core region 
can be well fitted by two narrow Gaussian functions characterising the 
absorption (FWHM $=110\pm16$\,km\,s$^{-1}$) and the emission 
(FWHM $=133\pm24$\,km\,s$^{-1}$) components. The expansion velocity measured 
in the absorption component amounts to $v_{\rm wind}=63\pm17$\,km\,s$^{-1}$ 
(see, e.g., Fig.~\ref{Fig_pcygni}). 

(4) As the luminosity of SN\,2018evt dropped, the strength of the Balmer 
lines relative to the underlying continuum increased between days $\sim 125$ 
and 240. The relative contribution by the blue wing to the total H$\alpha$ 
line emission decreased nonmonotonically through the end of our spectral 
series at day 365. The relative intensity of the red wing increased between 
days $\sim 125$ and 240 (see bottom panels of Figs.~\ref{Fig_ha_evolve2} 
and \ref{Fig_ha_evolve3}). 

The changes in line structure were accompanied by a shift of the central 
peak of the broad H$\alpha$ component from $-1200$\,km\,s$^{-1}$ to 
$\sim +100$\,km\,s$^{-1}$ between days 125 and 365. This shift can be 
understood as follows. The broad H$\alpha$ component is produced in a cold 
dense shell (CDS) within the region between the forward and reverse shocks 
(see Section~\ref{sec_interact} for more details). As the ejecta expand over 
time and become progressively optically thin, the occultation of the red-side 
emission of H$\alpha$ gradually decreases.
This results in an increase in the observed red-wing intensity and leads to 
a redward shift of the peak of the H$\alpha$ profile. Note that such a 
behaviour is different from the H$\alpha$ evolution at earlier phases 
reported for other interacting SNe. For instance, during days $\sim 75$--100, 
SN\,1997cy-like events often show a decreased intensity in the red wing 
and an increased intensity in the blue wing. These signatures identified at 
relatively early phases were interpreted as a result of the formation of new 
dust grains in the shocked material (see, e.g., \citealp{Lucy_etal_1989} for 
the first documented case of SN\,1987A and \citealp{Smith_etal_2009, 
Trundle_etal_2009, Fox_etal_2011, Smith_etal_2012}, 
and \citealp{Silverman_etal_2013} for SN\,1997cy-like events, 
and \citealp{Zhang_etal_2021} for the Type II SN\,2018hfm). 

(5) The continuum polarization of SN\,2018evt decreased substantially over 
time. The $B$- and $V$-band polarizations were both $\sim1.3$\% on day 140. 
The continuum polarization dropped from $\sim 0.6$\% to 0.3\% between days 
172 and 219. The polarization position angle remained constant within the 
uncertainties. That is, while the nonsphericity of the ejecta-CSM 
interaction region decreased with the recession of the photosphere into 
interior zones, the orientation of this asphericity did not change 
significantly on the plane of the sky 
(Figs.~\ref{Fig_iqu_ep1}--\ref{Fig_iqu_ep4}, Table~\ref{Table_pol}).

(6) The narrow central peak of the prominent H$\alpha$ emission line is 
almost completely unpolarized (Figs.~\ref{Fig_iqu_ep3} and \ref{Fig_ipth}). 
Such a depolarization probably occurred in an H-recombination zone. The 
residual low line-of-sight polarization level of $\sim 0.14$\% 
(Fig.~\ref{Fig_isp_ha}) has been adopted for the ISP correction. At all 
four epochs, the peak polarization in the wings is a factor of $\sim 2.5$
higher than the continuum polarization. The position angles across the 
emission profile are different from those measured in the continuum 
(Table~\ref{Table_pol}). 

Since the intrinsic polarization of the H$\alpha$ recombination core is zero, 
we attributed a line-of-sight polarization level of 0.14$\pm$0.08\% to the 
ISP (see, Section~\ref{sec_isp} and Appendix~\ref{app_isp}). In other words, 
if a substantial amount of scattering dust is present in the CSM, an apparent 
line polarization at the H$\alpha$ emission core would be expected, which is 
incompatible with the observed low line-of-sight polarization level. 
Therefore, we infer that the number density of dust grains in the volume 
within the first $\sim 200$\,days is probably insignificant. Furthermore, the 
asymmetry of the emission-line profiles is intrinsic and not only apparent 
owing to obscuration by dust. CSM with such a low dust content is consistent 
with the configuration of the circumstellar environment of the Type IIn 
SNe\,1997eg \citep{Hoffman_etal_2008} and 
2010jl \citep{Patat_etal_2011_sn2010jl}. 

(7) The polarization across the H$\alpha$ line was found to increase 
monotonically from the minimum at the emission peak toward both shorter and 
longer wavelengths. The polarized flux exhibits an enhancement in the blue 
wing relative to the red wing as shown in the bottom panels of 
Figure~\ref{Fig_ipth}. The peaks of the polarization are outside the 
intermediate emission component of H$\alpha$, suggesting that the polarized 
flux is contributed by the broad component while the intermediate component 
is likely to depolarize the emission. Such a structure was first seen in the 
H$\alpha$ polarization profile of the Type IIn 
SN\,1998S \citep{Leonard_etal_2000}. It can be explained by the fractional 
contribution of the flux from the polarized continuum increasing toward the 
edges of the depolarizing intermediate component.

An alternative mechanism that explains the broad component of H$\alpha$ is given by the line-emitting clouds and its large number of fragmented cloudlets. According to the Monte-Carlo calculation based on the late-time observation of the Type IIn SN\,2008iy, a large number of small, fragmented clouds ($N_{f}\gtrsim 10^{6}$) is required to account for the observed smoothness of the H$\alpha$ profile, compared to the number of the non-fragmented, shocked clouds ($N_{c}\sim 10^{3}$, \citealp{Chugai_2009, Chugai_2018, Chugai_2021}). 
We suggest that such a configuration is also compatible with the decreased 
polarization toward the emission center of the H$\alpha$ profile since the 
flux will become progressively dominated by recombination toward the 
lower velocities, which is intrinsically unpolarized. The reproduction of the 
line profile through numerical simulations will be essential to probe the 
detailed line-forming mechanisms.

Although similar polarization profiles have been identified in the Type IIn 
SNe\,1997eg \citep{Hoffman_etal_2008} and 
2010jl \citep{Patat_etal_2011_sn2010jl}, we do not see strong evidence of a 
significant difference between the polarization angle over the H$\alpha$ 
profile and the pseudocontinuum as shown by those other events. The 
polarization position angle across the H$\alpha$ profile also exhibits 
little wavelength dependence. Therefore, the continuum-emitting region and 
the H-rich component may share a similar axial symmetry. 

\section{Discussion} \label{sec_discussion}
\subsection{Comparison Between SNe\,IIn and SNe\,Ia-CSM~\label{sec_iin_iacsm}}
In Table~\ref{Table_iin_iacsm}, we briefly compare the general observational 
properties of SNe\,IIn and SNe\,Ia-CSM. The core difference between the two 
types of strongly interacting SNe is obviously whether the underlying 
explosion has a thermonuclear origin or is due to core collapse. The 
comparison suggests both similarities and discrepancies. For example, 
SNe\,IIn exhibit a bimodal distribution of their rise times, and SNe\,Ia-CSM 
are on average more luminous. The identification of strong 
\ion{He}{{\sc I}} $\lambda$5876 and \ion{O}{{\sc I}} $\lambda$7774 in the 
late phases of SNe\,IIn also differentiates the two classes. Moreover, it 
seems that the wind velocities inferred from the blueshift of narrow P~Cygni 
absorption features in SNe\,Ia-CSM fall into a narrow and low range ($\sim 50$--100\,km\,s$^{-1}$), while SNe\,IIn exhibit a wider range of wind velocity ($\sim 20$--800\,km\,s$^{-1}$).
However, the small sample size of long-term 
polarimetric temporal series of both SNe\,IIn and SNe\,Ia-CSM is not (yet) 
sufficient to deduce any time patterns. 

\begin{table*}
\begin{center}
\caption{Comparison of the general observational properties of SNe\,IIn and SNe\,Ia-CSM. \label{Table_iin_iacsm}}
\begin{small}
\begin{tabular}{c|c|c|c}
\hline
            &             & SNe\,IIn       & SNe\,Ia-CSM \\
\hline
Light Curve & Rise Time & Fast, 20$\pm$6\,d; slow 50$\pm$11\,d$^{[a]}$  & 20$\lesssim t_{\rm rise} \lesssim 40$\,d$^{[b]}$  \\
            &             & $20 \lesssim t_{\rm rise} \lesssim 50$\,d$^{[c]}$  & \\\cline{2-4}
            & Peak Mag    & $-20 \lesssim M_{r} \lesssim -17.8$\,mag$^{[a]}$, & $-21.3 \lesssim M_{R} \lesssim -19$\,mag$^{[b]}$ \\
            &             &  $-19 \lesssim M_{R} \lesssim -16$\,mag$^{[c]}$  &  \\
\hline
Spectra     & Early       &  Balmer lines + blue continuum$^{[d]}$  &  Balmer lines + SN\,1991T-like spectrum      \\
            &             &           &  strong \ion{Fe}{{\sc III}}, weak/no \ion{Ca}{{\sc II}} and \ion{Si}{{\sc II}}      \\\cline{2-4}
            & Late        & \ion{He}{{\sc I}} $\lambda$5876  &  Weak He features$^{[b]}$ \\
            &             & Prominent \ion{O}{{\sc I}} $\lambda$7774  & No strong evidence of O$^{[b]}$  \\
            & Wind Velocity  & $20 \lesssim v_{\rm wind} \lesssim 800$\,km\,s$^{-1}$ $^{[e]}$  & $50 \lesssim v_{\rm wind} \lesssim 100$\,km\,s$^{-1}$ $^{[f]}$  \\
\hline
Polarization & Continuum   & Around peak, 1.7\%$\lesssim p^{\rm Cont}\lesssim 3.0$\%$^{[g]}$  &  day 141, $p^{\rm Cont}\approx1.3$\%$^{[*]}$; $\gtrsim$ day 200, $\lesssim 0.5\%^{[h,*]}$    \\
            &             & Decrease over time$^{[i]}$  &   Decrease over time $^{[j,*]}$    \\\cline{2-4}
          & Position Angle   & Exhibit little time evolution  & Same as SN~IIn$^{[*]}$    
    \\\cline{2-4}
            & Balmer Lines & Depolarized at the narrow emission core, & Same as SN~IIn$^{[*]}$  \\
            &              & increase and exceed $p^{\rm Cont}$ toward outer wings &         \\
            &              & Higher polarization peak in the red wing$^{[k]}$  &   \\
            \cline{3-4}
            &              & Misaligned with the ejecta and He-rich CSM$^{[k]}$  &   No misalignment between H-rich CSM and ejecta$^{[*]}$ \\            
            \cline{2-4}
             & \ion{He}{{\sc I}}  & Misaligned with the H-rich CSM,  & No \ion{He}{{\sc I}} emission  \\
            &              & aligned with the ejecta$^{[k]}$  &   \\ 
\hline
\end{tabular}\\ 
{$^{[a]}$}{\citet{Nyholm_etal_2020}}, 
{$^{[b]}$}{\citet{Silverman_etal_2013}}, 
{$^{[c]}$}{\citet{Kiewe_etal_2012}}, 
{$^{[d]}$}{\citet{Filippenko_1997}}, 
{$^{[e]}$}{\citet{Salamanca_etal_1998, Fassia_etal_2001, Salamanca_etal_2002, Pastorello_etal_2002, Miller_etal_2010_08iy, Fransson_etal_2014, Inserra_etal_2014, Fox_etal_2015, Inserra_etal_2016, Andrews_etal_2017, Chugai_2019, Tartaglia_etal_2020, Taddia_etal_2020}}, 
{$^{[f]}$}{\citet{Kotak_etal_2004, Aldering_etal_2006, Dilday_etal_2012, Silverman_etal_2013_PTF11kx, Silverman_etal_2013}}, 
{$^{[g]}$}{\citet{Wang_etal_2004}}, 
{$^{[h]}$}{\citet{Leonard_etal_2000}}, 
{$^{[i]}$}{\citet{Hoffman_etal_2008}}, 
{$^{[j]}$}{\citet{Inserra_etal_2014}},
{$^{[k]}$}{\citet{Hoffman_etal_2008}}, 
{$^{[*]}$}{This work}.\\
\end{small}
\end{center}
\end{table*}

\subsection{The Structure of the SN-CSM Interaction Region} \label{sec_interact}
The high continuum polarization is also indicative of significant 
asphericity of the SN-CSM interaction region. Because both 
SN\,2002ic \citep{Wang_etal_2004} and SN\,2018evt exhibited considerable 
polarization at late times, major departures from spherical symmetry could 
be an intrinsic property of the SN\,1997cy-like events. Moreover, the late-time 
spectroscopic and spectropolarimetric properties of SNe\,Ia-CSM and SNe\,IIn 
exhibit considerable similarity, suggesting commonality in the configuration 
of the SN-CSM interaction region of the two classes. 

The polarization angle of a given emission component carries information 
about its geometric orientation. The dominant polarization axis, which is 
defined over the entire optical range, remained unchanged with time (see 
Table~\ref{Table_pol}). 

\textcolor{black}{One configuration that can produce a wavelength-independent polarization level and a time-invariant polarization position angle is an aspherical  ejecta-CSM interaction zone. 
As the ejecta expand homologously, their scattering opacity decreases as the column density declines. 
In SN\,2018evt, we observed monotonically decreasing levels of polarization across both the continuum and the broad H$\alpha$ wings, which is contradictory to the optically-thick regime ($\tau \textgreater 1$). 
This is because if $\tau$ is large, the reduction of multiple scattering will first lead to an increase of the continuum polarization until reaching the highest level when $\tau \approx 1$ before the polarization starts to decrease over time \citep{Hoeflich_1991}. 
For SN\,2018evt, the fact that the continuum polarization decreases from days 172 to 219 can be understood as the expansion of the SN ejecta leading to a continuous reduction in the scattering optical depth. 
Additionally, one picture that can qualitatively explain the polarization in the H$\alpha$ wings is the presence of an aspherical H-rich circumstellar envelope. 
The polarization may arise from electron scattering in the aspherical ejecta-CSM interaction zone. 
However, electron scattering cannot be the only source that shapes the broad H$\alpha$ component; Dopper broadening must play a significant role (see explanation below). 
As the CDS expands, the flux of the broad H$\alpha$ wings becomes progressively dominated by Doppler broadening while the contribution from electron scattering decreases, resulting in a decrease of the polarization in the broad H$\alpha$ wings. }

In the late phases of SN\,2018evt, the expansion speed of the CDS can be 
inferred from the FWHM of the broad H$\alpha$ 
component \citep{Dessart_etal_2015, Smith_2017}. The rationale is mainly based 
on the fact that the CDS has become transparent to the radiation from the 
inner ejecta. Simulations of the spectral line profiles suggest that, at early 
times, complete thermalisation is taking place in the CDS. This yields a high 
optical depth and accounts for the majority of line broadening through 
noncoherent scattering 
with thermal electrons \citep{Chugai_etal_2001, Dessart_etal_2009, 
Dessart_etal_2015}. 
As the CDS expands, thermalisation becomes incomplete over 
all depths in the CSM, and the profile of the broad component becomes 
progressively dominated by the broadening from the large-scale velocity 
of the CDS \citep{Dessart_etal_2009, Dessart_etal_2015, Taddia_etal_2020}. For 
instance, the electron-scattering optical depth drops below 2/3 after day 
$\sim$350, implying a weak frequency-redistribution mechanism at such late 
phases \citep{Dessart_etal_2015}. 


\textcolor{black}{
\citet{Taddia_etal_2020} proposed that the intermediate component of the H$\alpha$ line of the Type IIn SN\,2013L arises from the pre-ionised gas. 
Such a region in the unshocked dense CSM is also expanding at the same wind velocity as the narrow H$\alpha$ component ($\sim 100$\,km\,s$^{-1}$). 
The H$\alpha$ emission was broadened to form the intermediate component with FWHM $\approx 1000$\,km\,s$^{-1}$ in this optically thick ($\tau \textgreater 1$) region, 
while the narrow emission peak originates in the outer, optically thin ($\tau \textless 1$) CSM.
However, this picture may not be able to account for the H$\alpha$ profile observed in SN\,2018evt. 
The higher-resolution spectrum of SN\,2018evt at day 198 shows that the narrow H$\alpha$ component is clearly separated from the underlying intermediate component (see Figure~\ref{Fig_pcygni} and the upper-right panel of Figure~\ref{Fig_ipth}), indicating that the latter cannot be developed through broadening by electron scattering in the unshocked CSM expanding at $\sim 63$\,km\,s$^{-1}$.
}

\textcolor{black}{
An alternative scheme which may account for such a discrete central line profile was proposed by \citep{Chugai_2021} based on late-time observations of the Type IIn SN\,2008iy, which shows a narrow P~Cygni profile superimposed on an intermediate (FWHM $\approx 2000$\,km\,s$^{-1}$) component at day 702. 
The intermediate component can be interpreted to arise from a zone that contains shocked and fragmented circumstellar clouds \citep{Chugai_etal_1994}. 
These cloud have already been processed by the forward shock but have not been disturbed by the expanding SN ejecta, which corresponds to the CDS. 
The H$\alpha$-emitting gas inhabits the velocity spectrum covered by the intermediate component, 
from the high-velocity range that is similar to the shock speed which accelerates the gas in the CSM, to the low-velocity range that represents the region that has not yet fragmented but accelerated through the development of vortical turbulence in shear flows (the Kelvin-Helmholtz instability). 
The velocity range thus determines the main profile of the intermediate component. 
Additionally, the smooth H$\alpha$ profile requires a sufficient fragmentation of the shocked circumstellar clouds. 
}

\textcolor{black}{
The narrow component is formed in the unshocked wind outside the CDS. 
According to the modeling of SN\,2008iy at day 702, the wind speed is higher toward the inner region of the CSM owing to the acceleration by the expanding CDS. 
The pre-shock CSM around SN\,2008iy was accelerated to 145\,km\,s$^{-1}$ at the inner layers as indicated by the blueshifted absorption component, compared to the 45\,km\,s$^{-1}$ wind speed required at the outermost layers as constrained by the emission component \citep{Chugai_2021}. 
On the contrary, a constant wind speed cannot achieve a satisfactory fit to the emission component \citep{Chugai_2021}.
}

\textcolor{black}{
While a plausible modeling of SN\,2008iy at day 702 requires minimal contribution of the CDS and electron scattering, 
we note that our high-resolution spectropolarimetry obtained on day 198 is much earlier. 
At this relatively early phase, the polarization signatures demonstrate a nonnegligible contribution of electron scattering during our observations of SN\,2018evt. 
Evidence includes the temporal evolution of the continuum and the broad H$\alpha$ wing, the progressive depolarization toward the center of the H$\alpha$ line, and the peak in the polarization spectrum on day 198 around $-100$ to $-200$\,km\,s$^{-1}$, which is similar to the velocity of the narrow blueshifted absorption minimum. 
Therefore, the H$\alpha$ profile is likely to be developed through an interplay between electron scattering and the emitting fragmented circumstellar clouds. 
The former would produce polarization signals, while the latter would produce strong depolarization over the velocity spectrum covered by the cloud fragments since the emitting flux is from recombination rather than scattering. 
The latter becomes progressively more dominant over time as the ejecta and the CDS expand, causing a decrease of polarized flux in all spectral regions.
}

\textcolor{black}{
One concern about such an interplay scenario of SN\,2018evt is the small amount of variation of the width of the intermediate component from days 125 to 365 (see Figures~\ref{Fig_ha_evolve2} and \ref{Fig_ha_evolve3}). 
The width of the intermediate component is determined by the shock velocity ($v_{\rm sh}$) and density contrast between the cloud ($\rho_{\rm c}$) and intercloud gas ($\rho_{\rm ic}$): $v_{\rm c} \propto v_{\rm sh}/\sqrt{\rho_{\rm c}/\rho_{\rm ic}}$ \citep{Chugai_2019}. 
As the CDS expands and the shock decelerates, the characteristic speed of the gas in the shocked clouds would also decrease over time. 
On the other hand, we found that the central depolarization of the H$\alpha$ component favors an emission source in which recombination becomes progressively dominated toward the line center.  
The velocity coverage of the intermediate H$\alpha$ is also in good agreement with the range between the two polarization peaks identified across the entire H$\alpha$ profile (from $-2000$\,km\,s$^{-1}$ to $+3200$\,km\,s$^{-1}$; see Figure~\ref{Fig_ipth}). 
We remarked that our Gaussian decomposition and fitting of the H$\alpha$ profile of SN\,2018evt only serve as a qualitative description to the observations. 
The actually profiles originating from different zones may conspicuously differ from a Gaussian function. 
The detailed profile of each component and whether their temporal evolution can be understood under the framework of the fragmented shocked clouds scheme would require extensive numerical simulations. 
}


\begin{figure*}
\includegraphics[width=1.0\linewidth]{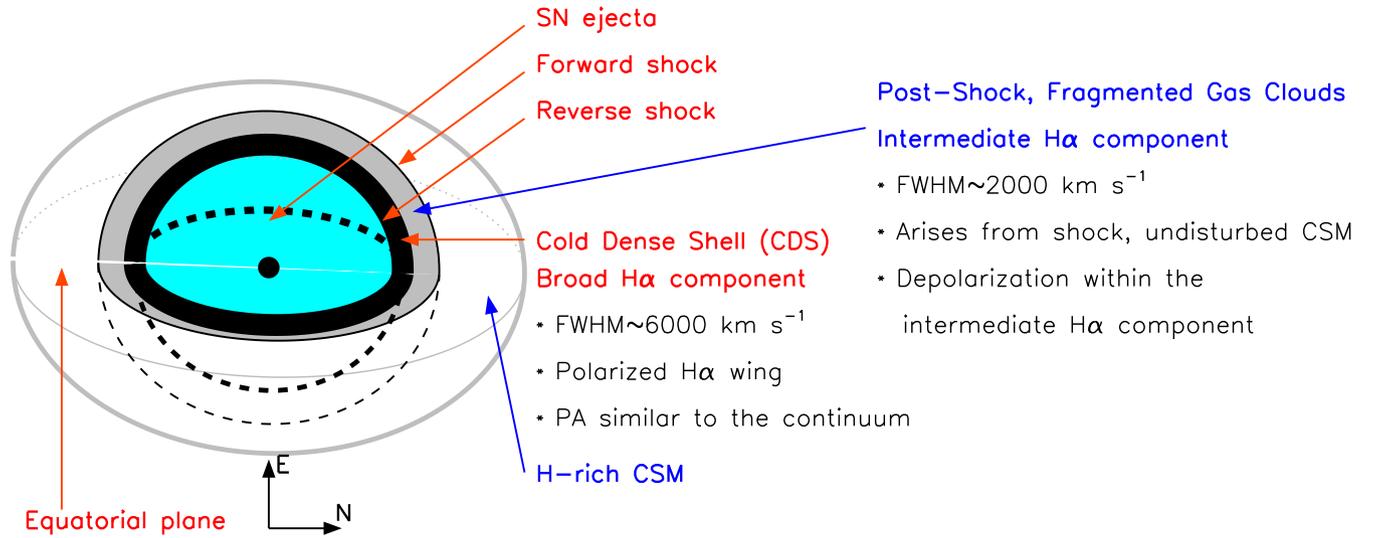}
\caption{Schematic sketch (not to scale) of the CSM geometry of SN\,2018evt viewed from a direction perpendicular to the plane of sky (at a phase of about 1\,yr), as explained in Section~\ref{sec_interact}. 
Owing to the lack of evidence to the contrary, the drawing assumes that the SN ejecta and CSM shell are concentric and not tilted with respect to one another.
\label{Fig_schem}
}
\end{figure*} 

In Figure~\ref{Fig_schem} we propose a schematic picture of the ejecta-CSM 
interaction in SN\,2018evt. Narrow, intermediate, and broad components of 
H$\alpha$ were first recognised in the spectrum of 
SN\,1988Z \citep{Filippenko_1991} and identified by \citet{Chugai_etal_1994}. 
The basic configuration has been adapted from the sketch of CSM-interacting 
SNe (for example, Type IIn and Type Ibn) as presented in  Figure~1 
of \citet{Smith_2017}. 

Figure~\ref{Fig_schem} illustrates the following. \\
(1) The very inner region consists of the SN ejecta which expand into the CSM. \\
(2) A forward shock wave was created when the SN exploded and expanded into 
the ambient CSM. Meanwhile, a reverse shock is propagating into the SN ejecta. 
\textcolor{black}{The CSM swept up by the forward shock was heated and expanded.} \\
(3) As this reverse shock is propagating inward and heating the inside ejecta, 
the forward shock moves into the H-rich CSM, creating a photoionised layer. 
According to the hydrodynamics of shock 
propagation \citep{Chevalier_etal_1994}, the SN ejecta between the forward and 
the reverse shocks produce a high-density zone and subsequently a low 
ionisation parameter. Therefore, the gas has a relatively low temperature and 
was partially ionised. The high-velocity SN ejecta collide with the CSM, 
resulting in a broad H$\alpha$ emission component \citep{Chugai_etal_1994}.\\
\textcolor{black}{
(4) Gas processed by the forward shock but not yet disturbed was heated and accelerated. 
A fraction of the gas was fragmented into numerous clumps, developing a broad velocity spectrum of the H$\alpha$ emitting region. 
Electron scattering would also contribute to the entire H$\alpha$ profile, as indicated by the temporal evolution of the polarization across the continuum and the H$\alpha$ wings, and it becomes progressively less significant as the polarized flux decreased over time.
}

From error-weighted least-squares fitting of the Stokes $Q-U$ measurements for 
epochs 1, 2, and 4 (see Fig.~\ref{Fig_qu} and Table.~\ref{Table_pol}), we 
estimate that $PA^{\rm Cont} \approx 89^{\circ}$. If the continuum polarization 
arises from an ellipsoidal CSM scattering zone \citep{Hoeflich_1991}, we 
conclude that the $PA$ of its semimajor axis is $\sim 179^{\circ}$ on the 
plane of the sky because it is perpendicular to the $PA^{\rm Cont}$ (see 
Fig.\,\ref{Fig_schem}). The continuum-emitting region and the H-rich component 
have not been distinguished and presented as one axial symmetry in the 
schematic diagram since no strong evidence of different $PA$ between the 
pseudocontinuum and across the H$\alpha$ profile has been identified. 

In the outer layer of the CSM, the shape 
of the region will be traced by the direction of the last scattering. In an 
atmosphere with Thomson scattering by free electrons, the level of 
polarization measured in SN\,2018evt at early times suggests considerable 
asphericity. For example, with a density distribution following an 
inverse-square law, $n(r) \propto r^{-2}$, and a close inner boundary of CSM 
(1\% of the radius of the semiminor axis of a hypothetical ellipsoidal CSM 
configuration), the continuum polarization ($p \approx 1.35$\%) at day 141 
indicates an axial ratio 0.75 \citep{Hoeflich_1991}. At days 195 and 219, as 
the ejecta are moving through the CSM, the polarized flux decreases and the 
estimated effective axial ratio becomes greater than 0.9. We note that these 
are only upper limits in the case when the SN is seen from the equator 
(perpendicular to the axis of symmetry). The true axis ratio could be 
smaller, which would yield a larger asphericity.  

A few weeks after the onset of the interaction between the SN ejecta and CSM, 
a $\sim 1$--2\% continuum polarization developed in the Type IIn SN\,201jl. 
At the same time, the wings became increasingly depolarized from the 
H$\alpha$ wings toward the rest wavelength of the line 
centre \citep{Patat_etal_2011_sn2010jl}.

In SN\,2018evt, an almost complete depolarization is also seen at the line 
centre of the H$\alpha$ profile, since the narrow H$\alpha$ core is formed by 
photons emitted by the outermost H-rich CSM that experiences no scattering by 
free electrons (see, e.g., Fig.~18 of \citealp{Dessart_etal_2015}). Such 
polarization signatures show a remarkable resemblance to SN\,2018evt at much 
later phases. A two-dimensional polarized radiative-transfer model for overluminous interacting SNe adopting a prolate CSM with a certain 
pole-to-equator density ratio can reproduce the observed polarization 
properties of SN\,2010jl \citep{Dessart_etal_2015}. \citet{Dragulin_etal_2016} 
have developed a semianalytic model for the interaction of spherically 
symmetric stellar winds with the pre-existing CSM. The exact geometry, density 
profile, and optical depth of the CSM require detailed modeling that is beyond 
the scope of this paper. Multidimensional radiation hydrodynamics simulations 
at different phases are essential to model the time-evolving polarization 
profile of SN\,2018evt and further reveal the detailed structures of 
interacting SNe. 

The above schematic as illustrated in Figure~\ref{Fig_schem} explains the 
presence of both a broad, polarized H$\alpha$ wing and an intermediate-width, 
depolarized H$\alpha$ component. 
\textcolor{black}{The overall configuration of the ejecta-CSM 
interaction is also consistent with the picture of cloud fragmentation proposed by \citet{Chugai_etal_2004}, \citet{Chugai_2019}, and \citet{Chugai_2021}. 
The formation of the entire H$\alpha$ profile may be due to the combination of the emission from the CDS, the shocked and fragmented clouds, and electron scattering.
}
Detailed modelling is required 
to produce a more precise picture of the SN ejecta expanding into the CSM. 

The polarization spectrum of the H$\alpha$ emission exhibits an asymmetric 
profile, and the polarization peak in the red wing is slightly higher than 
that in the blue wing at all four epochs (Table~\ref{Table_pol}). The reason 
is that the H$\alpha$ line emission is higher in the blue than in the red 
wing, owing to the underlying pseudocontinuum ($p$ has already been 
intensity-normalised). A more meaningful representation is provided by the 
polarized flux ($p \times I$) shown in the bottom panels of 
Figure~\ref{Fig_ipth}. At all four epochs, the polarized flux exhibits an 
enhancement in the blue wing relative to the red wing. Such a behaviour was 
also reported for the H$\alpha$ profile of the Type IIn 
SNe\,1997eg \citep{Hoffman_etal_2008} and 
2010jl \citep{Patat_etal_2011_sn2010jl}. The red-wing deficit in the 
polarized flux may be the result of the flux from the receding half of the 
scattering region being partially blocked by the approaching side, which can 
be explained by an ellipsoidal or disk-like CSM configuration that is not 
face-on to our line of sight.

\subsection{CSM Mass Estimate} \label{sec_mass}
In this section, we conduct a rough order-of-magnitude estimation of the 
mass of the CSM around SN\,2018evt. We especially focus on the late-time 
bolometric luminosity, for which we consider the kinetic energy of the SN 
ejecta as the source of the radiation energy. Therefore, the mass estimated 
below accounts for the CSM swept up by the expanding explosion shock wave. 
As discussed in Section~\ref{sec_interact}, because the effective axial 
ratio of the ellipsoidal CSM is $\sim 0.9$ around day 200, a steady, 
spherically-symmetric mass loss is adopted for simplicity.

The observed late-time luminosity is dominated by the kinetic energy of the 
SN explosion interacting with the CSM which can be written as 
$dE_{\rm k} = q_{r} \times 4 \pi r_{\rm sh}^2 dr_{\rm sh}$. Here $r_{\rm sh}$ 
denotes the radius of a shell of CSM reached by the forward shock with 
negligible width $dr_{\rm sh}\ll r_{\rm sh}$. The term $q_{r}$ gives the 
dynamic pressure, or the kinetic energy per unit volume, and can be expressed 
in terms of the fluid density of the CSM ($\rho_{\rm csm}$) and the flow 
velocity: $q_{\rm r} = (1/2) \rho_{\rm csm} v_{\rm sh}^2$. Thus, the 
bolometric luminosity becomes $L = \epsilon_{k} 
\frac{dE_{k}}{dt} = 2\pi \epsilon_{k} \rho_{\rm csm} r_{\rm sh}^2 v_{\rm sh}^3$. 
In this expression, $\epsilon_{k}$ is the conversion efficiency from kinetic 
energy to radiation, and $v_{\rm sh}$ is the flow speed. 

Assuming that the radial density structure of the CSM follows a power law as 
$\rho_{r} = \rho_{0} r_{\rm sh}^{-s}$, the mass-loss rate can be written as 
$\dot M = 4 \pi r_{\rm sh}^2 \rho_{\rm csm} v_{\rm wind}$, where 
$v_{\rm wind}$ is the velocity of the wind from the progenitor. We remark 
that for the sake of simplicity, our raw approximation is based on the 
assumption of a steady, spherically symmetric mass-loss profile. Such a 
configuration is in contradiction to the implied disk-like CSM concentration 
from polarimetry. Therefore, the estimated mass of the CSM is more likely an 
upper limit. More careful modeling is required to investigate the effect of 
the asymmetric CSM on its mass profile. 

In the case of steady mass loss ($s=2$), $\dot M$ becomes 
$4 \pi \rho_{0} v_{\rm wind}$, and the bolometric luminosity can be written 
as $L = \frac{\epsilon_{k}}{2} \frac{\dot M}{v_{\rm wind}} v_{\rm sh}^3$. 
Therefore, $\dot M = \frac{2L v_{\rm wind}}{\epsilon_{k} v_{\rm sh}^3}$, or 
\begin{equation}
\dot M = \frac{3.17}{\epsilon_{k}}
\bigg{(} \frac{L}{10^{42}\,{\rm erg\,s^{-1}}} \bigg{)}
\bigg{(} \frac{v_{\rm wind}}{100\,\rm {km\,s^{-1}}} \bigg{)}
\bigg{(} \frac{v_{\rm sh}}{1000\,\rm{km\,s^{-1}}} \bigg{)}^{-3} 
\times 0.1\,{\rm M}_{\odot}\,{\rm yr^{-1}}.
\end{equation}

Because of the presence of significant H$\alpha$ emission as early as day 
$-9$ (Fig.~\ref{Fig_spec}), the CSM has a very small inner radius, 
$R_{\rm in}$. On day $\sim 365$, when the last optical spectrum presented in 
this paper was obtained, the residuals of the two-Gaussian component fitting 
(Fig.~\ref{Fig_ha_evolve1}) suggest that the narrow P~Cygni H$\alpha$ profile 
from the CSM still persisted. A rough lower limit on the outer radius 
$R_{\rm out}$ of the shell can be placed as
\begin{equation}
R_{\rm out} \approx v_{\rm CDS} \times t_{\rm shock}. 
\end{equation}

Here, $v_{\rm CDS}$ is the velocity of the CDS 
at the time of the fast expansion of the SN ejecta into the CSM.

$t_{\rm shock}$ denotes the time elapsed since the SN explosion. Since no 
significant increase of the decline rate was observed in the late-time 
bolometric luminosity until $\sim 1$\,yr after the estimated maximum light 
($\sim 385$ days after shock breakout), we infer that the reverse shock is 
still crossing the ejecta and the forward shock is still effectively 
interacting with the H-rich CSM at $t_{\rm shock} \approx 385$ days. Adopting 
$v_{\rm CDS} = (6580 \pm 140)$\,km\,s$^{-1}$ (see Fig.~\ref{Fig_ha_evolve3}e), 
we estimate that $R_{\rm out} \gtrsim (2.19\pm0.05) \times 10^{16}$\,cm. 

Therefore, the lower limit on the duration of the mass-loss phase of a 
hypothetical companion becomes $t_{\rm ml} = \frac{R_{\rm out}}{v_{\rm wind}}$, 
or 
\begin{equation}
t_{\rm ml} = 31.7
\bigg{(} \frac{R_{\rm out}}{10^{16}\,{\rm cm}} \bigg{)}
\bigg{(} \frac{v_{\rm wind}}{100\,{\rm km\,s}^{-1}} \bigg{)}^{-1}\,{\rm yr}. 
\end{equation}
\noindent
The mass of the CSM can be estimated as $M_{\rm csm} = \dot M t_{\rm ml}$, or 
\begin{equation}
M_{\rm csm} = \frac{10.0}{\epsilon_{k}}
\bigg{(} \frac{L}{10^{42}\,{\rm erg\,s^{-1}}} \bigg{)}
\bigg{(} \frac{v_{\rm sh}}{1000\,\rm {km\,s^{-1}}} \bigg{)}^{-3} 
\bigg{(} \frac{R_{\rm out}}{10^{16}\,{\rm cm}} \bigg{)}\,{\rm M}_{\odot}. 
\end{equation}

Finally, we attempt a rough estimate of the mass-loss rate and the total mass 
of the CSM. Based on the observational properties of SN\,2018evt at day 
$\sim 365$ (when the last spectrum was obtained), and assuming that the mass 
loss matches a steady-state wind law ($\rho_{\rm csm} \propto r^{-2}$), we 
adopt log\,$L \approx 42.44 \pm 0.02$\,erg\,s$^{-1}$ from Table~\ref{Table_bolo_18evt},
$v_{\rm sh} \approx v_{\rm CDS} \approx 6580 \pm 140$\,km\,s$^{-1}$ 
(see, e.g., Fig.~\ref{Fig_ha_evolve3}e), and 
$v_{\rm wind}=63 \pm 17$\,km\,s$^{-1}$ as determined from the P~Cygni feature. 
This leads to a mass-loss rate of 
$\dot M \approx (1.9 \pm 0.5) \times 10^{-3}\,/\epsilon_{k}\,{\rm M}_{\odot}$\,yr$^{-1}$ and a total mass of the CSM of $M_{\rm csm} \approx (0.21\pm0.03)\,/\epsilon_{k}\,{\rm M}_{\odot}$. 
The efficiency factor is highly uncertain. 
Simulations by \citet{van_Marle_etal_2010} suggest an enhanced conversion 
efficiency with an increasing density of the circumstellar shell. For instance, 
a 15--30\% of maximum efficiency in converting the kinetic energy of the shock 
to bolometric luminosity can be reached for a circumstellar shell mass of 
10\,M$_{\odot}$. Therefore, we infer that SN\,2018evt experienced an order of 
$\approx 10^{-2}$\,M$_{\odot}$\,yr$^{-1}$ mass-loss rate and exhibit at least 
an order of $\approx$M$_{\odot}$ of CSM.


Obviously, the inferred age and mass of the CSM are highly dependent on the 
size of the dense circumstellar cloud, which is not well constrained by the 
current observations. However, a lower limit can be derived from the fact 
that the latest spectrum about 1\,yr after the estimated peak luminosity 
is still dominated by strong Balmer emission lines. The slowly declining 
pseudobolometric luminosity observed until day 368 (when the last photometric 
point was obtained; see Sec.~\ref{sec_lc}) also indicates that the intense 
mass loss building up the CSM started earlier than the SN explosion, namely 
at least $t_{\rm shock}^{\rm lim} \approx 385$ days before the last spectrum 
was obtained. Therefore, lower limits on the duration of the mass loss and 
the mass of the CSM yields $t_{\rm ml}^{\rm lim} \approx (110 \pm 30)$\,yr. 

SN\,2018evt emerged from solar conjunction in Dec. 2019. The latest LCO 
$Bg'Vr'i'$ photometry shows a convincing light-curve break between 2019-12-14 
and 2020-02-08. For example, the $g'$ and $r'$ light-curve decline rates 
have changed from $(0.588\pm0.005)\,(100\,{\rm day})^{-1}$ and $(0.558\pm0.009)\,(100\,{\rm day})^{-1}$
(between days 170 and 370) to $(1.11\pm0.11)\,(100\,{\rm day})^{-1}$ and
$(1.31\pm0.07)\,(100\,{\rm day})^{-1}$ (between days 480 and 533), respectively. Although the time evolution of 
the bolometric luminosity is at best poorly documented owing to the lack of 
NIR photometry, we tentatively estimate that a secondary break of the SN 
luminosity evolution occurred around $t'_{\rm shock} \approx 500$ days after the 
explosion. Therefore, the outer bound of the dense CSM becomes 
$R'_{\rm out} \approx (2.84 \pm 0.06) \times 10^{16}$\,cm. The corresponding duration of the mass loss and the amount of the CSM then become 
$t'_{\rm ml} \approx (143 \pm 39)$\,yr and 
$M'_{\rm csm} \approx (0.27\pm0.04) / \epsilon_{k}\,{\rm M}_{\odot}$, 
respectively. Observational properties of SN\,2018evt obtained at even later 
epochs will be presented in a separate paper (Wang, Lingzhi et al., in prep.). 

\subsection{Implications of Various Pre-explosion Mass-Loss Models} \label{sec_models}
The inferred mass of the CSM around SN\,2018evt of several, and possibly even 
a few tens, solar masses is in extreme contrast to the upper mass limit 
($\lesssim 0.03\,{\rm M}_{\odot}$; \citealp{Lundqvist_etal_2013}) in both SD 
and DD models for normal SNe\,Ia that do not exhibit Balmer lines. The 
following examines various mass-loss mechanisms but dismisses most of them. 

The mass-loss rates of massive (3--7\,M$_{\odot}$) AGB stars appears to be similar 
tothe above estimates for SN\,2018evt (typically in the range of $10^{-8}$ to 
$10^{-5}$\,M$_{\odot}$\,yr$^{-1}$, but reaching up to 
$>10^{-4}$\,M$_{\odot}$\,yr$^{-1}$ in the superwind phase; \citealp{Hofner_etal_2018}). 
According to simulations of the evolution of solar-metallicity stars, 
the mass of an AGB star can be as large as $\sim 11$\,M$_{\odot}$ with an 
H mass fraction of $\sim 70$\% \citep{Siess_2006}. However, the mechanism driving 
such intense mass loss within a short time window just prior to the terminal 
explosion of the companion WD as a SN remains unexplained. 

Another way of producing much-enhanced mass loss in a binary system is 
Roche-lobe overflow (RLOF) at low wind velocity ($\sim 50$\,km\,s$^{-1}$; 
\citealp{Mohamed_etal_2007}). When the companion fills its Roche lobe, 
the mass-transfer rate is drastically enhanced (by two orders of 
magnitude) with respect to a wind, leading to intensive mass loss 
strongly concentrated toward the orbital plane. Mass transfer by 
Roche-lobe overflow (RLOF) can efficiently strip a star of its H 
envelope \citep{Smith_2014}. \citet{Meng_etal_2017} proposed that a 
common envelope could be formed during such a RLOF phase. Moreover, the 
explosion of hybrid CONe WDs with a nondegenerate companion and within 
such a massive H-rich common envelope may explain most of the 
observational features as well as the rates of 
SNe\,Ia-CSM \citep{Meng_etal_2018, Soker_2019}. 

The wind velocity of SN\,2018evt is very similar to that reported for 
PTF11kx ($65 \pm 10$\,km\,s$^{-1}$, \citealp{Dilday_etal_2012}) but only 
$\sim 1/3$ of that determined for the other two well-studied objects 
showing remarkably similar photometric and spectroscopic evolution, namely 
SNe\,2012ca \citep{Inserra_etal_2014} and 1999E \citep{Rigon_etal_2003}, 
which were suggested to be core-collapse events. The low wind velocity of 
SN\,2018evt together with a substantial amount of mass loss is consistent 
with the lower velocity bound of a few tens of km\,s$^{-1}$ during 
outbursts of LBVs. However, the current modeling suggests an order of 
magnitude lower mass-loss rate, even with a super-Eddington wind that 
might be driven by SOME dynamical instability during the final accretion 
phase to the WD \citep{Meng_etal_2017, Meng_etal_2018}. Therefore, we 
consider the high mass-loss rate estimated for SN\,2018evt may not be 
compatible with this scenario. A symbiotic nova system also requires that 
the CSM is highly concentrated in the orbital plane to be consistent with 
the photometric behaviour \citep{Dilday_etal_2012}. The high kinetic energy 
and large amount of CSM as inferred from the broad H$\alpha$ emission and 
the slowly evolving photometric and spectroscopic features perhaps suggest 
that the CSM around SN\,2018evt may not have originated from multiple 
eruptions from a recurrent nova. 

\citet{Chevalier_2012} discussed the possibility that a compact object may 
spiral into the central core of its companion star through a common-envelope 
evolution, which may lead to the explosion of a luminous, long-lasting 
SN~IIn with massive CSM. This mechanism accounts for the delay by tens to 
hundreds of years in the SN explosion after the intense mass loss (a few 
$10^{-2}$--$10^{-1}$\,M$_{\odot}$\,yr$^{-1}$). This picture may also be 
compatible with a thermonuclear explosion if the compact object is a WD 
\citep{Livio_etal_2003}. Furthermore, the enhanced CSM during the 
common-envelope phase is concentrated toward the orbital plane of the binary 
\citep{Terman_etal_1995, Taam_etal_2000}, which is in agreement with the 
similar axial symmetry suggested by SN\,2018evt. Detailed modeling of 
SN\,2018evt will be particularly useful to test whether the common-envelope 
scheme would be able to account for (1) the polarization and its temporal 
evolution, and (2) massive CSM ($\sim$\,M$_{\odot}$), as inferred from 
observations. 

The identification of an LBV progenitor of the nearby Type IIn SN\,2005gl in 
pre-explosion {\it HST} images provides a strong clue for SN explosions of 
LBVs, and indicates heavy mass loss and intense interaction between ejecta 
and CSM \citep{Gal-Yam_etal_2007}. The inferred mass loss of SN\,2018evt, 
$\dot M \approx (1.9 \pm 0.5) \times 10^{-3}\,/\epsilon_{k}\,{\rm M}_{\odot}$\,yr$^{-1}$ 
at a speed of $v_{\rm wind} = 68\pm17$\,km\,s$^{-1}$ over a period of at least 
$t_{\rm ml} \gtrsim (110 \pm 30)$\,yr, or $t'_{\rm ml} \approx (143 \pm 39)$\,yr, may 
resemble LBV-like eruptions during about a century before explosion. A single 
LBV eruption may last from years to a few decades. For example, the 1890 
eruption of $\eta$~Car has ejected a total mass of 10--20\,M$_{\odot}$ within 
$\sim 20$\,yr \citep{Smith_etal_2003, Smith_Ferland_2007, Smith_Frew_2011}. 
A long duration of the CSM build-up may require a series of eruptions. The 
mass loss of SN\,2018evt may fall into the LBV giant-eruption regime in 
Figure~3 of \citet{Smith_2017}, which illustrates the mass-loss rate as a 
function of wind velocity. 


One extreme case of an SN that exploded inside a massive H-rich envelope is 
given by the Type IIn SN\,2006gy \citep{Smith_etal_2007}. Recently, 
\citet{Jerkstrand_etal_2020} proposed that emission lines at day $\sim 400$ 
of this luminous event (i.e., $M_{R}^{\rm Peak} \approx -21.8$\,mag; 
\citealp{Smith_etal_2007}) are from neutral iron. The large mass of 
ground-state iron ($\gtrsim 0.3$\,M$_{\odot}$) expanding at 1500\,km\,s$^{-1}$ 
is unlikely to be produced by a core-collapse explosion for which the 
expected kinetic energy of the ejecta is an order of magnitude higher. 
Therefore, \citet{Jerkstrand_etal_2020} proposed that SN\,2006gy can be 
understood as a typical SN~Ia hitting a dense shell of CSM. 
If the modelled iron mass of SN\,2006gy is correct, it would establish a case 
of an extensive CSM enrichment preceding a common-envelope evolution that 
leads to a SN~Ia explosion. However, compared to other events that have been classified as SNe\,Ia-CSM, SN\,2006gy has an order-of-magnitude brighter peak luminosity, 
a much broader light curve, and significantly different spectral evolution. 
Whether the progenitor systems of various types of SNe interacting 
with a massive CSM are so heterogeneous is still pending observational tests.

\section{Summary} \label{sec_summary}

We reported the results of our photometric, spectroscopic, and polarimetric 
follow-up observations of the SN\,1997cy-like SN\,2018evt from about 100\,days to $\sim 1$\,yr after the estimated peak luminosity. We identified an early 
break in the pseudobolometric luminosity around day 170, followed by a major 
further acceleration of the decline one year after peak luminosity. Based on 
a steady mass-loss wind profile, we infer that SN\,2018evt exploded inside a 
massive circumstellar cloud with a lower mass limit $M'_{\rm csm} \approx (0.21\pm0.03)\,/\epsilon_{k}\,{\rm M}_{\odot}$, which may result from a mass-loss rate of 
$\dot M \approx (1.9 \pm 0.5) \times 10^{-3}\,{\rm M}_{\odot}$\,yr$^{-1}$ at a speed of $v_{\rm wind} = 68\pm17$\,km\,s$^{-1}$ over a period 
of at least $t_{\rm ml} \gtrsim (110 \pm 30)$\,yr. 

The polarization properties of SN\,2018evt also indicate some geometric 
similarities to SNe~IIn. For example, a high level of continuum 
polarization ($\sim 2$--3\%) and a significant depolarization at the central 
core of the Balmer lines have also been observed in the Type IIn SNe\,1997eg
\citep{Hoffman_etal_2008}, 1998S \citep{Leonard_etal_2000}, and 2010jl 
\citep{Patat_etal_2011_sn2010jl}. These features suggest that SNe~Ia-CSM and 
SNe~IIn alike possess substantial amounts of CSM with major deviations 
from spherical symmetry. Furthermore, multi-epoch spectropolarimetry shows 
that the level of the continuum polarization can decrease between about two 
weeks and 100 days after the explosion \citep{Hoffman_etal_2008}. No strong 
evidence of misalignment between the symmetry axes of the SN ejecta and the 
CSM has been identified in our observations. 

Furthermore, the polarization of the prominent H$\alpha$ profile consists of 
a polarized broad wing and a depolarized intermediate core. The former comes 
from the photons emitted by the high-velocity regions in the inner cold dense 
shell, and the latter arises from shocked, fragmented, emitting gas clouds that 
have not been disturbed by the SN ejecta. High-resolution spectropolarimetry 
has been crucial in unveiling these and other properties of SN\,2018evt. Not 
only does polarimetry achieve low-order spatial resolution regardless of 
angular size at any distance as long as the flux is sufficient but, at high 
spectral resolution, it can reveal structures in line profiles that, in 
total-flux spectra of SNe, are completely washed out by the extreme Doppler 
broadening. 

The formation of a very massive CSM structure within just a century before 
the explosion is the central challenge for any attempt to model SN\,2018evt 
and similar events involving either core-collapse or thermonuclear explosions. 
LBV eruptions as well as common-envelope evolution seem to be candidate 
mechanisms, and both may be realised by nature. LBV eruptions can only be 
connected to core-collapse events, whereas common-envelope phases may precede 
both thermonuclear and core-collapse explosions albeit with some preference 
for the former. 

The observational properties of SN\,2018evt exhibit a strong resemblance 
to the known SN\,1997cy-like events. However, the exact nature of this event 
may still remain unclear. The high and evolving polarization signal 
measured from SN\,2018evt can safely rule out any homogeneous mass-loss 
procedure and the presence of massive spherical shells of the CSM. A 
significant equator-to-polar mass-loss process of the progenitor system 
or the pre-existence of an aspherical protoplanetary nebula is favored. 

High-resolution spectropolarimetry will continue to be an important tool to 
characterise the configuration of the massive CSM, which accounts for most 
of the energy output of SN\,2018evt-like events. Additionally, spectroscopy 
at very late phases, when the optical depth of the interaction zone becomes 
sufficiently low, will identify unique fingerprints of the explosion physics. 
Observations and modelling of the line species, the ionisation states, and 
the line profiles should respectively add to the understanding of the 
explosion core's composition, physical conditions, and dynamics, thereby 
further elucidating the nature of these special events.

\section*{Acknowledgements}
We would like to thank the referee Dr. Nikolai N. Chugai for his careful scrutiny which resulted in very helpful, constructive suggestions that significantly improved the paper. 
Melissa L. Graham provided helpful discussion and assistance with follow-up 
observation planning. 
We are grateful to the European Organisation for Astronomical Research in 
the Southern Hemisphere (ESO) for the generous allocation of observing time. 
We especially thank the staff of the Paranal Observatory for 
their proficient and motivated support of this project in service mode. 
The polarimetric studies in this work are based on observations collected 
at ESO's La Silla Paranal Observatory under programme IDs 0102.D-0163(A) and 2102.D-5031. Some spectra presented in this work are based on observations collected as part of the extended Public ESO Spectroscopic Survey for 
Transient Objects (ePESSTO, \citealp{Smartt_etal_2015}), program 199.D-0143(M). 
GROND observations at La Silla were performed as part of program 102.A-9099. Some of the funding for GROND (both hardware as well as personnel) was generously provided by the Leibniz Prize to Prof. G. Hasinger (DFG grant HA 1850/28-1). 
We acknowledge usage of observations of SNe\,2018evt and \,2012ca made with the Las 
Cumbres Observatory (LCO) global telescope network. Other observations of SN\,2012ca were secured at ESO's La Silla Paranal Observatory as part of the PESSTO survey (ESO programs 188.D-3003 and 191.D-0935), as well as with the Panchromatic Robotic Optical Monitoring and Polarimetry Telescope (PROMPT) through CNTAC proposal CN2012A-103, with the Australian National University 2.3\,m Telescope, and with the {\it Neil Gehrels Swift Observatory}.
PyRAF, PyFITS, and STSCI$\_$PYTHON are products of the Space Telescope Science Institute, which is operated by the Association of Universities for Research in
Astronomy, Inc., under National Aeronautics and Space Admininistration (NASA) contract NAS5-26555. 
We also use data from the European Space Agency (ESA) mission
{\it Gaia} (\url{https://www.cosmos.esa.int/gaia}), processed by the {\it Gaia}
Data Processing and Analysis Consortium (DPAC,
\url{https://www.cosmos.esa.int/web/gaia/dpac/consortium}). 
Funding for the DPAC has been provided by national institutions, in particular the institutions participating in the {\it Gaia} Multilateral Agreement. 
This research has made use of NASA's Astrophysics Data System Bibliographic Services;
the SIMBAD database, operated at CDS, Strasbourg, France;
the NASA/IPAC Extragalactic Database (NED), which 
is operated by the Jet Propulsion Laboratory, California Institute of Technology, 
under contract with NASA.
The Infrared Telescope Facility is operated by the University of Hawaii under contract NNH14CK55B with NASA.

The research of Y.\ Yang has been supported through a Benoziyo Prize Postdoctoral Fellowship and the Bengier-Winslow-Robertson Fellowship. 
M.\ Bulla acknowledges support from the Swedish Research Council (Reg. no. 2020-03330).
T.\ W.\ Chen is grateful for funding from the Alexander von Humboldt Foundation and the EU Funding under Marie Sk\l{}odowska-Curie grant H2020-MSCA-IF-2018-842471.
A.\ V.\ Filippenko's group at U.C. Berkeley acknowledges generous support from the TABASGO Foundation, the Christopher R. Redlich fund, the Miller Institute for Basic Research in Science (in which A.V.F. was a Miller Senior Fellow), Sunil Nagaraj, Landon Noll, Gary and Cynthia Bengier, Clark and Sharon Winslow, Sanford Robertson, and many additional donors. 
L.\ G.\ acknowledges financial support from the Spanish Ministerio de Ciencia e Innovaci\'on (MCIN), the Agencia Estatal de Investigaci\'on (AEI) 10.13039/501100011033, and by the European Social Fund (ESF) "Investing in your future" under the 2019 Ram\'on y Cajal program RYC2019-027683-I and the PID2020-115253GA-I00 HOSTFLOWS project. 
A.\ Gal-Yam’s research is supported by the EU via ERC grant 725161, the ISF GW excellence center, an IMOS space infrastructure grant and BSF/Transformative and GIF grants, as well as The Benoziyo Endowment Fund for the Advancement of Science, the Deloro Institute for Advanced Research in Space and Optics, The Veronika A. Rabl Physics Discretionary Fund, Minerva, Yeda-Sela, and the Schwartz/Reisman Collaborative Science Program;  A.G.-Y. is the incumbent of the The Arlyn Imberman Professorial Chair.
D.\ Hiramatsu was supported by NSF grants AST-1313484 and AST-1911225, 
as well as by NASA grant 80NSSC19kf1639. 
P.\ H\"{o}flich acknowledges support from the NSF project
``Signatures of Type Ia Supernovae, New Physics, and Cosmology,'' grant 
AST-1715133. 
E.\ Y.\ Hsiao and M.\ Shahbandeh acknowledge support provided by NSF grant 
AST-161347 and the Florida Space Research Program. 
The research of J.\ Maund is supported through a Royal Society University Research Fellowship. 
C.\ McCully, and G.\ Hosseinzadeh were supported by NSF grant AST-1313484. 
C.\ Pellegrino was supported by NSF grant AST-1911225. 
Time-domain research by D.\ J.\ Sand is supported by NSF grants AST-1821987, 
1813466, and 1908972, and by the Heising-Simons Foundation under grant \#2020-1864. 
The supernova research by Lifan Wang is supported by NSF award AST-1817099 and HST-GO-14139.001-A.
Lingzhi Wang is sponsored in part by the Chinese Academy of Sciences 
(CAS), through a grant to the CAS South America Center for Astronomy 
(CASSACA) in Santiago, Chile. 
X.\ Wang is supported by the National Natural Science Foundation 
of China (NSFC grants 11178003 and 11325313). 
J.\ C.\ Wheeler is supported by NSF grant AST-1813825. 
The LCO team is supported by NSF grants AST-1911225 and AST-1911151.

\section*{Data Availability}

 The reduced data used in this work may be shared upon request to Yi Yang (yiyangtamu@gmail.com). 


\bibliographystyle{mnras}

\section*{Affiliations (Continued)}
\noindent
{\it
$^{21}$Department of Astronomy, University of Texas, Austin, TX 78712, USA \\
$^{22}$Purple Mountain Observatory, Chinese Academy of Sciences, Nanjing 210008, China \\
$^{23}$The Raymond and Beverly Sackler School of Physics and Astronomy, Tel Aviv University, Tel Aviv 69978, Israel \\
$^{24}$Steward Observatory, University of Arizona, 933 North Cherry Avenue, Tucson, AZ 85721-0065, USA \\
$^{25}$Gemini Observatory/NSF's NOIRLab, 670 N. A'ohoku Place, Hilo, Hawai'i, 96720, USA \\
$^{26}$Institute of Space Sciences (ICE, CSIC), Campus UAB, Carrer de Can Magrans, s/n, E-08193 Barcelona, Spain \\
$^{27}$Institut d’Estudis Espacials de Catalunya (IEEC), E-08034 Barcelona, Spain \\
$^{28}$Physics Department and Tsinghua Center for Astrophysics (THCA), Tsinghua University, Beijing, 100084, People's Republic of China \\
$^{29}$Beijing Planetarium, Beijing Academy of Science and Technology, Beijing, 100044, People's Republic of China 
}




\appendix

\section{Construction of the Pseudobolometric Light Curve} \label{app_bolo}
Based on the absence of 
significant spectral evolution from days 125 to 365, we consider that a 
single late-time spectrum adequately represents the major spectral features 
of SN\,2018evt during the late phases for which we obtained photometry. The 
VLT/FORS2 spectrum of SN\,2018evt at epoch 4 (day 219) was used to 
characterise the shape of the spectral energy distribution (SED) in the 
optical. An NIR spectrum obtained on day 262 by the NASA Infrared Telescope 
Facility (IRTF; \citealp{Rayner_etal_2003}) served to represent the spectral 
shape in the NIR. The details of the NIR spectral properties of SN\,2018evt 
at late phases will be discussed in a forthcoming paper  (Lingzhi Wang, 
et al., in prep).

We dereddened the optical and NIR spectra of SN\,2018evt adopting 
$E(B-V)^{\rm MW}_{\rm 18evt} = 0.051$\,mag and a Galactic $R_V = 3.1$ extinction law 
\citep{Cardelli_etal_1989}, scaled the NIR spectrum, and tied it to the red 
end of the optical spectrum to compensate for the different phases of 
observation. This procedure creates a composite optical-NIR spectral 
template. We then performed photometry on this template to obtain 
synthesised magnitudes in the $Bg'Vr'i'JHK$ bands. Thereafter, a warping 
procedure was applied to the template to match the difference between the 
synthesised magnitudes and the actual photometry of SN\,2018evt in the AB 
system after correcting for Galactic extinction. Finally, two 
pseudobolometric luminosity estimates were obtained at each photometric 
epoch by integrating the warped spectrum over two wavelength ranges, namely 
3870--9000\,\AA\ for the optical and 3870--23,200\,\AA\ for the optical-NIR.

\begin{figure}
\includegraphics[width=1.0\linewidth]{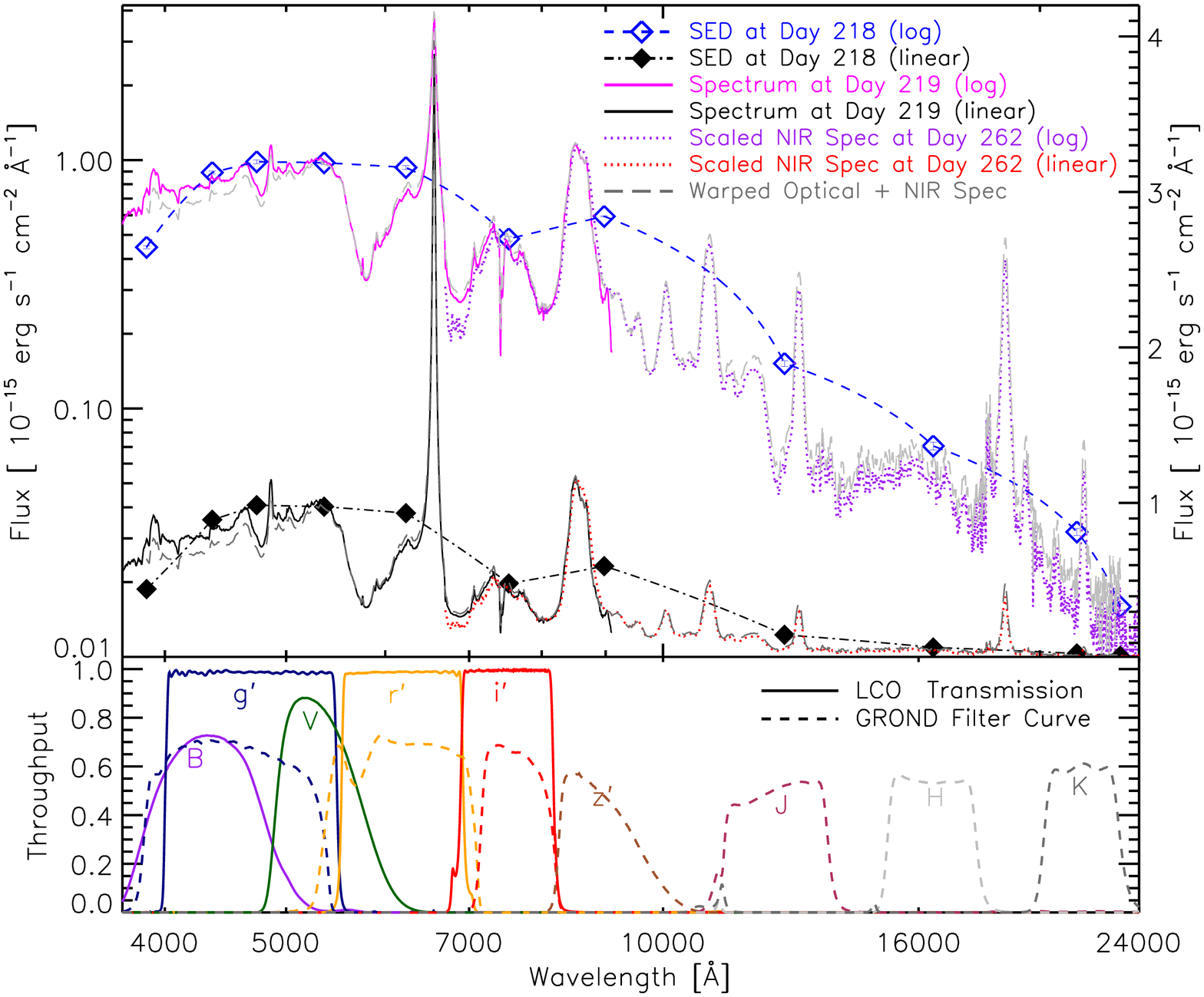}
\caption{The SED constructed for SN\,2018evt. In the upper panel, black 
diamonds show the flux from the photometry in each bandpass on day 218 at 
their pivotal wavelengths. The black dot-dashed line outlines the SED 
obtained by connecting the bandpass monochromatic fluxes. The black solid 
line and the red-dotted line represent the flux-calibrated VLT spectrum on 
day 219 and the scaled NIR spectrum on day 262, respectively. The composite 
optical-NIR spectral template (grey long-dashed line) was warped to match 
the photometry. The blue, purple, and grey lines and symbols present the 
same quantities on a logarithmic scale as indicated by the legend. The lower 
panel depicts the associated bandpass throughput curves. Notice the different 
shapes between the LCO and the GROND $g'$, $r'$, and $i'$ bandpasses.
\label{Fig_sed}
}
\end{figure}

The SED of SN\,2018evt is illustrated in Fig.~\ref{Fig_sed}. 
The wavelength-weighted mean flux densities for each bandpass were computed 
as $\langle F(\lambda) \rangle = \frac{\int \lambda F_{\lambda} T(\lambda) d\lambda}{\int \lambda T(\lambda) d\lambda}$, as in the STMAG system described by \citet{Koornneef_etal_1986}. The calculated mean flux densities are also 
shown in Fig.~\ref{Fig_sed}. Here $F_{\lambda}$ and $T_{\lambda}$ represent 
the flux density in units of erg\,cm$^{-1}$\,s$^{-1}$ \AA$^{-1}$ and 
dimensionless bandpass throughput at a given wavelength $\lambda$, respectively. Since the 
normalisation was applied to each bandpass individually, the absolute scale 
of the filter throughput will not affect the computed weighted mean flux 
density. A more detailed description of the steps is given by
\citet{Yang_etal_2018_latelc} and Appendix C of \citet{Yang_etal_2020}. 

\section{Fitting the Balmer line profile} \label{app_balmer_fit}
\begin{figure}
\includegraphics[width=1.0\linewidth]{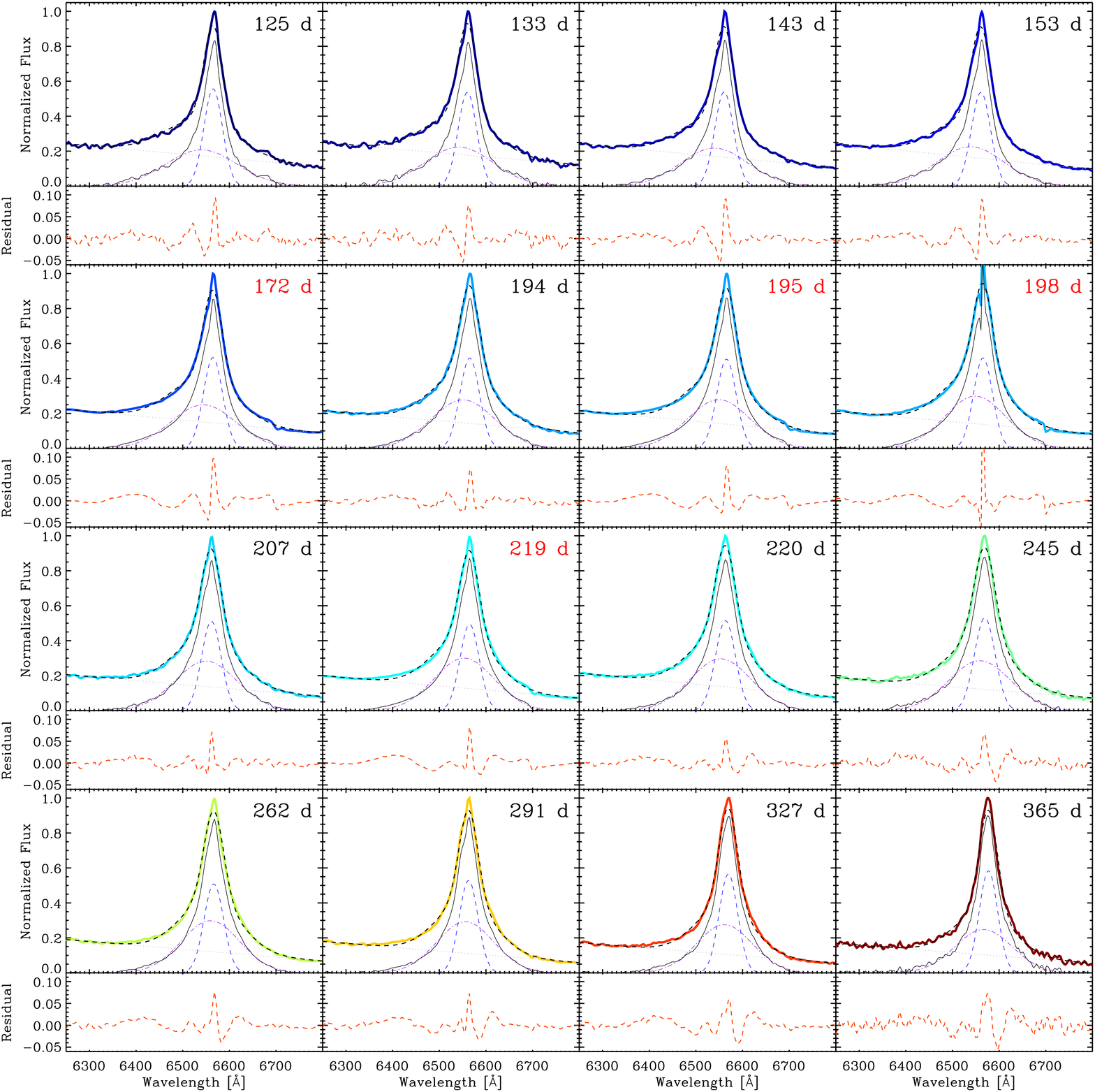}
\caption{Two-Gaussian component fitting of the late-time H$\alpha$ profile 
of SN\,2018evt. All spectra were resampled to the pixel size of the FLOYDS/LCO spectra, $\sim 1.7$\,\AA. In each subpanel, grey 
dotted, purple dot-dashed, and blue dashed lines represent the fits to the 
pseudocontinuum and to the broad and intermediate Gaussian components, 
respectively. The black dashed line presents the sum of these three 
components which tracks the fitted total-flux spectrum as shown by the 
colour-coded, solid curves. The lower subpanels display the corresponding 
residuals.}. 
\label{Fig_ha_evolve1}
\end{figure}

To better quantify the H$\alpha$ profile and probe its temporal evolution, we 
fitted the data as follows. First, we assigned the blue limit 
($\lambda_{\rm blue}^{\rm H\alpha} = 6330$\,\AA) and the red limit 
($\lambda_{\rm red}^{\rm H\alpha} = 6700$\,\AA) of the region of interest by 
visually inspecting the data. Second, we determined the pseudocontinuum by 
fitting a first-order polynomial to the wavelength ranges 6250--6330\,\AA\ 
and 6700--6800\,\AA. The limits were chosen to avoid any apparent spectral 
features and telluric absorption. Finally, we subtracted the pseudocontinuum 
and fitted the spectrum with a multiple Gaussian function, 
$f_c(\lambda) = \sum_{i=1}^n A_{i} \  {\rm exp} 
(- (\lambda_{i} - \lambda_{i}^{\rm rest})^2/(2 \sigma_{i}^2))$. 
The fitting parameters are the central wavelength in the rest frame 
($\lambda_i$), the width ($\sigma_i$), and the peak height ($A_i$) of 
each component. The results for two Gaussians can be seen in 
Figure~\ref{Fig_ha_evolve1}. 

The VLT flux spectrum at epoch 3 (day 198; also included in 
Fig.~\ref{Fig_ha_evolve1}) has a higher resolution than the other spectra. 
It clearly reveals that the narrow H$\alpha$ core exhibits a P~Cygni 
profile, which is unresolved in our other spectra, indicating the existence 
of a dense and optically thick CSM, into which the SN ejecta are expanding. 
Because the above two-Gaussian component fitting does not account for the 
contributions from this P~Cygni profile, overt residuals up to $\sim 6$\% 
arose around the line centre. However, the overall residuals across the 
wings ($\lesssim 2$\%) and the RMS value over the entire fitting range 
($\lesssim 1.5$\%) are both small. Therefore, we suggest that outside the 
central region dominated by the very narrow P~Cygni profile, the late-time 
H$\alpha$ emission of SN\,2018evt is satisfactorily described by a 
combination of a broad and an intermediate Gaussian component. 

\section{Estimation of the Interstellar Polarization} \label{app_isp}
Determining the minimal polarization in a spectral line is difficult, 
especially when noise and limited spectral resolution combine. The measurements 
will be biased toward higher levels if the line is not spectrally resolved. 
Our VLT spectropolarimetry at epoch 3 was conducted with a spectral resolution 
$R \approx 2140$ (3\,\AA\ FWHM) at a sampling of 0.73\,\AA\,pixel$^{-1}$. Such a 
relatively high spectral resolution enables a more thorough investigation of 
the ISP than is possible from low-resolution observations (see, e.g., 
\citealp{Wang_etal_2004}). 

\begin{figure}
\includegraphics[width=1.0\linewidth]{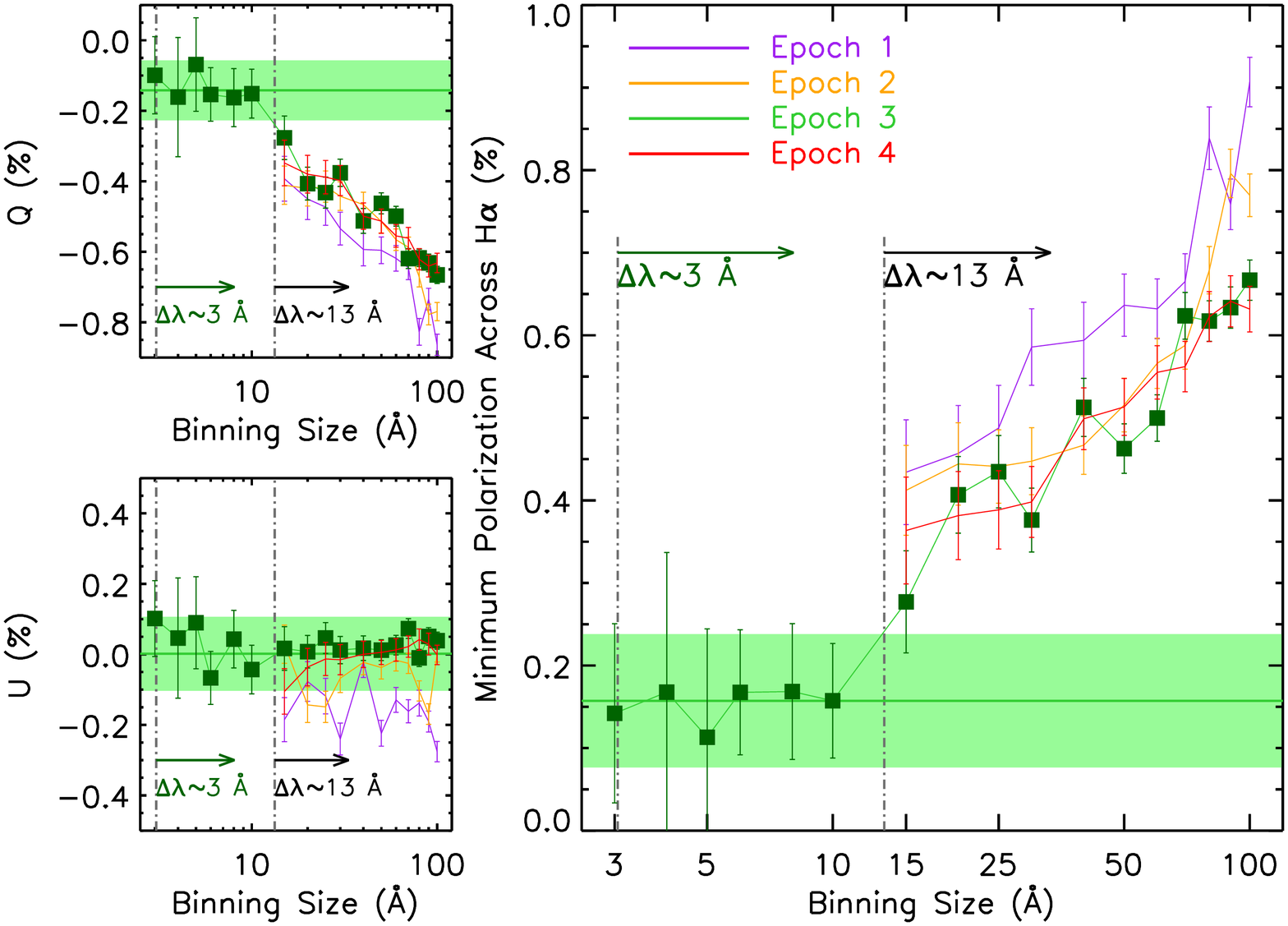}
\caption{The minimum polarization measured across the H$\alpha$ emission peak 
as a function of bin size (right panel). In each panel, vertical dot-dashed 
lines indicate the size of the resolution element in the observations, i.e., 
$\Delta \lambda \approx 13$\,\AA\ for epochs 1, 2, and 4, and 
$\Delta \lambda \approx 3$\,\AA\ for epoch 3. The small scatter exhibited by the 
dark-green filled squares in all panels suggests that the polarization 
measured with bin sizes of 3--10\,\AA\ corresponds to the actual ISP level 
if SN\,2018evt is intrinsically unpolarized in the narrow H$\alpha$ emission 
peak. The weighted mean value and the errors estimated in the main text are 
indicated by the green shaded areas. 
\label{Fig_isp_ha}
}
\end{figure}

In order to quantify the effect of spectral resolution on the measured 
polarization, we determined the minimum polarization across the H$\alpha$ 
emission core in the Stokes spectra using various bin sizes. We chose a number 
of bin sizes $\Delta \lambda > 13.3$\,\AA\ for the 300V grism 
configuration ($R \approx 440$) applied at epochs 1, 2, and 4, and 
$\Delta \lambda > 3.0$\,\AA\ for the 1200R grism setup used at epoch 3. 
This operation was performed on the flux spectra obtained at each half-wave 
retarder-plate angle. 
The minimum polarization levels across the H$\alpha$ emission core determined 
with different $\Delta \lambda$ are presented in Figure~\ref{Fig_isp_ha}. As 
shown in the right panel, the minimum polarization levels at epoch 3 tend to 
be constant for $3 \leq \Delta \lambda \leq 10$\,\AA. A sharp increase from 
15\,\AA\ and larger implies that the true level of the ISP is represented by the 
polarization binned to 3 to 10\,\AA.

We calculated means of the Stokes $Q$ and $U$ parameters over the range of 
bin sizes that minimise the polarization across the narrow emission peak in 
H$\alpha$, weighting each spectral element by the inverse square of its 
1$\sigma$ error. The arbitrarily chosen bin sizes were 3, 4, 5, 6, 8, and 
10\,\AA. Uncertainties were estimated by adding the associated errors in the 
error-weighted mean, the error-weighted standard deviation, and the 
uncertainty in the 10\,\AA\ measurement in quadrature. The first two terms 
can be found as Equations 17 and 18 (respectively) 
of \citet{Yang_etal_2018_pol}.  In this way, we derived 
$q_{\rm ISP} = -0.14 \pm 0.08$\% and $u_{\rm ISP} = 0.00 \pm 0.10$\%, yielding 
$p_{\rm ISP} = 0.14 \pm 0.08$\%, and 
{\it PA}$_{\rm ISP} = 89\fdg{6}\pm15\fdg{4}$. For the estimation of the 
uncertainties, the covariance matrix of the Stokes parameters has been 
taken into account following Appendix F of \citet{Montier_etal_2015}.

\begin{table*}
\caption{LCO $Bg'Vr'i'$ photometry of SN\,2018evt. 
\label{Table_phot}}
\begin{tiny}
\begin{tabular}{c|cc|cc|cc|cc|cc}
\hline
Phase$^a$ & MJD & $B$ (mag)   & MJD & $V$ (mag)   & MJD & $g'$ (mag)  & MJD & $r'$ (mag)  & MJD & $i'$ (mag)  \\
\hline
124.328 & 58476.328 & 16.967(013) & 58476.332 & 16.362(011) & 58476.336 & 16.588(008) & 58476.344 & 16.178(009) & 58476.348 & 16.308(013) \\
124.328 & 58476.328 & 16.980(013) & 58476.336 & 16.361(012) & 58476.340 & 16.591(009) & 58476.344 & 16.182(010) & 58476.348 & 16.313(014) \\
134.688 & 58486.688 & 16.976(010) & 58486.691 & 16.377(009) & 58485.309 & 16.599(012) & 58485.309 & 16.131(013) & 58485.312 & 16.241(019) \\
134.688 & 58486.688 & 17.018(011) & 58486.691 & 16.382(008) & 58486.691 & 16.613(006) & 58485.312 & 16.145(014) & 58485.316 & 16.244(021) \\
139.688 & 58491.688 & 17.008(010) & 58491.691 & 16.392(017) & 58486.695 & 16.598(007) & 58486.695 & 16.166(007) & 58486.699 & 16.303(009) \\
139.688 & 58491.688 & 16.989(017) & 58491.691 & 16.376(024) & 58491.691 & 16.634(027) & 58486.699 & 16.166(006) & 58486.703 & 16.314(011) \\
144.656 & 58496.656 & 16.994(022) & 58496.660 & 16.396(013) & 58491.695 & 16.625(013) & 58491.695 & 16.245(031) & 58491.699 & 16.332(018) \\
144.660 & 58496.660 & 17.062(018) & 58496.664 & 16.400(015) & 58496.664 & 16.641(010) & 58491.699 & 16.203(025) & 58491.703 & 16.289(013) \\
149.254 & 58501.254 & 17.057(016) & 58501.258 & 16.399(012) & 58496.668 & 16.617(010) & 58496.668 & 16.120(009) & 58496.672 & 16.339(014) \\
149.258 & 58501.258 & 17.043(018) & 58501.262 & 16.426(011) & 58501.262 & 16.623(009) & 58496.672 & 16.161(008) & 58496.672 & 16.317(014) \\
154.398 & 58506.398 & 17.086(036) & 58506.402 & 16.470(024) & 58501.266 & 16.649(009) & 58501.266 & 16.208(009) & 58501.270 & 16.385(013) \\
154.402 & 58506.402 & 17.148(036) & 58506.406 & 16.388(022) & 58506.406 & 16.601(016) & 58501.270 & 16.204(010) & 58501.270 & 16.395(013) \\
163.285 & 58515.285 & 17.062(009) & 58515.289 & 16.457(011) & 58506.410 & 16.617(016) & 58506.410 & 16.192(016) & 58506.414 & 16.374(025) \\
163.289 & 58515.289 & 17.066(010) & 58515.293 & 16.466(011) & 58515.293 & 16.658(022) & 58506.410 & 16.213(015) & 58506.414 & 16.357(025) \\
163.301 & 58515.301 & 17.102(010) & 58515.305 & 16.468(011) & 58515.305 & 16.702(009) & 58515.309 & 16.239(011) & 58515.312 & 16.419(017) \\
163.301 & 58515.301 & 17.104(011) & 58515.305 & 16.500(008) & 58515.309 & 16.694(008) & 58515.312 & 16.233(011) & 58515.316 & 16.423(015) \\
177.207 & 58529.207 & 17.137(009) & 58529.207 & 16.515(009) & 58529.211 & 16.749(007) & 58529.215 & 16.266(008) & 58529.219 & 16.510(013) \\
177.207 & 58529.207 & 17.106(011) & 58529.211 & 16.509(009) & 58529.215 & 16.765(007) & 58529.215 & 16.276(009) & 58529.219 & 16.518(014) \\
190.559 & 58542.559 & 17.210(016) & 58542.562 & 16.653(009) & 58542.566 & 16.834(006) & 58542.570 & 16.335(007) & 58542.570 & 16.611(021) \\
190.559 & 58542.559 & 17.208(014) & 58542.562 & 16.615(009) & 58542.566 & 16.832(007) & 58542.570 & 16.312(008) & 58542.574 & 16.579(025) \\
197.281 & 58549.281 & 17.312(012) & 58549.285 & 16.643(010) & 58549.285 & 16.877(008) & 58549.289 & 16.372(010) & 58549.293 & 16.605(012) \\
197.281 & 58549.281 & 17.311(012) & 58549.285 & 16.635(009) & 58549.289 & 16.869(008) & 58549.293 & 16.351(010) & 58549.293 & 16.635(013) \\
204.855 & 58556.855 & 17.279(019) & 58556.859 & 16.646(014) & 58556.859 & 16.882(014) &  &  &  &  \\
204.855 & 58556.855 & 17.309(018) & 58556.859 & 16.659(014) & 58556.863 & 16.894(014) &  &  &  &  \\
211.004 & 58563.004 & 17.264(027) & 58563.008 & 16.680(021) & 58563.008 & 16.892(019) & 58563.012 & 16.365(020) & 58563.016 & 16.708(027) \\
211.004 & 58563.004 & 17.264(027) & 58563.008 & 16.626(022) & 58563.012 & 16.893(022) & 58563.016 & 16.381(022) & 58563.016 & 16.656(030) \\
217.863 & 58569.863 & 17.362(015) & 58569.867 & 16.718(016) & 58569.867 & 16.975(018) & 58569.871 & 16.435(021) & 58569.875 & 16.724(023) \\
217.863 & 58569.863 & 17.357(014) & 58569.867 & 16.726(018) & 58569.871 & 16.955(016) & 58569.875 & 16.419(016) & 58569.875 & 16.745(020) \\
225.406 & 58577.406 & 17.424(015) &           &             &           &             &           &             &            &  \\
225.410 & 58577.410 & 17.438(015) &           &             &           &             &           &             &            &  \\
227.109 & 58579.109 & 17.460(016) & 58579.109 & 16.829(011) & 58579.113 & 17.057(009) & 58579.117 & 16.519(011) & 58579.121 & 16.826(015) \\
227.109 & 58579.109 & 17.478(014) & 58579.113 & 16.810(011) & 58579.117 & 17.050(008) & 58579.117 & 16.485(009) & 58579.121 & 16.857(013) \\
233.484 & 58585.484 & 17.498(012) & 58585.488 & 16.864(010) & 58585.492 & 17.091(007) & 58585.496 & 16.562(007) & 58585.500 & 16.911(013) \\
233.488 & 58585.488 & 17.520(011) & 58585.492 & 16.851(009) & 58585.496 & 17.083(006) & 58585.496 & 16.564(008) & 58585.500 & 16.886(013) \\
242.098 & 58594.098 & 17.529(022) & 58594.102 & 16.911(016) & 58594.105 & 17.128(015) & 58594.105 & 16.558(011) & 58594.109 & 16.907(018) \\
242.098 & 58594.098 & 17.577(025) & 58594.102 & 16.873(016) & 58594.105 & 17.131(014) & 58594.109 & 16.561(011) & 58594.113 & 16.940(019) \\
244.219 & 58596.219 & 17.591(013) & 58596.223 & 16.904(009) & 58596.227 & 17.137(008) & 58596.230 & 16.591(008) & 58596.234 & 16.971(012) \\
244.219 & 58596.219 & 17.552(013) & 58596.223 & 16.893(011) & 58596.230 & 17.172(008) & 58596.234 & 16.590(007) & 58596.238 & 16.879(012) \\
252.832 & 58604.832 & 17.548(017) & 58604.836 & 16.877(020) & 58604.840 & 17.166(023) & 58604.848 & 16.593(023) & 58604.848 & 16.937(033) \\
252.836 & 58604.836 & 17.578(018) & 58604.840 & 16.960(026) & 58604.844 & 17.173(021) & 58604.848 & 16.636(027) & 58604.852 & 16.956(029) \\
260.566 & 58612.566 & 17.685(009) & 58612.570 & 17.013(009) & 58612.574 & 17.269(009) & 58612.582 & 16.716(010) & 58612.586 & 17.067(013) \\
260.570 & 58612.570 & 17.696(009) & 58612.574 & 17.013(009) & 58612.578 & 17.253(009) & 58612.582 & 16.719(010) & 58612.586 & 17.083(012) \\
269.367 & 58621.367 & 17.755(032) & 58621.371 & 17.137(071) & 58621.379 & 17.304(031) & 58621.379 & 16.740(016) & 58621.383 & 17.116(025) \\
269.367 & 58621.367 & 17.706(029) &           &             &           &             & 58621.383 & 16.754(016) & 58621.387 & 17.140(029) \\
277.508 & 58629.508 & 17.784(009) & 58629.512 & 17.092(009) & 58629.516 & 17.346(008) & 58629.520 & 16.800(008) & 58629.523 & 17.172(013) \\
277.508 & 58629.508 & 17.770(010) & 58629.512 & 17.098(009) & 58629.520 & 17.350(008) & 58629.523 & 16.804(009) & 58629.527 & 17.178(013) \\
285.727 & 58637.727 & 17.788(010) & 58637.734 & 17.151(011) & 58637.738 & 17.382(009) & 58637.742 & 16.795(009) & 58637.746 & 17.186(016) \\
285.730 & 58637.730 & 17.786(009) & 58637.734 & 17.149(011) & 58637.738 & 17.384(010) & 58637.742 & 16.801(009) & 58637.746 & 17.207(015) \\
313.066 & 58665.066 & 17.987(011) & 58665.070 & 17.338(011) & 58665.074 & 17.571(009) & 58665.082 & 17.021(011) & 58665.086 & 17.414(015) \\
313.070 & 58665.070 & 17.980(011) & 58665.074 & 17.335(010) & 58665.078 & 17.574(009) & 58665.082 & 17.027(010) & 58665.086 & 17.422(015) \\
328.719 & 58680.719 & 18.035(024) & 58680.727 & 17.409(019) & 58680.730 & 17.619(014) & 58680.738 & 17.083(013) & 58680.746 & 17.488(019) \\
328.723 & 58680.723 & 18.079(025) & 58680.730 & 17.374(017) & 58680.734 & 17.643(014) & 58680.742 & 17.090(013) & 58680.750 & 17.509(021) \\
346.996 & 58698.996 & 18.200(014) & 58699.004 & 17.535(012) & 58699.012 & 17.754(008) & 58699.020 & 17.233(012) & 58699.023 & 17.672(018) \\
347.000 & 58699.000 & 18.218(013) & 58699.008 & 17.535(012) & 58699.016 & 17.733(009) & 58699.020 & 17.230(010) & 58699.027 & 17.597(018) \\
367.746 & 58719.746 & 18.292(016) & 58719.754 & 17.678(017) & 58719.762 & 17.840(010) & 58719.770 & 17.322(010) & 58719.773 & 17.753(018) \\
367.750 & 58719.750 & 18.287(017) & 58719.758 & 17.672(015) & 58719.766 & 17.861(009) & 58719.770 & 17.314(009) & 58719.777 & 17.803(021) \\
\hline
\end{tabular}\\
{$^a$}{Days after $B$-band maximum on MJD 58352.} 
\end{tiny}
\end{table*}

\begin{table*}
\caption{GROND $grizJHK$ photometry of SN\,2018evt. 
\label{Table_grond_phot}}
\begin{scriptsize}
\begin{tabular}{c|c|cccc|ccc}
\hline
Phase$^a$ & MJD & $g$ (mag)   & $r$ (mag)  & $i$ (mag) & $z$ (mag) & $J$ (mag)  & $H$ (mag) & $K$ (mag) \\
\hline
141.324 & 58493.324 & 16.522(004) & 16.227(002) & 16.322(004) & 15.822(004) & 15.548(029) & 15.208(033) & 14.986(022) \\
152.293 & 58504.293 & 16.569(019) & 16.250(008) & 16.359(008) & 15.842(009) & 15.554(028) & 15.353(038) & 15.029(032) \\
161.340 & 58513.340 & 16.520(010) & 16.203(007) & 16.308(008) & 15.782(009) & 15.549(026) & 15.232(020) & 15.105(036) \\
168.305 & 58520.305 & 16.621(007) & 16.310(005) & 16.420(008) & 15.882(007) & 15.701(021) & 15.303(025) & 15.268(032) \\
174.387 & 58526.387 & 16.648(009) & 16.315(004) & 16.462(006) & 15.889(005) & 15.716(021) & 15.411(025) & 15.165(040) \\
182.352 & 58534.352 & 16.694(015) & 16.350(010) & 16.521(013) & 15.940(009) & 15.787(043) & 15.438(049) & 15.059(056) \\
188.207 & 58540.207 & 16.737(007) & 16.377(004) & 16.558(006) & 15.966(007) & 15.692(037) & 15.603(044) & 15.333(038) \\
196.305 & 58548.305 & 16.770(007) & 16.425(008) & 16.583(010) & 16.011(007) & 15.887(047) & 15.463(045) & 15.371(036) \\
211.215 & 58563.215 & 16.885(019) & 16.495(009) & 16.686(010) & 16.090(009) & 15.947(035) & 15.716(040) & 15.681(041) \\
219.227 & 58571.227 & 16.910(006) & 16.540(005) & 16.730(006) & 16.135(006) & 15.999(029) & 15.778(040) & 15.629(035) \\
228.352 & 58580.352 & 16.973(004) & 16.577(004) & 16.796(005) & 16.171(004) & 16.035(030) & 15.817(045) & 15.699(034) \\
234.211 & 58586.211 & 17.001(007) & 16.622(007) & 16.839(006) & 16.231(007) & 16.018(035) & 15.979(067) & 15.686(038) \\
238.230 & 58590.230 & 17.019(011) & 16.642(010) & 16.861(010) & 16.249(008) &             &             &             \\
256.176 & 58608.176 & 17.132(004) & 16.725(004) & 16.971(005) & 16.344(005) &             &             &             \\
272.094 & 58624.094 & 17.235(006) & 16.820(004) & 17.070(004) & 16.463(004) & 16.274(021) & 16.103(023) & 15.982(032) \\
287.125 & 58639.125 & 17.344(006) & 16.913(008) & 17.186(007) & 16.568(006) & 16.264(050) & 16.252(069) & 15.986(097) \\
297.000 & 58649.000 & 17.380(011) & 16.967(004) & 17.235(006) & 16.623(005) & 16.431(018) & 16.277(020) & 16.208(045) \\
303.027 & 58655.027 & 17.383(006) & 16.976(008) & 17.320(009) & 16.677(005) & 16.354(042) & 16.369(056) & 16.351(102) \\
314.027 & 58666.027 & 17.503(004) & 17.091(001) & 17.375(003) & 16.752(004) & 16.627(025) & 16.505(028) & 16.286(041) \\
324.996 & 58676.996 & 17.570(010) & 17.164(006) & 17.449(008) & 16.848(008) & 16.775(033) & 16.481(034) & 16.220(063) \\
354.012 & 58706.012 & 17.753(006) & 17.360(003) & 17.686(005) & 17.025(004) & 16.839(036) &             & 16.313(046) \\
369.008 & 58721.008 & 17.848(003) & 17.464(003) & 17.777(004) & 17.129(004) & 16.970(026) & 16.734(030) & 16.377(057) \\
\hline
\end{tabular}\\
{$^a$}{Days after $B$-band maximum on MJD 58352.}
\end{scriptsize}
\end{table*}

\begin{table}
\caption{ZTF $g$ and $r$ photometry of SN\,2018evt. 
\label{Table_ztf_phot}}
\begin{scriptsize}
\begin{tabular}{ccccc}
\hline
Phase$^a$ & Filter & MJD & AB Mag & Error \\
\hline
104.563 & $r$ & 58456.563 & 16.61 & 0.16 \\
105.550 & $r$ & 58457.550 & 16.31 & 0.02 \\
105.555 & $r$ & 58457.555 & 16.27 & 0.02 \\
105.561 & $r$ & 58457.561 & 16.37 & 0.03 \\
105.566 & $r$ & 58457.566 & 16.37 & 0.04 \\
112.555 & $r$ & 58464.555 & 16.32 & 0.03 \\
112.560 & $r$ & 58464.560 & 16.30 & 0.03 \\
112.565 & $r$ & 58464.565 & 16.23 & 0.03 \\
112.570 & $r$ & 58464.570 & 16.37 & 0.05 \\
113.558 & $r$ & 58465.558 & 16.31 & 0.03 \\
113.563 & $r$ & 58465.563 & 16.28 & 0.03 \\
113.568 & $r$ & 58465.568 & 16.32 & 0.04 \\
116.540 & $r$ & 58468.540 & 16.24 & 0.02 \\
124.541 & $r$ & 58476.541 & 16.22 & 0.04 \\
129.544 & $r$ & 58481.544 & 16.22 & 0.03 \\
135.498 & $g$ & 58487.498 & 16.58 & 0.03 \\
140.499 & $r$ & 58492.499 & 16.24 & 0.04 \\
140.534 & $g$ & 58492.534 & 16.55 & 0.03 \\
151.502 & $r$ & 58503.502 & 16.29 & 0.06 \\
154.507 & $g$ & 58506.507 & 16.61 & 0.04 \\
160.522 & $r$ & 58512.522 & 16.29 & 0.11 \\
172.436 & $g$ & 58524.436 & 16.64 & 0.04 \\
188.336 & $r$ & 58540.336 & 16.42 & 0.04 \\
188.440 & $g$ & 58540.440 & 16.81 & 0.03 \\
191.481 & $g$ & 58543.481 & 16.83 & 0.05 \\
205.360 & $r$ & 58557.360 & 16.48 & 0.04 \\
208.397 & $r$ & 58560.397 & 16.49 & 0.04 \\
216.335 & $r$ & 58568.335 & 16.59 & 0.10 \\
221.403 & $r$ & 58573.403 & 16.52 & 0.04 \\
230.272 & $r$ & 58582.272 & 16.59 & 0.04 \\
230.353 & $g$ & 58582.353 & 17.03 & 0.05 \\
233.273 & $g$ & 58585.273 & 17.03 & 0.05 \\
233.314 & $r$ & 58585.314 & 16.59 & 0.03 \\
236.314 & $g$ & 58588.314 & 17.13 & 0.08 \\
236.361 & $r$ & 58588.361 & 16.74 & 0.09 \\
242.252 & $g$ & 58594.252 & 17.05 & 0.06 \\
242.335 & $r$ & 58594.335 & 16.65 & 0.05 \\
245.293 & $g$ & 58597.293 & 17.12 & 0.04 \\
245.376 & $r$ & 58597.376 & 16.65 & 0.05 \\
249.232 & $g$ & 58601.232 & 17.12 & 0.04 \\
255.247 & $r$ & 58607.247 & 16.70 & 0.04 \\
255.295 & $g$ & 58607.295 & 17.17 & 0.04 \\
265.196 & $g$ & 58617.196 & 17.21 & 0.04 \\
265.272 & $r$ & 58617.272 & 16.76 & 0.04 \\
280.286 & $r$ & 58632.286 & 16.84 & 0.04 \\
282.233 & $g$ & 58634.233 & 17.31 & 0.05 \\
282.234 & $g$ & 58634.234 & 17.31 & 0.05 \\
282.273 & $r$ & 58634.273 & 16.88 & 0.05 \\
282.286 & $r$ & 58634.286 & 16.85 & 0.04 \\
285.189 & $r$ & 58637.189 & 16.89 & 0.04 \\
285.231 & $g$ & 58637.231 & 17.34 & 0.05 \\
288.190 & $g$ & 58640.190 & 17.35 & 0.05 \\
288.210 & $r$ & 58640.210 & 16.90 & 0.05 \\
291.191 & $g$ & 58643.191 & 17.40 & 0.04 \\
291.232 & $r$ & 58643.232 & 16.91 & 0.05 \\
297.233 & $r$ & 58649.233 & 17.01 & 0.05 \\
300.190 & $g$ & 58652.190 & 17.48 & 0.04 \\
300.274 & $r$ & 58652.274 & 17.02 & 0.05 \\
316.172 & $r$ & 58668.172 & 17.09 & 0.05 \\
316.190 & $g$ & 58668.190 & 17.58 & 0.04 \\
319.206 & $r$ & 58671.206 & 17.12 & 0.04 \\
325.170 & $g$ & 58677.170 & 17.55 & 0.05 \\
325.190 & $r$ & 58677.190 & 17.10 & 0.05 \\
328.192 & $g$ & 58680.192 & 17.63 & 0.07 \\
328.235 & $r$ & 58680.235 & 17.20 & 0.05 \\
\end{tabular}\\
{$^a$}{Days after $B$-band maximum on MJD 58352.} \\
\end{scriptsize}
\end{table}

\begin{table}
\caption{Pseudobolometric luminosities of SN\,2018evt. \label{Table_bolo_18evt}}
\begin{scriptsize}
\begin{tabular}{ccc}
\hline
Phase$^a$ & log\,$L^b$ (Opt--NIR)        & log\,$L^b$ (Opt) \\
(Day)     & [log\,(erg\,s$^{-1}$)] & [log\,(erg\,s$^{-1}$)]  \\
\hline
124.3 & 42.960(021) & 42.791(019) \\
134.7 & 42.948(021) & 42.783(019) \\
139.7 & 42.939(021) & 42.777(019) \\
144.7 & 42.935(021) & 42.776(019) \\
149.3 & 42.927(021) & 42.769(019) \\
154.4 & 42.922(021) & 42.762(020) \\
163.3 & 42.920(021) & 42.757(019) \\
177.2 & 42.889(021) & 42.732(019) \\
190.6 & 42.855(021) & 42.700(019) \\
197.3 & 42.843(021) & 42.686(019) \\
204.9 & 42.824(021) & 42.672(019) \\
211.0 & 42.811(021) & 42.664(019) \\
217.9 & 42.793(021) & 42.648(019) \\
225.4 & 42.776(021) & 42.628(019) \\
227.1 & 42.772(021) & 42.623(019) \\
233.5 & 42.757(021) & 42.605(019) \\
242.1 & 42.737(021) & 42.584(019) \\
244.2 & 42.735(021) & 42.583(019) \\
252.8 & 42.722(021) & 42.572(019) \\
260.6 & 42.701(021) & 42.550(019) \\
269.4 & 42.678(022) & 42.526(020) \\
277.5 & 42.661(021) & 42.509(019) \\
285.7 & 42.649(021) & 42.498(019) \\
313.1 & 42.573(021) & 42.434(019) \\
328.7 & 42.520(021) & 42.378(019) \\
347.0 & 42.486(021) & 42.350(020) \\
367.8 & 42.437(021) & 42.303(020) 
\end{tabular}\\
{$^a$}{Days after $B$-band maximum on MJD 58352.} \\
{$^b$}{Uncertainty in the distance not included.} \\
\end{scriptsize}
\end{table}

\begin{table}
\caption{H$\alpha$ luminosity of SN\,2018evt. \label{Table_ha_18evt}}
\begin{scriptsize}
\begin{tabular}{cc}
\hline
Phase$^a$ & log\,$L_{H\alpha}^b$ \\
(Day)     & [log\,(erg\,s$^{-1}$)]  \\
\hline
124.6  & 41.5331(0192) \\
132.7  & 41.4965(0192) \\
142.6  & 41.5204(0192) \\
152.6  & 41.5194(0194) \\
172.3  & 41.5897(0191) \\
194.4  & 41.5483(0191) \\
195.3  & 41.5345(0191) \\
207.5  & 41.5388(0192) \\
219.2  & 41.5371(0190) \\
220.5  & 41.5171(0190) \\
244.7  & 41.4835(0191) \\
262.4  & 41.4351(0190) \\
291.4  & 41.3919(0191) \\
327.3  & 41.3149(0192) \\
365.4  & 41.2207(0192) \\
\end{tabular}\\
{$^a$}{Days after $B$-band maximum on MJD 58352.} \\
{$^b$}{Uncertainty in the distance not included.} \\
\end{scriptsize}
\end{table}


\bsp	
\label{lastpage}
\end{document}